\documentclass{ws-ijmpc}
\usepackage{float}
\usepackage{graphicx}
\usepackage{caption}
\usepackage{subcaption}
\usepackage[super]{cite} 
\usepackage[breaklinks]{hyperref}
\hypersetup{colorlinks,urlcolor=black,citecolor=black,linkcolor=black,filecolor=black}
\usepackage{breakurl}
\usepackage{anysize}
\marginsize{3cm}{3cm}{3cm}{3cm}
\begin{document}
	
\markboth{Sourav Chowdhury , Suparna Roychowdhury \& Indranath Chaudhuri}
{Universality and herd immunity threshold : Revisiting the SIR model
	for COVID-19}

\catchline{}{}{}{}{}

\title{{Universality and herd immunity threshold : Revisiting the SIR model
		for COVID-19}
}

\author{Sourav Chowdhury\thanks{email: chowdhury95sourav@gmail.com}\qquad Suparna Roychowdhury\thanks{email: suparna@sxccal.edu}\qquad Indranath Chaudhuri\thanks{email: indranath@sxccal.edu}}

\address{Department of Physics, St. Xavier's College (Autonomous)\\
30 Mother Teresa Sarani, Kolkata-700016, West Bengal, India}

\maketitle

\begin{history}
\received{Day Month Year}
\revised{Day Month Year}
\end{history}

\begin{abstract}
	COVID-19 pandemic has been raging all around the world for almost
	a year now, as of November 1, 2020. In this paper, we try to analyze the variation of the
	COVID-19 pandemic in different countries in the light of some
	modifications to the SIR model. The SIR model was modified by taking
	time-dependent rate parameters. From this modified SIR model, the
	basic reproduction number, effective reproduction number, herd immunity,
	and herd immunity threshold is redefined. The re-outbreak of the
	COVID-19 is a real threat to various countries. We have used
	the above-mentioned quantities to find the reasons behind the re-outbreak
	of this disease. Also, the effectiveness of herd immunity to prevent
	an epidemic has been analyzed with respect to this model. We have also
	tried to show that there are certain universal aspects in the spread
	and containment of the disease in various countries for a short period of time. 
	Finally, we have also analyzed the current pandemic situation in India and have attempted to 
	discuss the possibilities in order to predict its future behavior using our model.

\end{abstract}

\ccode{PACS Nos.:}

\section{Introduction}

	Nowadays, the COVID-19 pandemic is spreading throughout various countries. Due to this pandemic, a total of 213 countries have been affected, 47 million people are already infected and 1.2 million people have died worldwide, as of November 1, 2020.\cite{worldometers}
	Also in India, a total of 8.2 million people have been infected and 0.12 million people died, during this period. \cite{worldometers}. The worrying thing is that these numbers are still increasing on a daily basis. Also, some countries like Italy, Russia, Germany have faced a re-outbreak of this disease. COVID-19 pandemic is not only affecting public health but also affecting social health, world economy, and countries' economy. Therefore, it is very
	important to understand the behavior of this pandemic to prevent such widespread damage. 
	
	In this article, we mainly take the basic susceptible-infected-recovered (SIR) model \cite{basic_SIR}.
	SIR model is a well-known and widely
	used model. The SIR model and its successors like
	SEIR, SIRS, SEIRD, SEIAR have been used before in various epidemics like influenza, dengue, Ebola \cite{SEIR_ebola, Ebola_afr, SIRS_influenza, Rep_No_Influenza, SEIAR_influenza, SIR_dengue, cont_bio_model_measles}. These models are also being used to predict and understand the behavior of the COVID-19 pandemic.\cite{SIR_COVID, Time_dep_SIR_Undetected, SEIR_nonlin, SIRS_m, SIRS_Ro, SEIRD_covid, Dickman, SEIAR_basic, SEIAR_m} There are already various papers that are present
	on this topic,\cite{SIR_COVID, Time_dep_SIR_Undetected, SEIR_nonlin, SIRS_m, SIRS_Ro, SEIRD_covid, Dickman, SEIAR_basic, SEIAR_m, ESTRADA20201, perra2020nonpharmaceutical,Akamatsu2020.05.19.20107524,Subir_Das}
	still we have tried to understand this pandemic from a different point of view.
	
	Here, we modify the basic SIR model by taking the time-dependent rate
	parameters, (i) Infection rate $(\beta)$, (ii) Removed
	rate $(\gamma)$. From this modified SIR model we re-define
	the four important quantities, which are, (i) Basic reproduction number
	($R_{0}$), (ii) Effective reproduction number ($R_{i}$), (iii) Herd
	immunity of the population ($HI_{i}$), (iv) Herd immunity threshold
	of the population ($HIT$). 
	
	The herd immunity of a population ($HI_{i}$) is the fraction of immune
	people, against a disease, present in there. Herd immunity
	of a population is a key to decrease an epidemic. Herd immunity can
	increase in two different ways, (i) naturally, by increasing the number
	of infected people, and (ii) artificially, by vaccines.
	
	Nowadays when an epidemic starts in a country, the administration
	of that country usually intervenes and imposes various restrictions in the form of partial or full lockdowns as in the COVID-19 case. As a result
	of this, the herd immunity of the population is forced to remain low.
	 Despite the low herd immunity, it is also seen that, as in the case of COVID-19, the number of infected cases reached some peak value and then started to fall off in many countries. This requires a different explanation from the ones that have been previously proposed.
	We have tried to explain this aspect with the modified SIR model and have used the data of COVID-19 from various countries
	to verify them. The data used for this analysis is taken
	from ``The Humanitarian Data Exchange'' \cite{Covid_data}. 
	
	However, such low herd immunity of a population can bring a resurgence to the epidemic. This phenomenon is often
	termed in literature as reinfection or re-outbreak. Various authors have addressed
	this issue earlier\cite{Hasty_red_lockdown,time_var_rep_no}.
	Here, we have tried to discuss this fact with the help of
	Effective reproduction number ($R_{i}$), Herd immunity of the population
	($HI_{i}$), and Herd immunity threshold of the population ($HIT$). 
	
	As we have mentioned earlier, governments 
	resort to several methods to mitigate a pandemic-like situation, as seen from the present COVID-19 situation.
	Thus the behavior and the dynamics of the COVID-19
	pandemic could have common features in different countries. We have
	tried to address this aspect by searching for universal features in
	the parameters of our model.
	
	The paper is organized as follows: Section 2 deals with a discussion of the basic and modified SIR models along with the reproduction numbers, herd immunity, and herd immunity threshold. Section 3 consists of various simulations of the modified SIR model by considering imaginary parameter values and conditions. In section 4 we try to find the time variation of the model parameters in light of the COVID-19 data for several countries. In section 5 we present a brief study of the COVID-19 behavior in India and the last section, we present our concluding remarks.

\section{Model description}

Here we discuss a simple epidemic model which is known as SIR model.\cite{basic_SIR,keeling2011modeling} If we assume that the population is homogeneously mixed then the model equations can be written as,

\begin{equation}
	\frac{dS}{dt}=-\beta\frac{S}{N}I\qquad S(0)=S_{0}\label{eq1}
\end{equation}

\begin{equation}
	\frac{dI}{dt}=\beta\frac{S}{N}I-\gamma I\qquad I(0)=I_{0}\label{eq2}
\end{equation}

\begin{equation}
	\frac{dR}{dt}=\gamma I\qquad R(0)=0\label{eq3}
\end{equation}

$S(t)=$ Number of susceptible people at time $t$.

$I(t)=$ Number of active cases or the number of infectious
people at time $t$.

$R(t)=$ Number of removed people at time $t$. Removed
persons are those who were previously infected but now recovered or
dead.

$N=$ Total population in a community or a region or a country. 

$\beta=$ Infection rate.

$\gamma=$ Removed  rate of the infected people.

Hence, $S(t)+I(t)+R(t)=N$= Total
number of people in a population.

$I_{tot}(t)$ denotes the total number or the cumulative number of infected cases at time $t$. So, $I_{tot}(t)=I(t)+R(t)$.
Hence the rate of increase of $I_{tot}(t)$ can be written
as,

\begin{equation}
\frac{dI_{tot}}{dt}=\frac{dI}{dt}+\frac{dR}{dt}=\beta\frac{S}{N}I\label{eq4}
\end{equation}

Also, we know that $S(t)=N-I(t)-R(t)=N-I_{tot}(t)$.
Hence,

\begin{equation}
	\frac{dI_{tot}}{dt}=\beta I\left(1-\frac{I_{tot}}{N}\right)\label{eq5}
\end{equation}

This equation looks like the logistic equation of population growth.
However, the main difference is that the increment rate of the total
infected cases depends on the number of active cases rather than on
itself.

 An epidemic will start to spread if the number of infected people in
	a population starts to increase with time, which means,
	\begin{equation}
		\frac{dI}{dt}>0\label{eq2a}.
	\end{equation}
	Hence, from equation \ref{eq2} we can write,
	\begin{equation}
		\frac{dI}{dt}=\beta\frac{S}{N}I-\gamma I>0\label{eq2b}.
	\end{equation}
	Similarly, an epidemic will start to decrease when,
	\begin{equation}
		\frac{dI}{dt}=\beta\frac{S}{N}I-\gamma I<0\label{eq2c}.
	\end{equation}
	Also, at the peak of an epidemic,
	\begin{equation}
		\frac{dI}{dt}=\beta\frac{S}{N}I-\gamma I=0\label{eq2d}.
\end{equation}

\subsection{Reproduction number}

The reproduction number tells us about the average number of susceptible
people who can be infected by an infectious individual. 

Basic reproduction number $R_{0}$ is a very important quantity that tells us whether an epidemic will start or not.
$R_{0}$ can be defined as the number
of secondary infections due to a single primary infected individual
at the initial time of an epidemic. 

If at any time $t$, $S(t)$ and $I(t)$ are
the numbers of susceptible and infected people respectively then the
number of people who will be infected newly is $\beta\frac{S}{N}I$
per unit time. Now initially the number of infected people in
a population is very less, so we can assume $S(0)\approx N$.
Hence initially an infected people can infect $\beta$ number of susceptible
people per unit time. The removed rate is $\gamma$, hence on average,
an infected person will remain infected for $\frac{1}{\gamma}$ unit
of time. So, in this period, an infected person can spread the disease
to $\frac{\beta}{\gamma}$ number of people. Hence this is the number
of secondary infections due to a single infected individual. Thus
$R_{0}=\frac{\beta}{\gamma}$.

So, from equation \ref{eq2b} we can say that initially, an epidemic will start to grow if, 
\begin{equation}
\frac{\gamma}{\beta}<\frac{S(0)}{N}\qquad(I\neq0)\label{eq6}
\end{equation}

or,
\begin{equation}
\frac{\beta}{\gamma}>\frac{N}{S(0)}\approx1\label{eq7}
\end{equation}

or,
\begin{equation}
R_{0}=\frac{\beta}{\gamma}>1\label{eq8}
\end{equation}

So, if $R_{0}>1$ then a disease will start to spread, and if $R_{0}<1$then
the spread of the disease will die down.

$R_{0}$ describes only the initial behavior but not the late-time behavior of an epidemic. 
So, we introduce another important quantity which is known as
the instantaneous reproduction number or effective reproduction number
and denoted as, $R_{i}(t)$. At any time $t$, the number
of susceptible people can be infected by a single infected individual
is $\beta\frac{S}{N}$ per unit time. Hence on average, a single infected
individual can infect a total of $\frac{\beta}{\gamma}\frac{S}{N}$
number of people until that individual is removed. Thus, 
\begin{equation}
R_{i}(t)=\frac{\beta}{\gamma}\frac{S(t)}{N}\label{eq9}
\end{equation}

Now if we assume that the infection rate $\beta$ and recovery rate
$\gamma$ are constants then, $R_{0}=\frac{\beta}{\gamma}$ and $R_{i}(t)=R_{0}\frac{S(t)}{N}$.

As, $S(t)\leq N$ , so, $R_{i}(t)\leq R_{0}$.
So as time goes by, $R_{i}(t)$ decreases from $R_{0}$. 

Like the previous way, we can find that an epidemic will increase
if $R_{i}(t)>1$ and will decrease if $R_{i}(t)<1$.

At the peak of an epidemic, the effective reproduction number,
	$R_{i}(t)=1$ (from equation \ref{eq2d}).

\subsection{Herd immunity}

Immune people in a population are those who are currently infected
($I$) and who were infected, which means the removed people ($R$).
So, at any time $t$ total number of immune people in a population
is $I(t)+R(t)=I_{tot}(t)$. 

Herd immunity in a population is defined as the fraction of the population
which is immune to a disease.\cite{herd_immunity01,herdimmunity02} So, to stop the spread of the disease
there have to be a minimum fraction of immune people in a population
and which is called the "Herd immunity threshold ($HIT$)''.

From equation \ref{eq2c}, we can say that an epidemic will start to decrease when,
\begin{equation}
	\frac{\gamma}{\beta}>\frac{S}{N}\label{eq11}
\end{equation}
. 

Now fraction of immune people at time $t$ is 
\begin{equation}
\frac{I(t)+R(t)}{N}=\frac{N-S(t)}{N}=HI_{i}(t)\label{eq12}
\end{equation}

$HI_{i}(t)$ is the instantaneous herd immunity of a population
at any time $t$. So from equation \ref{eq11}, we can write,

\begin{equation}
1-\frac{\gamma}{\beta}<1-\frac{S}{N}=HI_{i}(t)\label{eq13}
\end{equation}

Hence an epidemic will stop if,

\begin{equation}
HI_{i}(t)>1-\frac{\gamma}{\beta}=HIT\label{eq14}
\end{equation}

Hence herd immunity threshold defined as,

\begin{equation}
HIT=1-\frac{\gamma}{\beta}
\end{equation}

If $\beta$ and $\gamma$ are constant of time then we can write,

\begin{equation}
HIT=1-\frac{\gamma}{\beta}=1-\frac{1}{R_{0}}\label{eq15}
\end{equation}

So, if $HI_{i}(t)<HIT$ then the epidemic will increase
and will decrease if, $HI_{i}(t)>HIT$. 

So, at the peak of an epidemic, $HI_{i}(t)=HIT$ (from equation \ref{eq2d}).

\subsection{Modifications to the SIR model}

In this section, we present certain modifications to the SIR model. The spread of an epidemic depends on the infectivity of a disease which is an universal property of the virus. However, it also depends on the social behavior and social interaction of a region, which can vary with time. Also, when
an epidemic starts, governments in those regions try to contain the
spread of the disease by locking that region down or part of it, making rules to prevent infections (like maintaining social
distancing, wearing masks, and spreading awareness among the people).
Hence the rate of infection $\beta$ will not be constant always.
Here we assume that the rate of infection $\beta$ is a time-dependent
quantity.

Also removed rate $\gamma$ can be time-dependent. It is mentioned earlier that increment of the removed rate 
depends not only on the recovery of infected people, also on the death rate or fatality rate of the disease.  
So, in general, the removed rate can be changed if the recovered rate or death rate is changed.
Initially, knowledge among the doctors and scientists is limited
for a new disease. But as time goes by the knowledge about that disease
will increase and this may help to improve the recovered rate. 
On the other hand, there are factors such as medical infrastructure, infectivity of the disease etc.
which directly affect the recovered rate and the death rate. Thus it is difficult to disentangle these
two effects in the current purview of our model and it would be only possible to predict the total increment of the removed rate as a function of both.
We would be interested to extend this model and study these two effects separately in future to
 better predict the behaviors of the deceased rate and the recovered rate.

Due to these modifications, the relation of basic reproduction number becomes $R_{0}=\frac{\beta(0)}{\gamma(0)}$
because the basic reproduction number is a quantity that is measured
initially. Also, the instantaneous reproduction number or effective
reproduction number is given by, $R_{i}(t)=\frac{\beta(t)}{\gamma(t)}\frac{S(t)}{N}$.

As we discussed in subsection 2.1, an epidemic will increase if $R_{i}(t)>1$
and will decrease if $R_{i}(t)<1$. At the peak of the
epidemic $R_{i}(t)=1$.

The instantaneous herd immunity of a population, $HI_{i}(t)$
is given by, $HI_{i}(t)=1-\frac{S(t)}{N}$.
Now, for the basic SIR model, herd immunity threshold ($HIT$) is a constant
in time since $\beta$ and $\gamma$ are constant quantities.
But here herd immunity threshold is a time-dependent quantity and
is given by, $HIT(t)=1-\frac{\gamma(t)}{\beta(t)}$.
Hence the maximum possible value of herd immunity threshold is $HIT_{max}=1-\left(\frac{\gamma\left(t\right)}{\beta\left(t\right)}\right)_{min}$.

As we discussed in subsection 2.2, an epidemic will increase if $HI_{i}(t)<HIT(t)$
and will decrease if $HI_{i}(t)>HIT(t)$. At the peak of
the epidemic $HI_{i}(t)=HIT(t)$.

From the basic SIR model, we can describe only one way to prevent
an epidemic in terms of herd immunity and herd immunity threshold.
This is,
\begin{itemize}
	\item An epidemic can be prevented by increasing herd immunity ($HI_{i}(t)$)
	of the population than the herd immunity threshold value ($HIT$).
\end{itemize}
However, in the modified SIR model, the herd immunity threshold is
a time-dependent quantity. So, we can describe two different ways
to prevent an epidemic. The first way is same as the previous one. The second way is,
\begin{itemize}
	\item An epidemic can be prevented by decreasing the herd immunity threshold
	($HIT(t)$) to the current herd immunity of the population ($HI_{i}(t)$).
\end{itemize}
Herd immunity threshold ($HIT$) can be reduced by decreasing the
infection rate ($\beta(t)$) or increasing the removed rate ($\gamma(t)$).

\subsection{Estimating $\beta(t)$ and $\gamma(t)$ from
	the SIR model}

From equation \ref{eq2} and equation \ref{eq3} we can write, 

\begin{equation}
	\beta\left(t\right)=\frac{N}{I\left(N-I-R\right)}\left(\frac{dI}{dt}+\frac{dR}{dt}\right)\label{eq17}
\end{equation}

where, 

$I$ = Number of infected or active cases at time $t$.

$R$ = Number of removed cases at time $t$.

From equation \ref{eq3} we can write,

\begin{equation}
	\gamma(t)=\frac{1}{I}\frac{dR}{dt}\label{eq18}
\end{equation}

Hence, from equations \ref{eq17} and \ref{eq18}, we can estimate
$\beta(t)$ and $\gamma(t)$. From these time
dependent rate parameters, we can easily estimate the variation of $HIT(t)$
and $R_{i}(t)$.

\section{A thought experiment}

Let a region initially contain a total of $N=10000$ people where a disease
starts to spread with infection rate $\beta=0.7\ \textrm{da\ensuremath{y^{-1}}}$.
Also, let the removed rate for this disease be $\gamma=0.15\  \textrm{da\ensuremath{y^{-1}}}$.
Hence, the basic reproduction number $R_{0}$ has a value, $R_{0}\approx4.6667>1$,
implying an epidemic will start to grow in this region.

\begin{figure}[h]
	\begin{tabular}{cc}
		\includegraphics[scale=0.26]{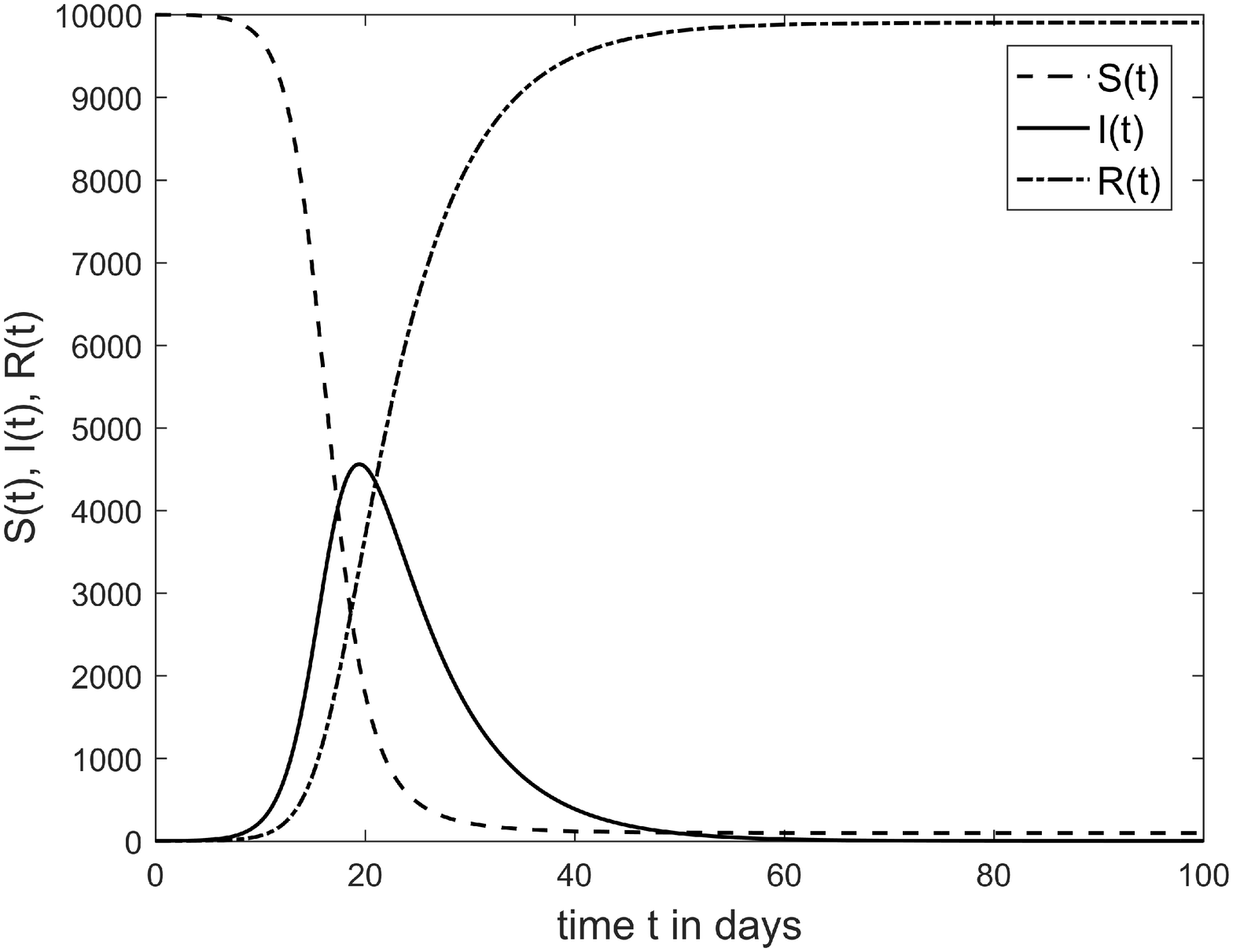}&
		\includegraphics[scale=0.26]{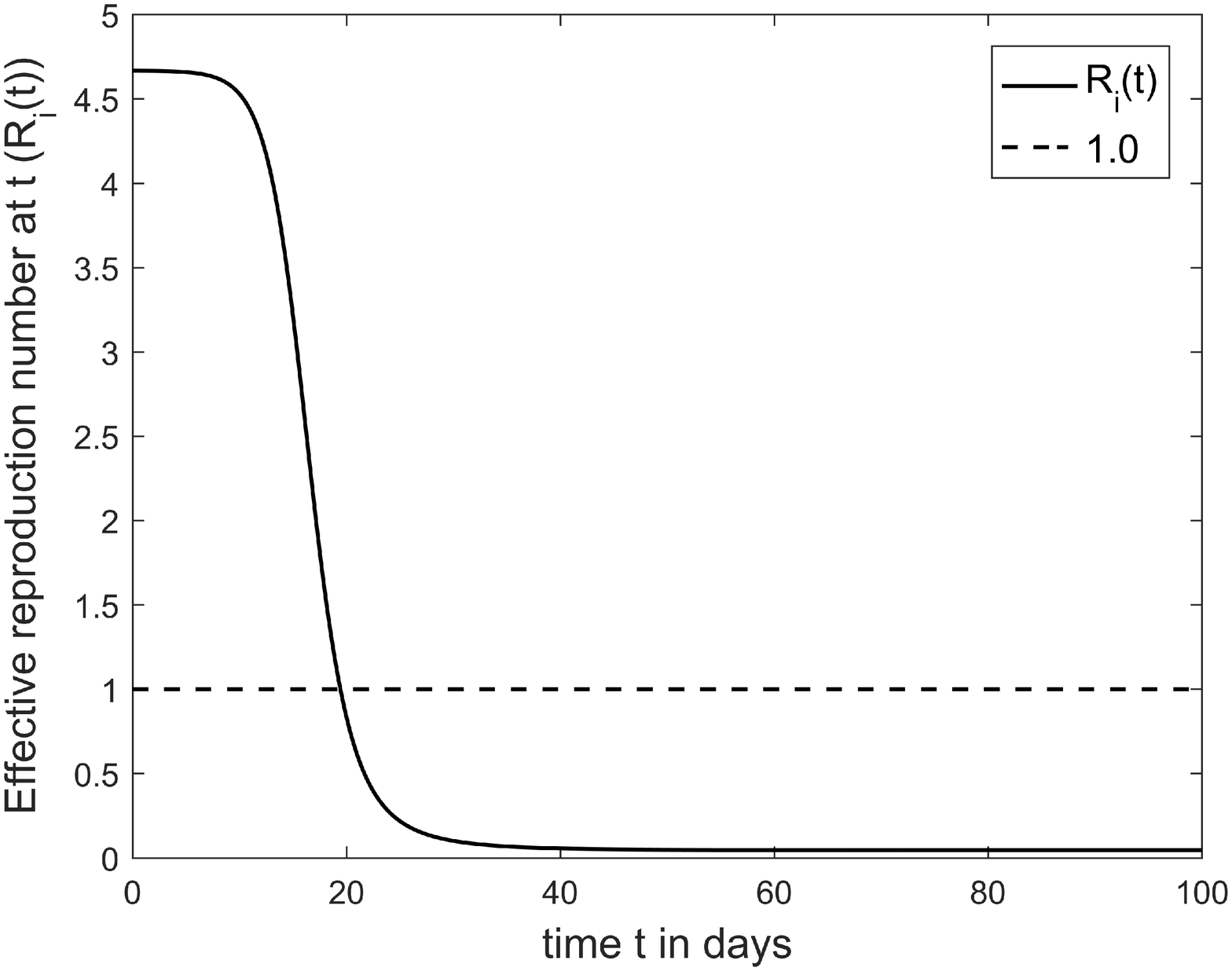}
	\end{tabular}
\centerline{\includegraphics[scale=0.28]{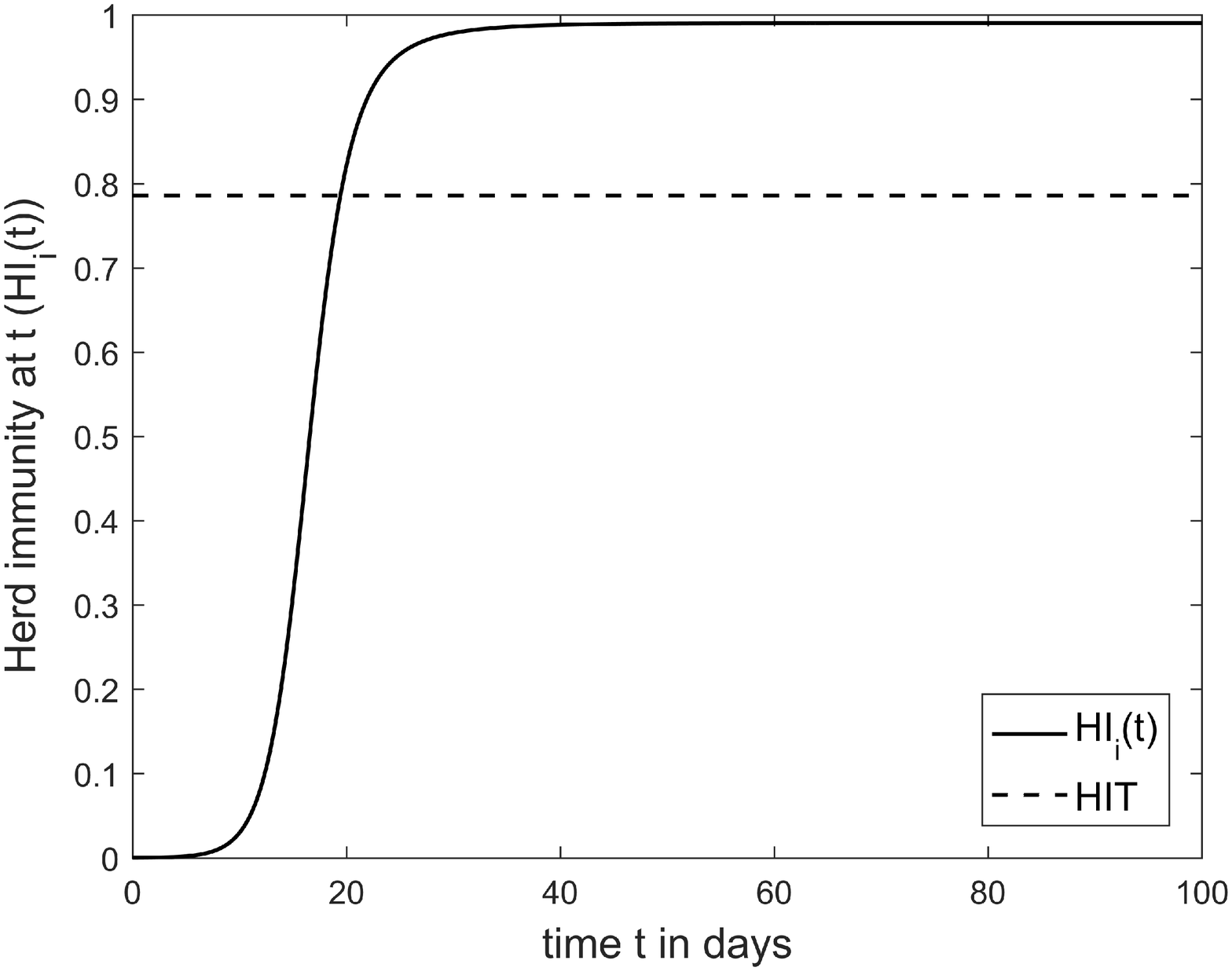}}
\caption{{\small Simulation of the SIR model. The left picture of row one shows
how $S(t)$, $I(t)$, and $R(t)$
is varying with time. The right picture of row one shows how effective reproduction
number $R_{i}(t)$ varies with time. The picture of row two
shows us the variation of herd immunity $HI_{i}(t)$ with
time.}\label{fig1}}
\end{figure}

Let there initially be
 only one infected person in the region, which
means $I(0)=1$. Thus, the initial number of 
susceptible people is $S(0)=N-I(0)=9999$ and removed people is $R(0)=0$.
 If we simulate the SIR system with the above-mentioned values of the parameters and the initial
conditions, we get the results which are shown in Figure \ref{fig1}. 

In this system, the herd immunity threshold has value $HIT$= 0.7857.
Thus, when 78.57\% of people of this population are immune from the disease, the epidemic will start to decrease. From figure
\ref{fig1}, we further see that the effective reproduction number
$R_{i}(t)$ is reduced from $R_{0}$ to unity, and then
it is reduced further. Similarly, we can see that herd immunity $HI_{i}(t)$
is increased from zero to $HIT$, and then it is increased further and has
a value greater than $HIT$. Recalling that an epidemic will start
to decrease after $I(t)$ attains the peak (as previously
discussed). Let, at $t=t_{p}$,
$I(t)$ attains its peak. So, from figure \ref{fig1}
we can see that at $t_{p}$, $R_{i}(t_{p})=1.0$ and $HI_{i}(t_{p})=HIT=0.7857$.
Here, $t_{p}=19.47$ days. Hence, approximately 20 days are needed
to start to decrease the spread of the disease. At the end of the
epidemic, which means, at $t\rightarrow\infty$, the number of susceptible
persons left in the population is $S(\infty)\approx97$.
This is very small compared to the whole population. Also at, $t\rightarrow\infty$, $R_{i}(\infty)=0.0457$ and $HI_{i}(\infty)=0.9902$.
Hence, at the end of the epidemic, the herd immunity of the population
is increased than the $HIT$ which has the value 0.7857 and
the effective reproduction number reduced from one. Thus,
there is no chance of a re-outbreak in this population.

So, we can see that almost all of the people are infected once during
this epidemic. If the fatality rate of this disease is very low then technically fewer people will die. But the problem is that the large number of infected cases will create pressure on the medical facilities, which can make the situation worse.
Thus many people can die because of not getting proper treatment. But if the fatality rate is large then the situation will be very grave. In this situation, the
government and local authorities will intervene and will put some
restrictions and safety rules to prevent the epidemic. Due to these
restrictions and rules, the infection rate will decrease and thus
fewer people will be infected during this epidemic.

Let, at time $t=t_{I}=10$ government starts to intervene. Here we
consider two cases of the government interventions,
\begin{enumerate}
\item Intervention-1 : At, $t_{I}=10$, $\beta$ decreases from $0.7\ \textrm{da\ensuremath{y^{-1}}}$
to $0.25\ \textrm{da\ensuremath{y^{-1}}}$. Means, $\beta(t)=0.7\ \textrm{da\ensuremath{y^{-1}}}$
when $t<t_{I}$ and $\beta(t)=0.25\ \textrm{da\ensuremath{y^{-1}}}$
when $t\geq t_{I}$.
\item Intervention-2 : At, $t_{I}=10$, $\beta$ decreases from $0.7\ \textrm{da\ensuremath{y^{-1}}}$
to $0.35\ \textrm{da\ensuremath{y^{-1}}}$. Means, $\beta(t)=0.7\ \textrm{da\ensuremath{y^{-1}}}$
when $t<t_{I}$ and $\beta(t)=0.35\ \textrm{da\ensuremath{y^{-1}}}$
when $t\geq t_{I}$.
\end{enumerate}
\begin{figure}[h]
	\begin{tabular}{cc}
		\includegraphics[scale=0.26]{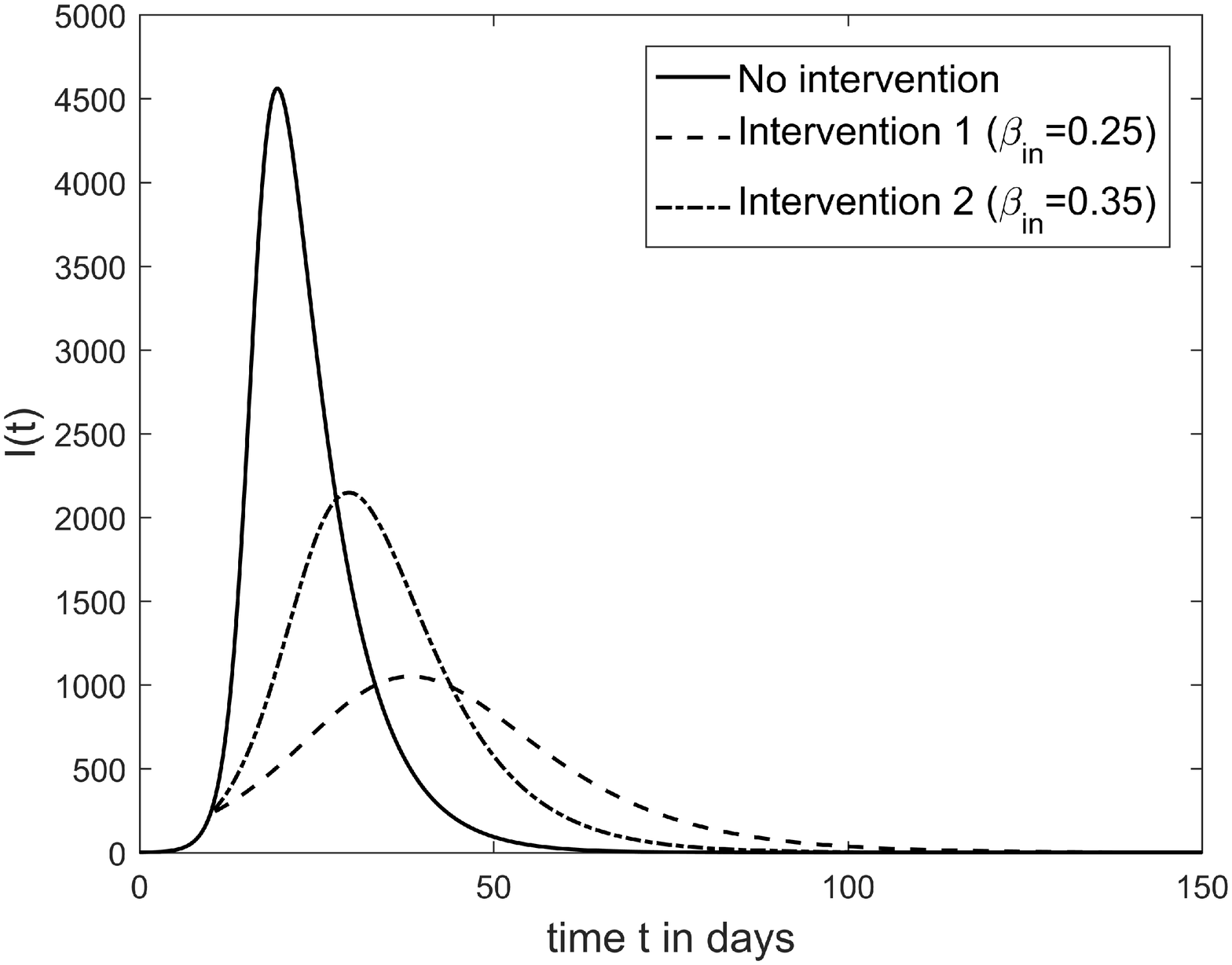}&
		\includegraphics[scale=0.26]{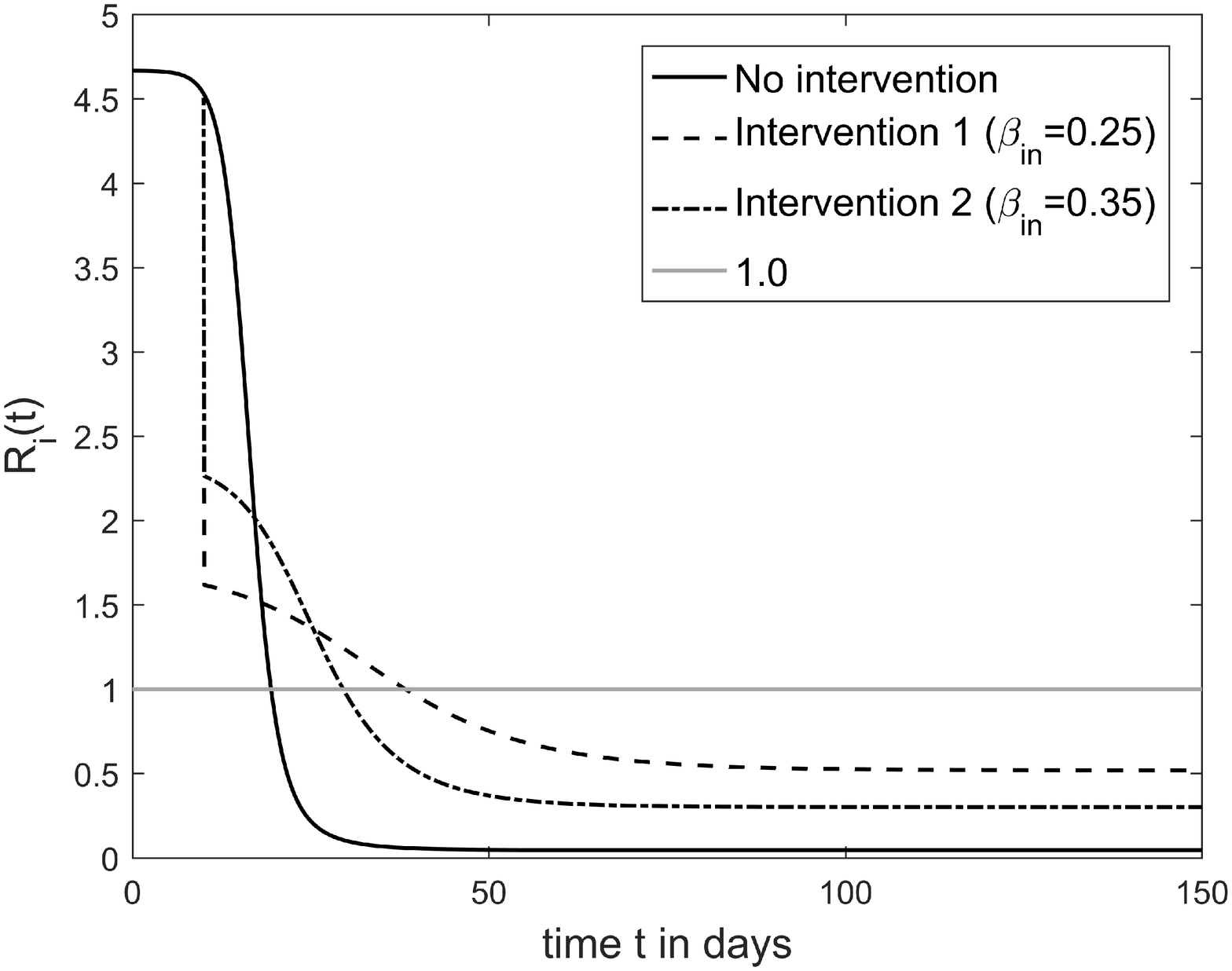}
	\end{tabular}	
\vspace*{8pt}
\caption{{\small Plots for the number of active cases $I(t)$ and the effective
reproduction number $R_{i}(t)$ for different interventions.
The left plot is $I(t)$ vs. $t$ plot for three different
interventions. The right plot is $R_{i}(t)$ vs. $t$ plot
for three different models for interventions.}\label{fig2}}
\end{figure}

From the $I(t)$ vs. $t$ plot of figure \ref{fig2} we
can see that due to the government's intervention, the peaks are flattened
than the no intervention case. Also, it is seen in figure \ref{fig2}
that due to government's intervention at $t=t_{I}=10$, the curve
of $R_{i}(t)$ decreases fast as compared to no intervention.

We found that at the end of the epidemic, the number of susceptible
people left in the population is approximately 3111 and 1290 for intervention-1
and intervention-2 respectively. So, interventions help to decrease
the total number of infections. Also from figure \ref{fig2}, we
can see that if the intervention is more strict then the epidemic
will take more time to reach the peak.

\begin{figure}[h]
\begin{tabular}{cc}
	\includegraphics[scale=0.26]{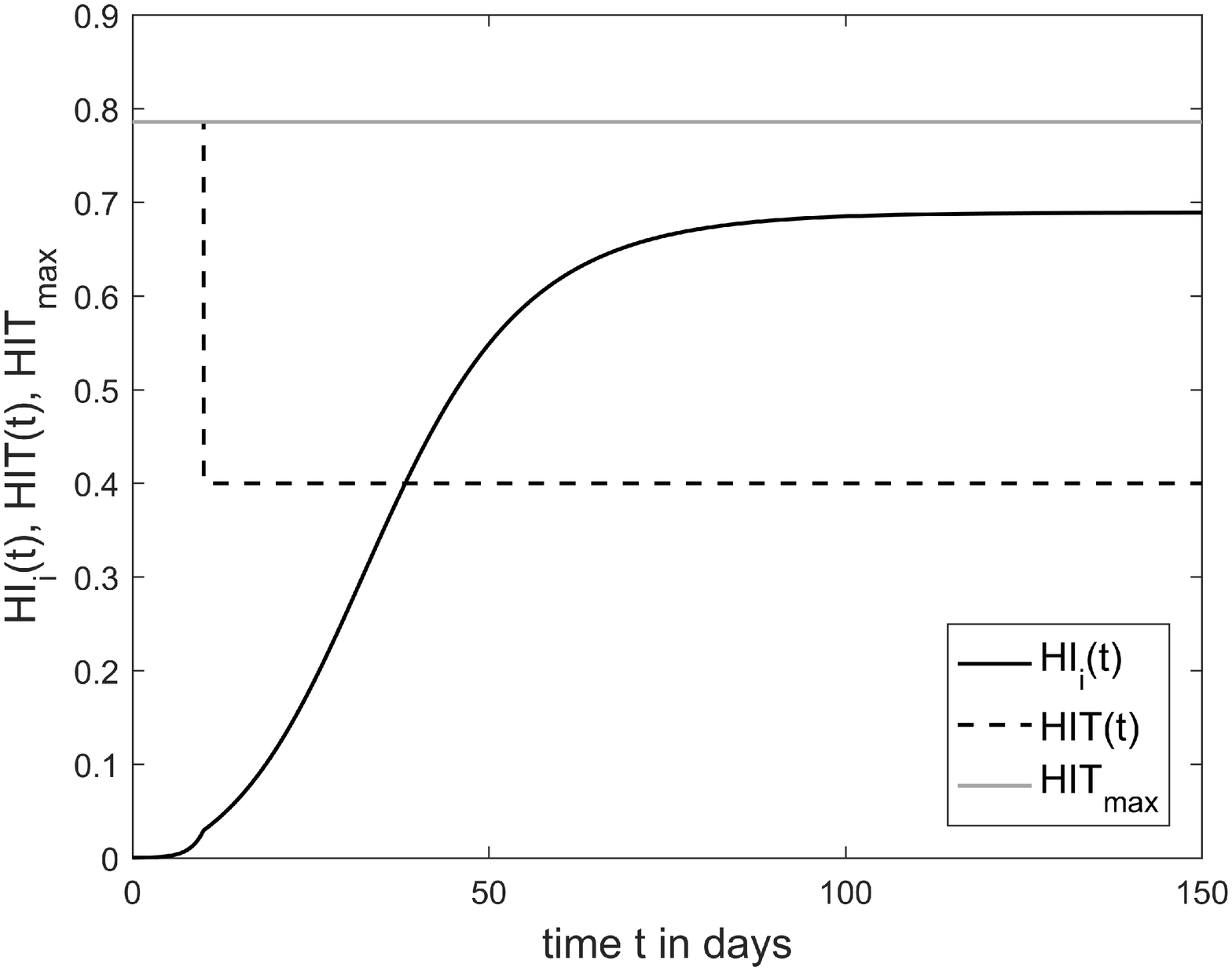}&
	\includegraphics[scale=0.26]{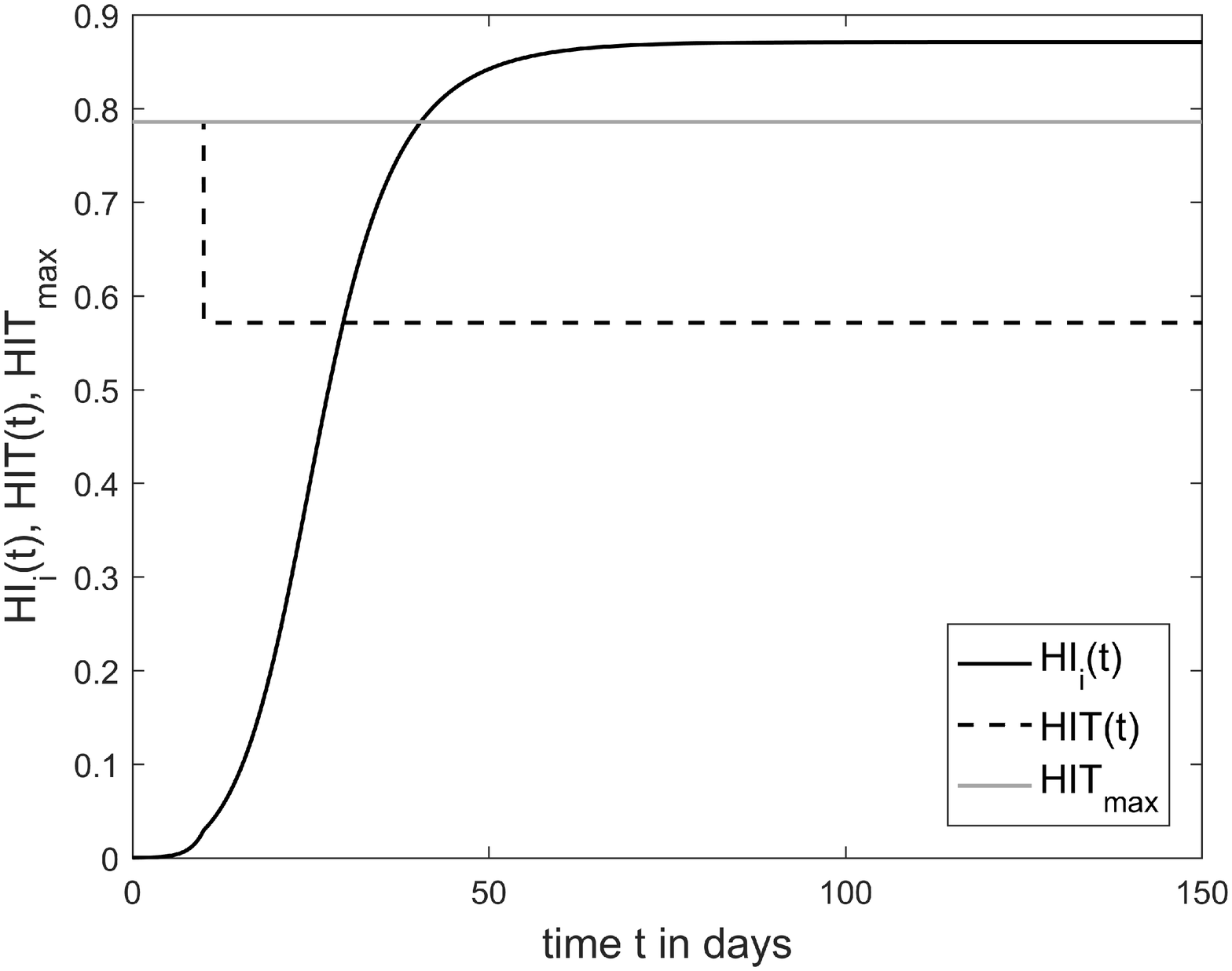}
\end{tabular}	
\vspace*{8pt}
\caption{{\small The plot of the variation of herd immunity $HI_{i}(t)$
with time for intervention-1 (left) and intervention-2 (right)}.\label{fig3}}
\end{figure}

Figure \ref{fig3} shows the variation of herd immunity $HI_{i}(t)$
for the two interventions. But we can see that there are some basic
differences between these herd immunity plots. For both cases,
the maximum value of herd immunity threshold is $HIT_{max}=0.7857$.
Due to the intervention, the herd immunity threshold value is reduced
from the maximum value to 0.4 and 0.5714 for intervention-1 and intervention-2
respectively. Here, the reduced value of the herd immunity threshold due to the intervention is represented by $HIT_{in}$. 
As, the difference between the $HIT_{max}$ and $HIT_{in2}$ for intervention-2
is small, thus the herd immunity at the end of the epidemic ($HI_{i}(\infty)$)
not just goes beyond the $HIT_{in2}$ also it crosses $HIT_{max}$.
But for intervention-1, the herd immunity of the population crosses
$HIT_{in1}$ but could not able to cross the $HIT_{max}$. 

\begin{figure}[h]
	\begin{tabular}{cc}
		\includegraphics[scale=0.26]{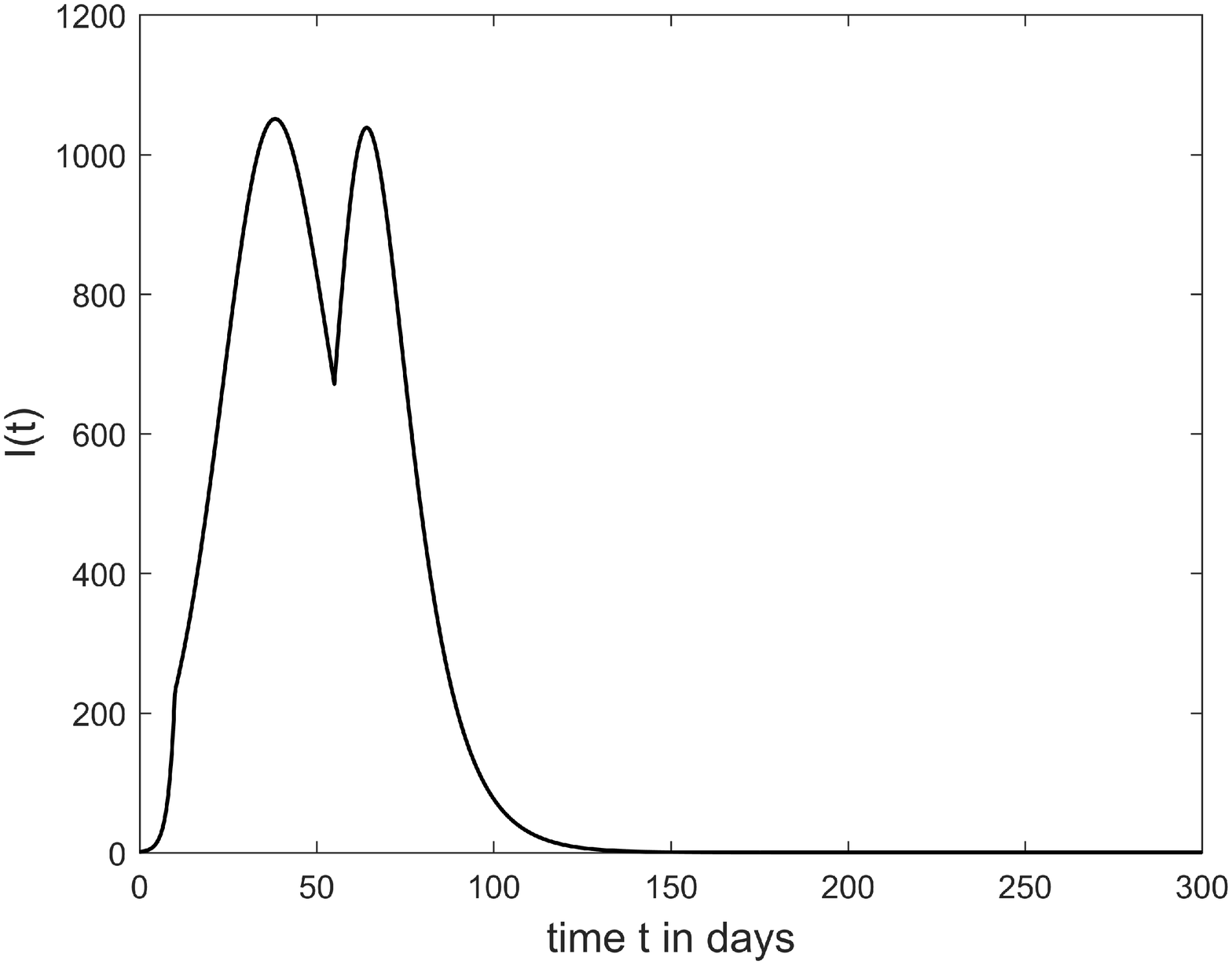}&
		\includegraphics[scale=0.26]{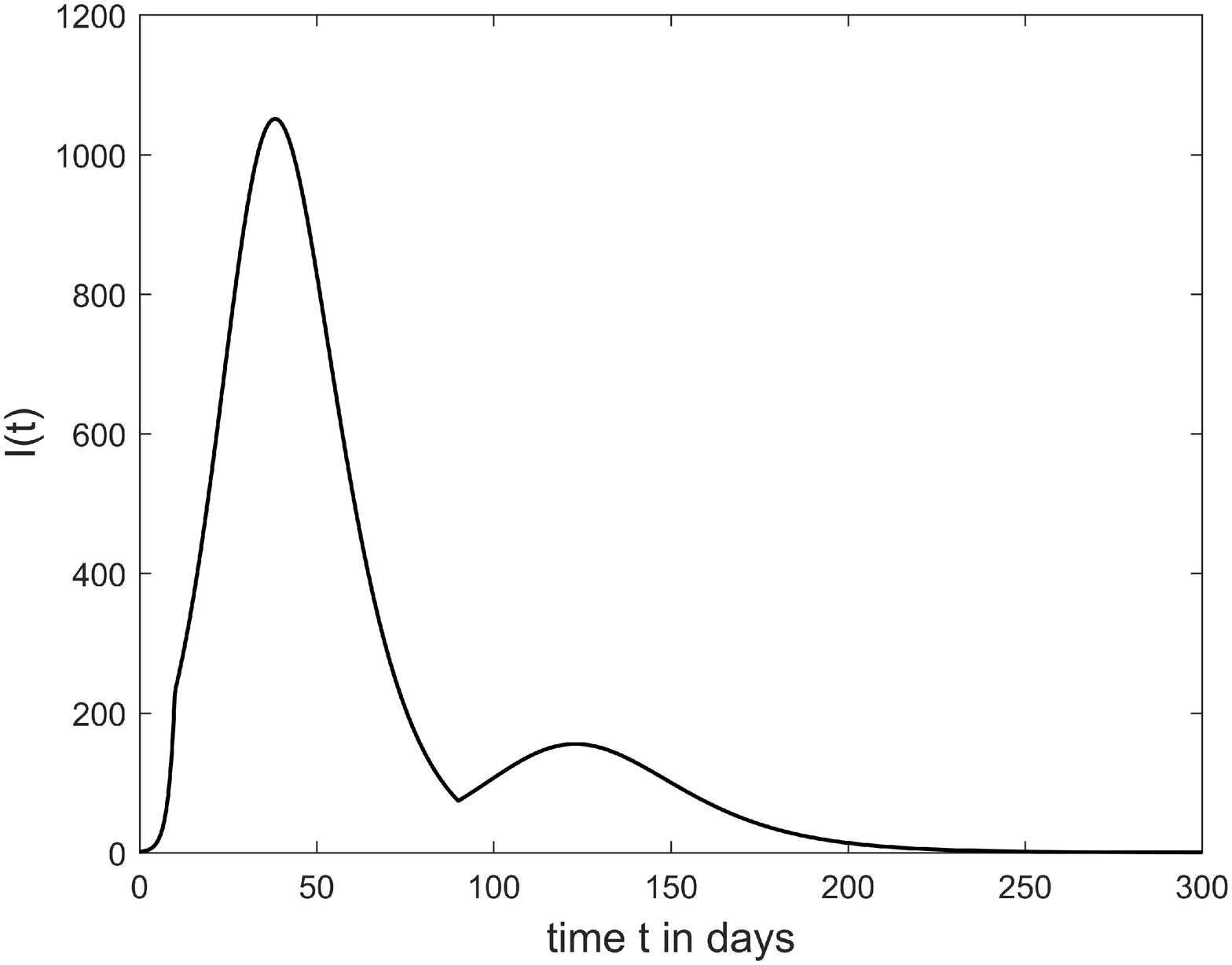}\\
		\includegraphics[scale=0.26]{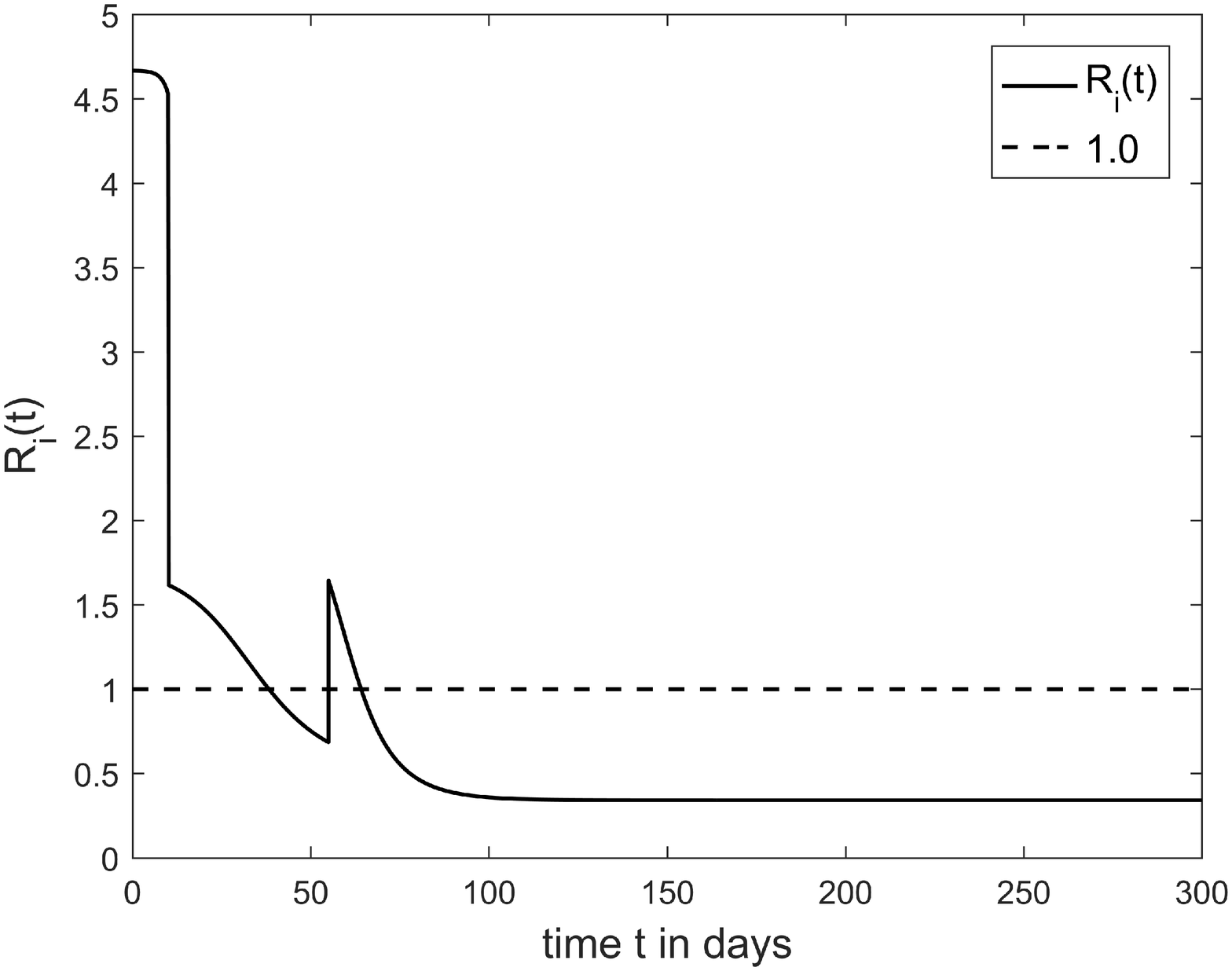}&
		\includegraphics[scale=0.26]{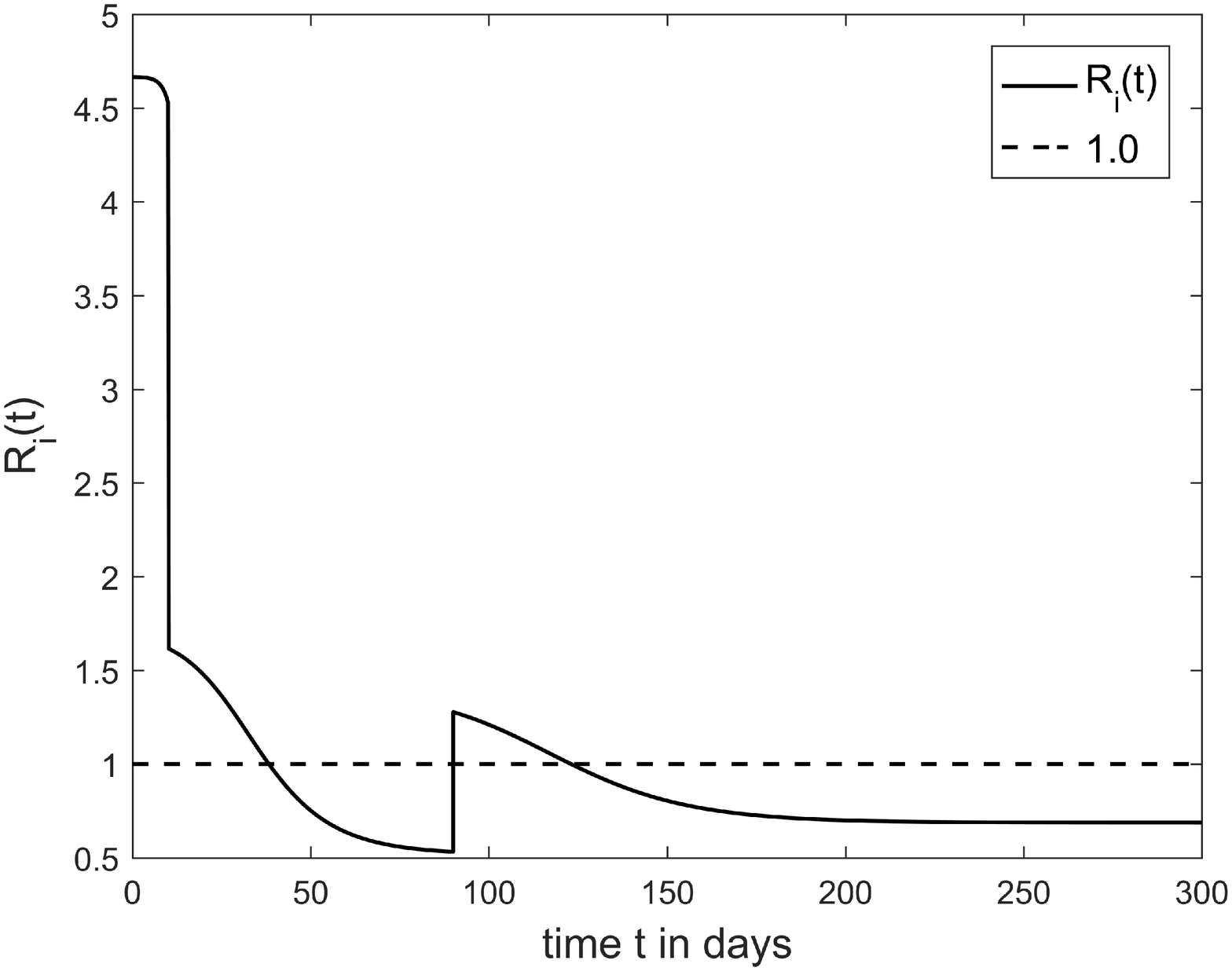}\\
		\includegraphics[scale=0.26]{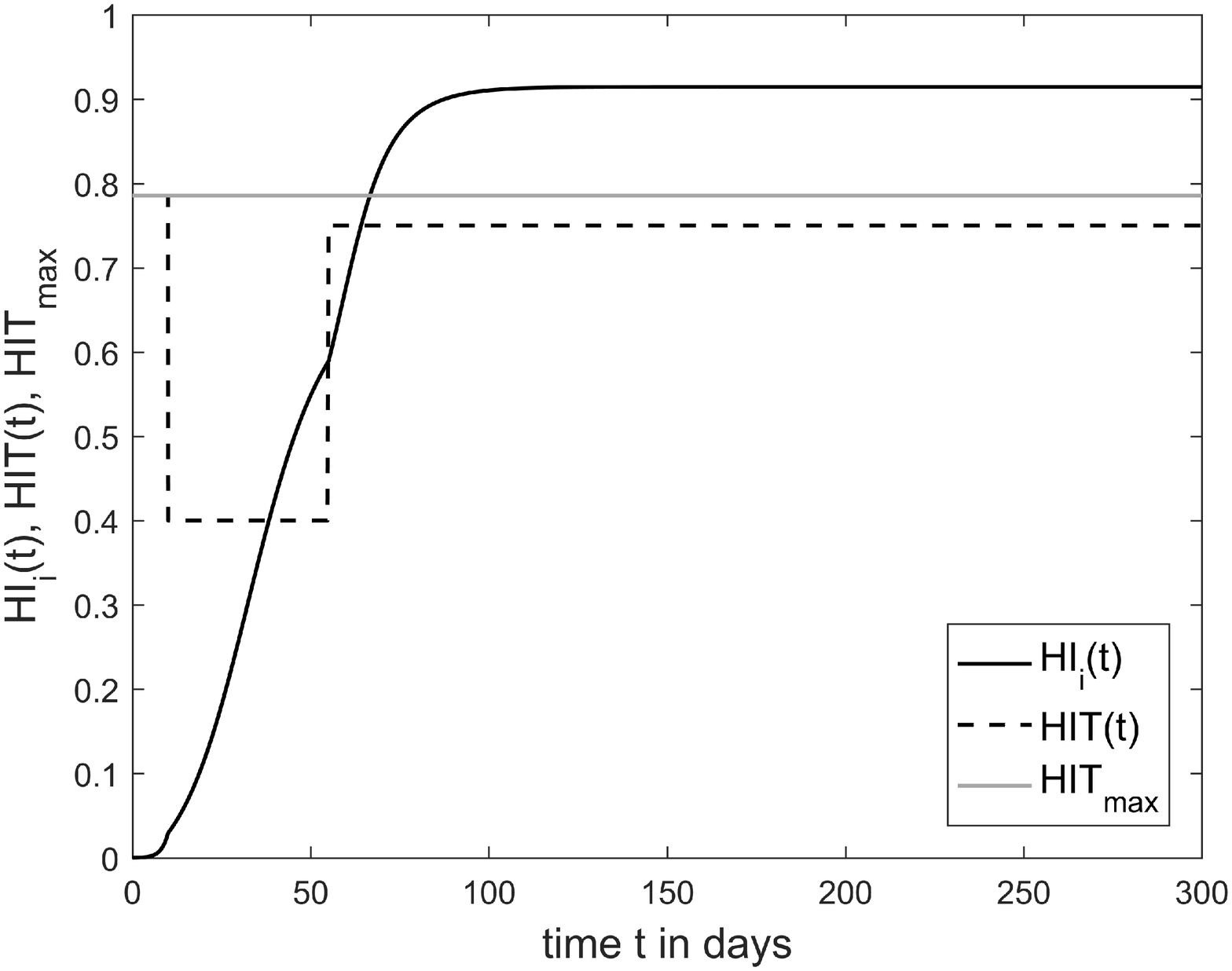}&
		\includegraphics[scale=0.26]{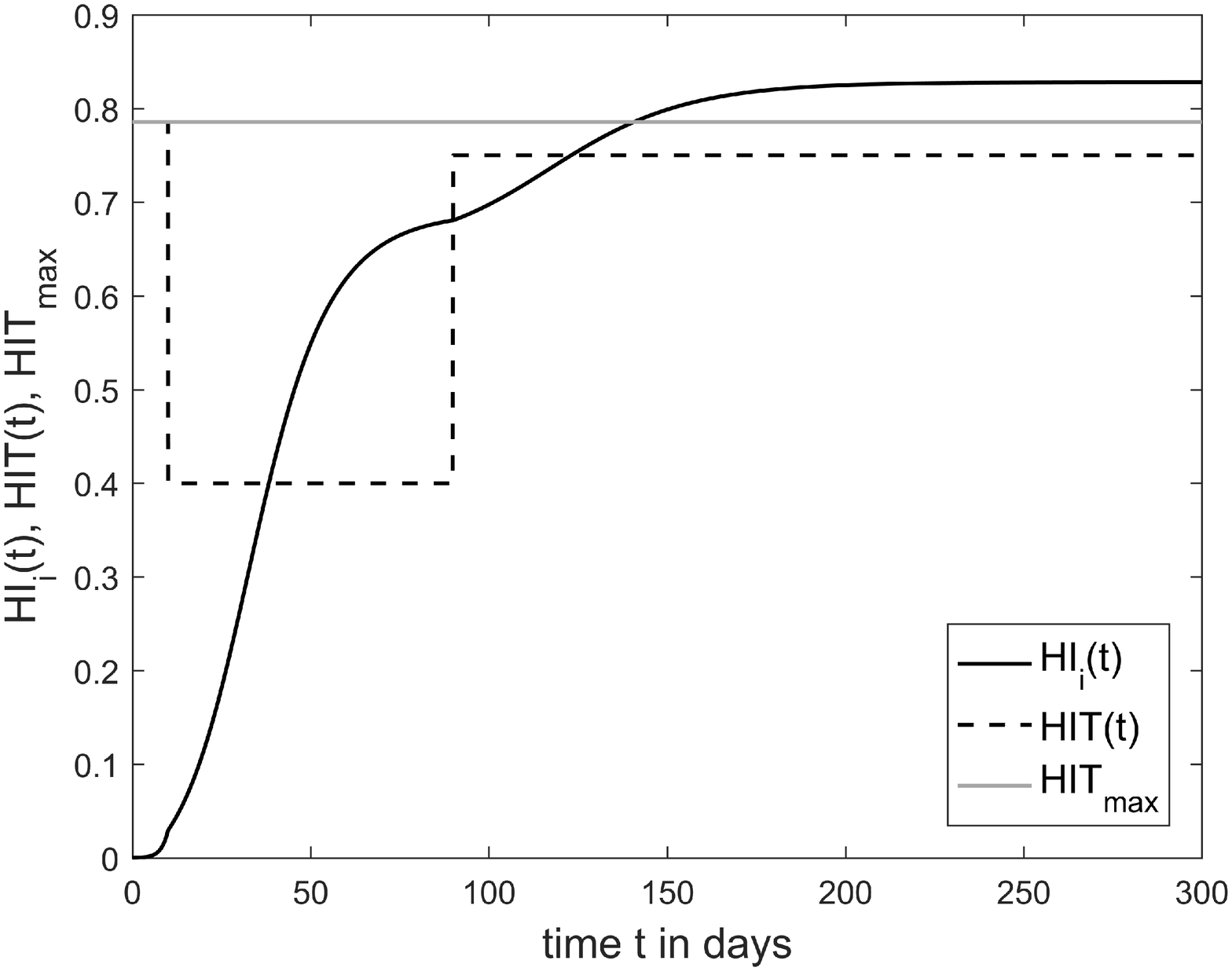}
	\end{tabular}
	\centering
	\caption{{\small Two possible examples of re-outbreak of the epidemic due to negligence
		of safety rules by the population or easing restriction by the government.
		The first column is an example where negligence starts at $t_{N}=55$
		and the second column is an example where negligence starts at $t_{N}=90$.
		The top row shows how $I(t)$ is varying with time.
		The middle row shows how effective reproduction number $R_{i}(t)$
		vary with time. The bottom row shows us the variation of herd immunity
		$HI_{i}(t)$ with time}.\label{fig4}}
\end{figure}

\subsection{Re-outbreak possibilities for interventions}

\subsubsection*{For intervention-1}

For intervention-1, the herd immunity of the population $HI_{i}(\infty)$
does not cross the $HIT_{max}$, so there can be a re-outbreak of
the epidemic because of the increment of $\beta$.
 $\beta$ can be increased due to the negligence of the
 safety rules by the population or maybe easing the restrictions by the government.
We assume that negligence starts at $t_{N}$. So, there can be two possibilities : (a) $\beta$ increases to such an amount that $HIT(t_{N})>HI_{i}(\infty)$,
(b) $\beta$ increases to such an amount that $HIT(t_{N})<HI_{i}(\infty)$.

Figure \ref{fig4} shows the two possible scenarios of the first
case. As, in this case $HIT(t_{N})>HI_{i}(\infty)$,
thus there will always be a chance of re-outbreak of the epidemic.
Here we consider two examples and it has been assumed that the negligence starts at $t_{N}=55$ and $t_{N}=90$ respectively.  Also, let $\beta$ increased from 0.25 to 0.6 in both examples.

From figure \ref{fig4} we can see that the second peak occurs at
$t_{p2}=64.12$ and $t_{p2}=123$ for the two consecutive examples.
After the second outbreak, we can see that the herd immunity of the
population goes beyond the $HIT_{max}$ for both examples. 

\begin{figure}[h]
	\begin{tabular}{cc}
		\includegraphics[scale=0.26]{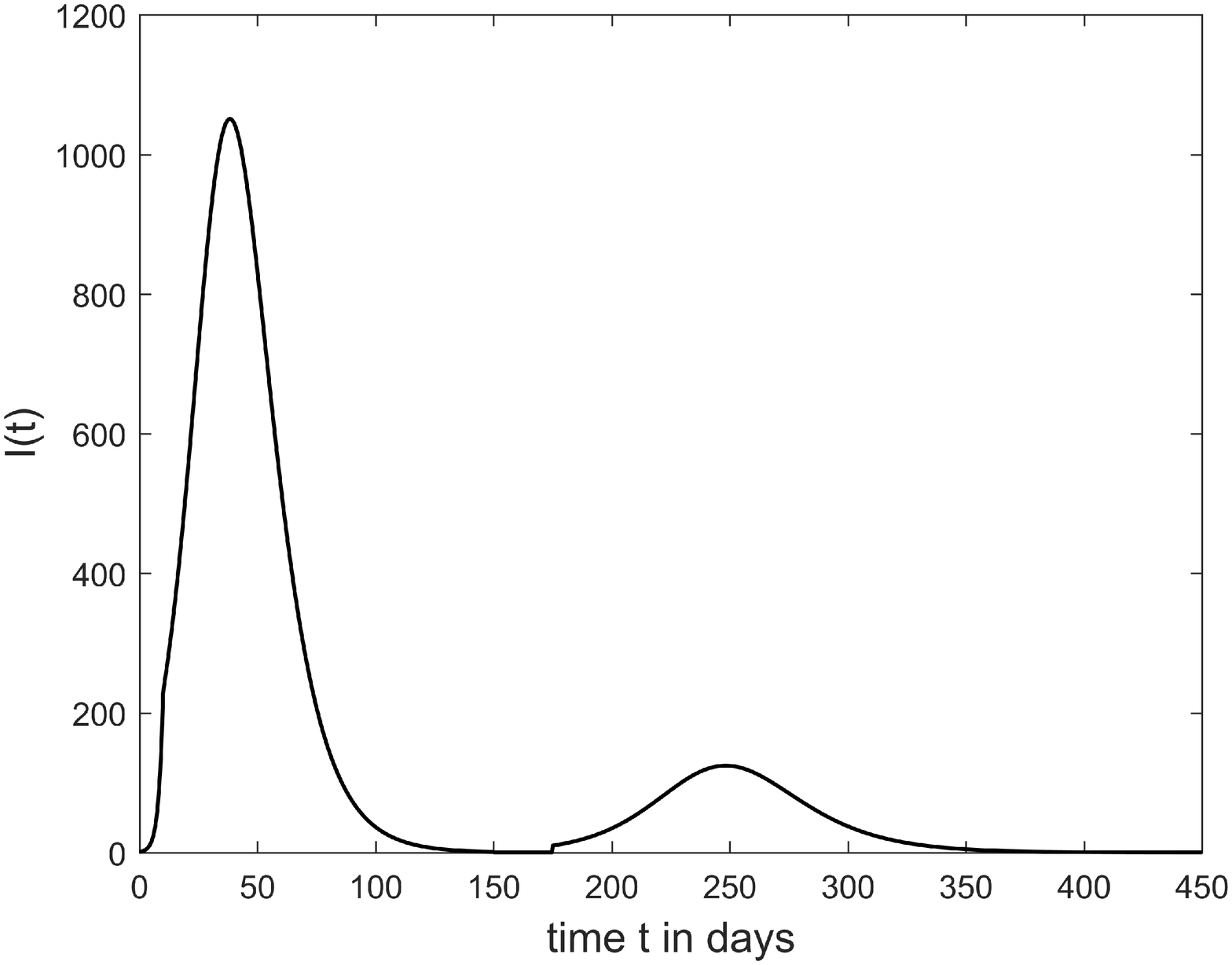}&
		\includegraphics[scale=0.26]{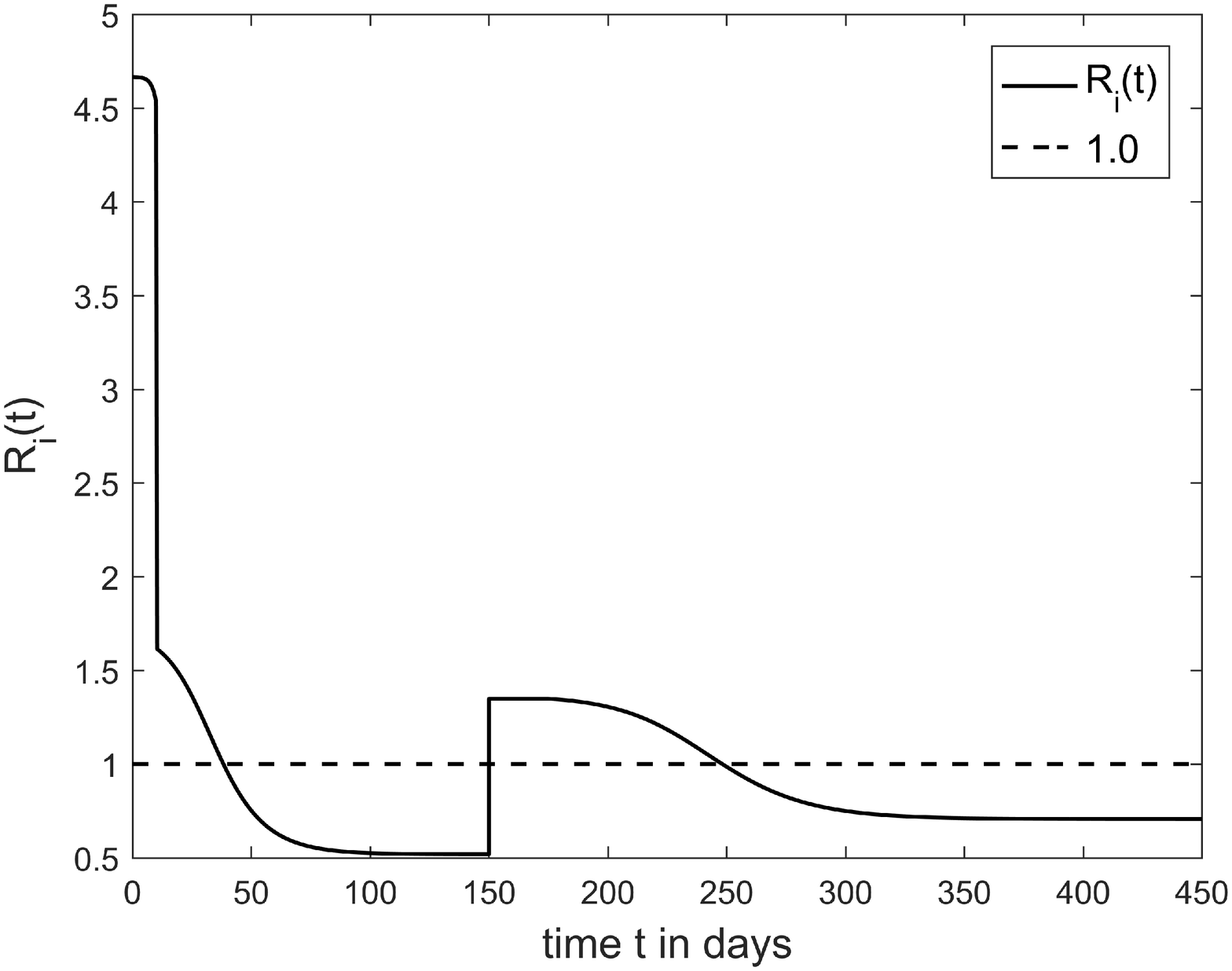}
	\end{tabular}
	\centerline{
		\includegraphics[scale=0.28]{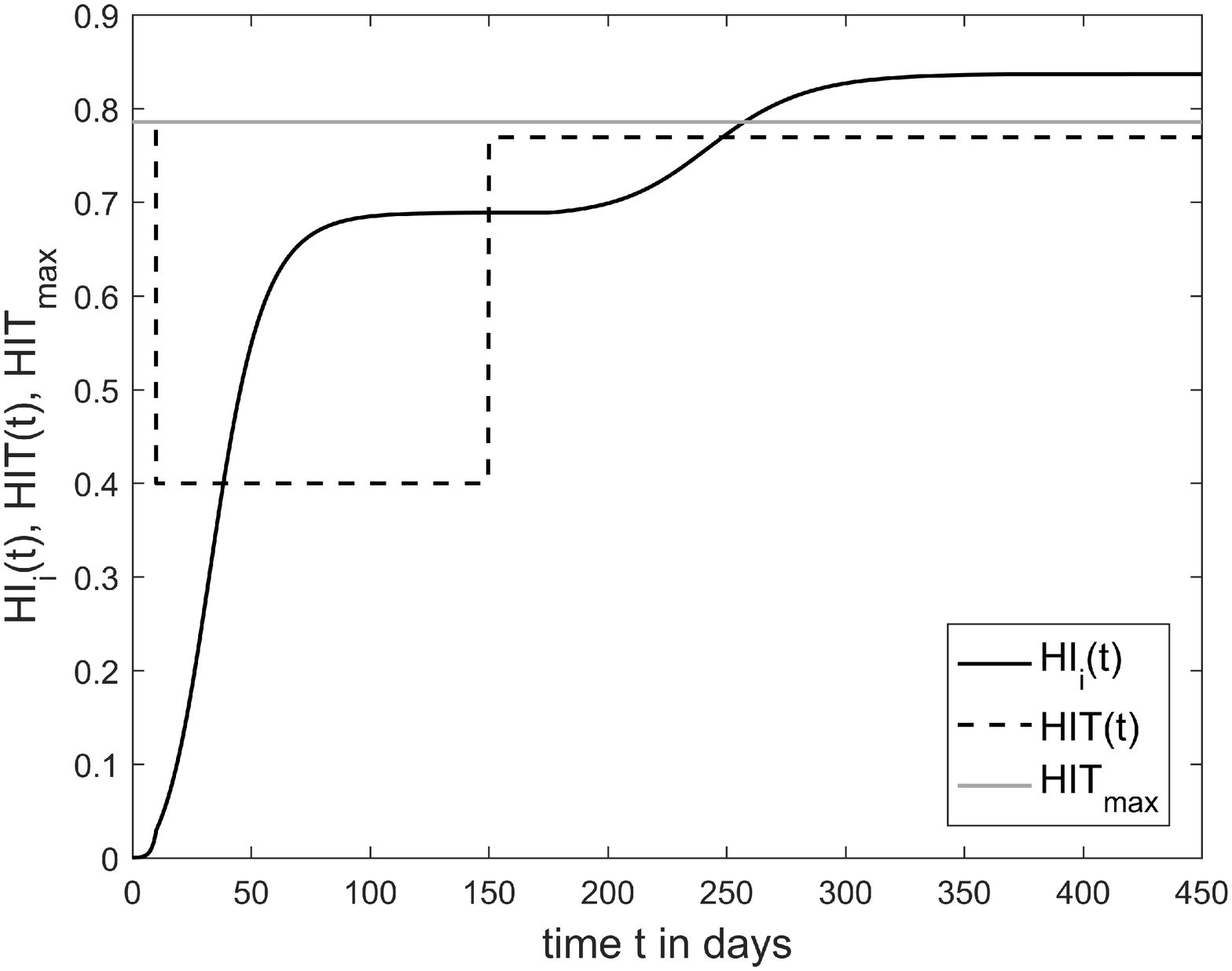}}
	\caption{{\small Re-outbreak for the intervention-1. The left picture of row one shows how $I(t)$
		is varying with time. The right picture of row one shows how effective reproduction
		number $R_{i}(t)$ vary with time. The picture of row two shows
		us variation  of herd immunity $HI_{i}(t)$ and herd immunity threshold
		$HIT(t)$.}\label{fig5}}
\end{figure}

Here, we have to remember that the accurate value of $HIT_{max}$ of an epidemic is impossible to find.
Because $HIT_{max}$ not only depends on the virus characteristics, it also depends on the social behavior of a region.
Also, $HIT_{max}$ can change from country to country
 and between various mutation states of a virus. Hence, after
these outbreaks, there will be no chance of re-outbreak again. For
	this example, herd immunity reaches $HIT_{max}$ after the second outbreak.
	But there can be a system where herd immunity reaches $HIT_{max}$
	after several outbreaks. Also, it is hard to say
	when a re-outbreak will start because re-outbreak starts whenever
	the population will start to neglect the rules or the government will
	lift the restrictions.

Here, a re-outbreak can also occur when the epidemic is almost at
the end ($I(t)\approx0$). Of course, this
re-outbreak cannot occur due to the internal infected people. However, if
somehow a small number of susceptible people become infected (maybe
for migration, from the tourists, or by some external carrier) then
a re-outbreak can again start. Figure \ref{fig5} shows an example
of this kind. Here we assume that the negligence starts at $t_{N}=150$
and the $\beta$ value increased to 0.65. But at the time $t_{N}$,
$I(t_{N})\approx0$ thus no outbreak can possible. Let,
at time $t_{f}=175$, ten susceptible people are somehow infected.
Hence $I(t_{f})=10$. Because of this small disturbance,
we can see that a second outbreak occurs and the peak of  it is at $t_{p2}=248.2$. So, we can say, until a population
crosses the $HIT_{max}$ there will always be a chance of re-outbreak.

\begin{figure}[h]
	\begin{tabular}{cc}
		\includegraphics[scale=0.26]{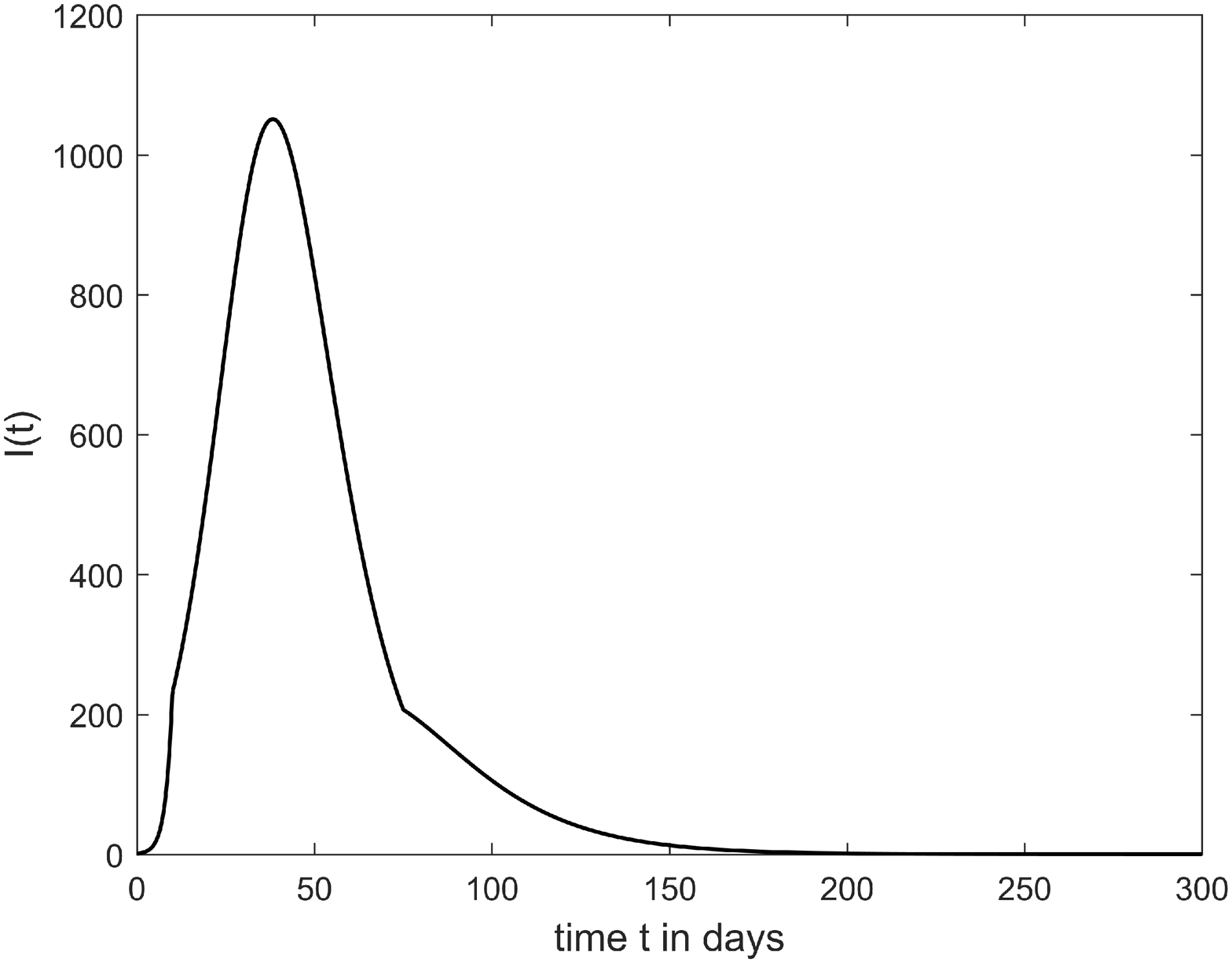}&
		\includegraphics[scale=0.26]{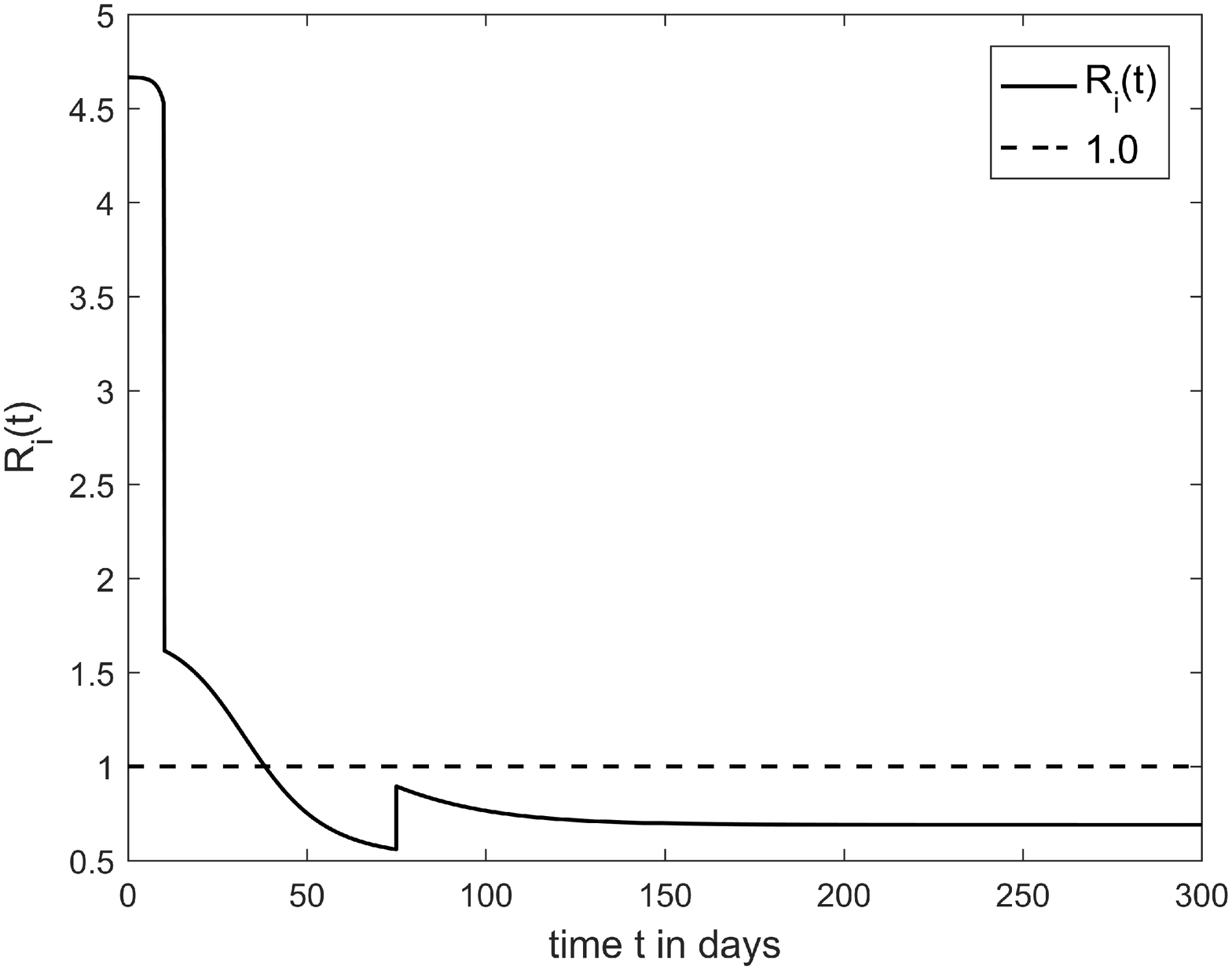}
	\end{tabular}
	\centerline{\includegraphics[scale=0.28]{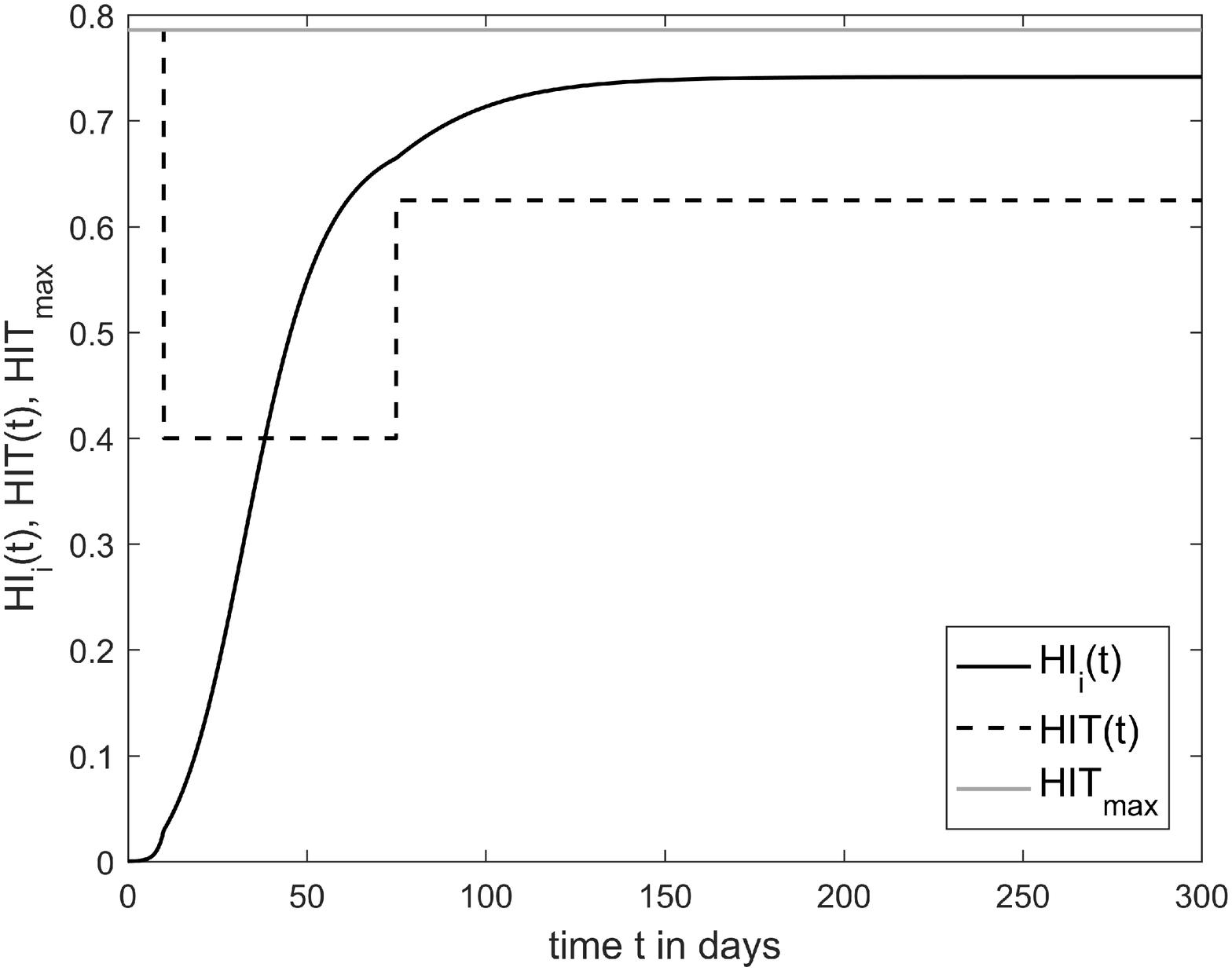}}
	\caption{{\small No re-outbreak for intervention-1. The left picture of row one shows how $I(t)$
		is varying with time. The right picture of row one shows how effective reproduction
		number $R_{i}(t)$ varies with time. The picture of row two
		shows us variation of herd immunity $HI_{i}(t)$ and herd immunity threshold $HIT(t)$.}\label{fig6}}
\end{figure}

Figure \ref{fig6} shows an example of the second case. Now, for
this second case we assume that $t_{N}$ is a little bit far from
the peak time. At this time herd immunity of the population $HI_{i}(t_{N})\approx HI_{i}(\infty)$.
As, for the second case, $HI_{i}(\infty)>HIT(t_{N})$
hence, $HI_{i}(t_{N})>HIT(t_{N})$. So, there
is no possibility of re-outbreak. But if $t_{N}$ is close to $t_{p}$
then there can be a re-outbreak if $HI_{i}(t_{N})<HIT(t_{N})<HI_{i}(\infty)$.
Here we assume that $t_{N}=75$ and from this time $t_{N}$, $\beta$
increased from 0.25 to 0.4.

\subsubsection*{For intervention-2}

For intervention-2 the herd immunity of the population very quickly crosses $HIT_{max}$ after crossing $HIT_{in}$. So, there is no chance of
re-outbreak.

\begin{figure}[h]
\begin{tabular}{cc}
\includegraphics[scale=0.26]{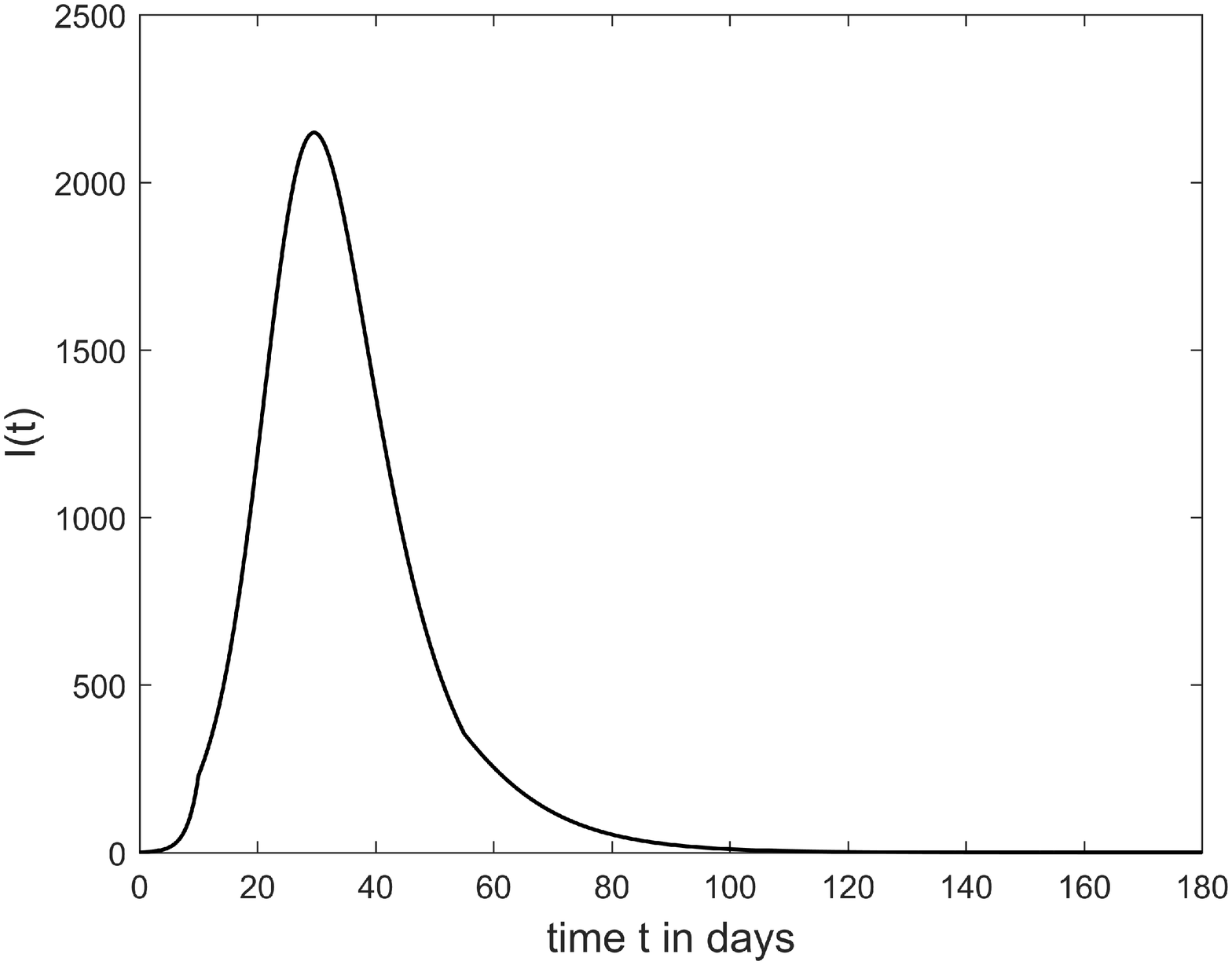}&
\includegraphics[scale=0.26]{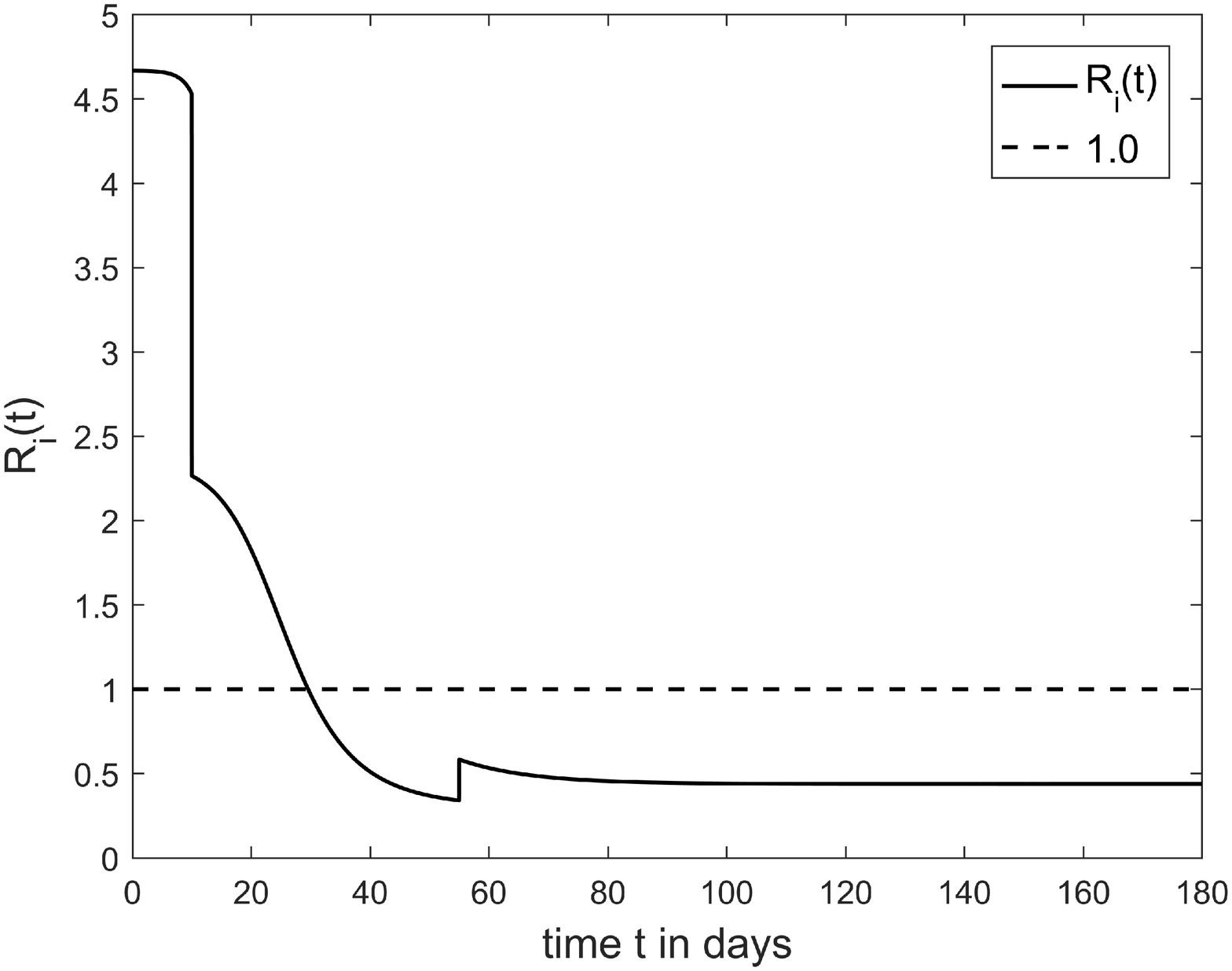}
\end{tabular}
\centerline{\includegraphics[scale=0.28]{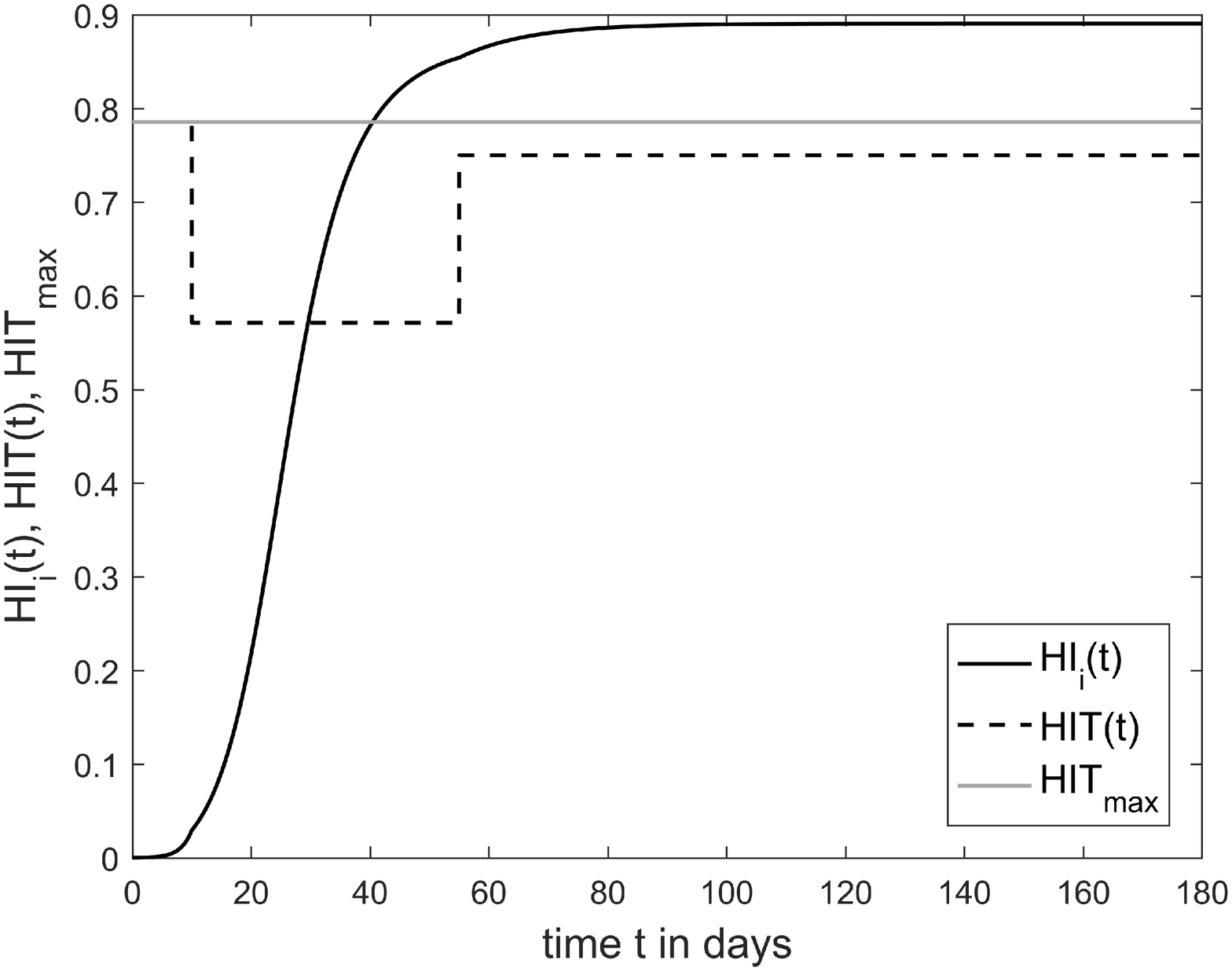}}
\caption{{\small No re-outbreak for intervention-2. The left picture of row one shows how $I(t)$
is varying with time. The right picture of row one shows how effective reproduction
number $R_{i}(t)$ vary with time. The picture of row two shows
us variation  of herd immunity $HI_{i}(t)$ and herd immunity threshold $HIT(t)$.\label{fig7}}}
\end{figure}

Here we have assumed that the negligence time is $t_{N}=55$ and due to this negligence $\beta$ increases to 0.6.
Figure \ref{fig7} shows that there is no increment of active cases, which means no re-outbreak.

\subsection{Summary}

So, the differences between the two interventions are
\begin{enumerate}
	\item As intervention-1 is stronger than the intervention-2, thus fewer
	people are infected for intervention-1 than the intervention-2 during
	this epidemic.
	\item For intervention-2, herd immunity of the population crosses the maximum
	herd immunity threshold value but for intervention-1, herd immunity
	of the population does not cross the maximum herd immunity threshold
	value.
\end{enumerate}
So, we can see that there can be various types of interventions, different
from strictness. Let, reduced value of the infection rate is represented
by $\beta_{in}$. If $\beta_{in}$ is close to the $\beta_{max}$
then we will say that intervention is a loose one. However, if $\beta_{in}$
lies a little bit far from $\beta_{max}$, then we will say the intervention
is a strict one.

Now for both no intervention case and loose intervention case, we saw that the herd immunity of the population crosses the maximum herd immunity threshold. But fewer people are infected in the loose intervention case than the no intervention case.
So, we can say that if the government and local authorities decide
	to fight an epidemic by increasing the herd immunity of the population,
	then loose intervention will be a good option than the no intervention
	one.

Now for the strict intervention case, we saw that number of infections
is small compared to the others. But problem is that there will always
chances of  re-outbreak in the population when strict interventions
are lifted or loosen (As it is impossible to maintain strict interventions
for a very long time). 

So, from our model, we can say that, if the goal is to decrease the number of infections then strict intervention will be a good option. Also, if the goal is to increase the herd immunity of the population beyond the maximum herd immunity threshold value then loose intervention will be a good option. However, we have to remember that in real life situation the herd immunity and maximum herd immunity depends on many factors. So, these interventions can have different impacts on the population and might not work well.

\section{Our model and Data from Various Countries}
In this section, we take COVID-19 pandemic
data \cite{Covid_data} of various countries and fit that data by a piece-wise continuous function. From these fitted functions, we calculate the infection rate ($\beta(t)$), removed rate ($\gamma(t)$), and herd immunity
of the population ($HI_{i}(t)$). Also, we calculate the herd immunity
threshold ($HIT(t)$) and effective reproduction number ($R_{i}(t)$)
from fitted active cases ($I$), removed cases ($R$), and estimated
infection rate $(\beta$), removed rate ($\gamma$). Here data are
taken from that date when the recovered cases and deaths both start
to increase steadily with time (except for Brazil). There are large discontinuities in
 Brazil's data at the initial time and around the end (01/11/2020) of the data. Thus we have rejected the initial part of the data. 
 We have retained the end part of the $I$ and $R$ data. However, the end part of the active cases data has not been used for the analysis.
Table \ref{ta1} shows such range of dates for various countries. For
a country, the first date of the range is considered as the initial
time t=0.

\begin{table}[h]
	\tbl{Range of the dates of data taken for various countries.}
	{\begin{tabular}{@{}ccc@{}} \toprule
			Country\hphantom{000} & Data were taken from & Data were taken till \\ \colrule
			USA\hphantom{000000} & 18/03/2020 & 1/11/2020 \\
			India\hphantom{00000} & 24/03/2020 & 1/11/2020 \\
			Peru\hphantom{000000} & 29/03/2020 & 1/11/2020 \\
			Russia\hphantom{0000} & 29/03/2020 & 1/11/2020 \\
			Colombia\hphantom{00} & 29/03/2020 & 1/11/2020 \\
			Bangladesh & 11/04/2020 & 1/11/2020 \\
			Italy\hphantom{00000} & 18/03/2020 & 1/11/2020 \\ 
			Brazil\hphantom{0000} & 15/04/2020 & 1/11/2020\\ \botrule
		\end{tabular} \label{ta1}}	
\end{table}

\begin{figure}[H]
	\begin{tabular}{cc}
		\includegraphics[scale=0.25]{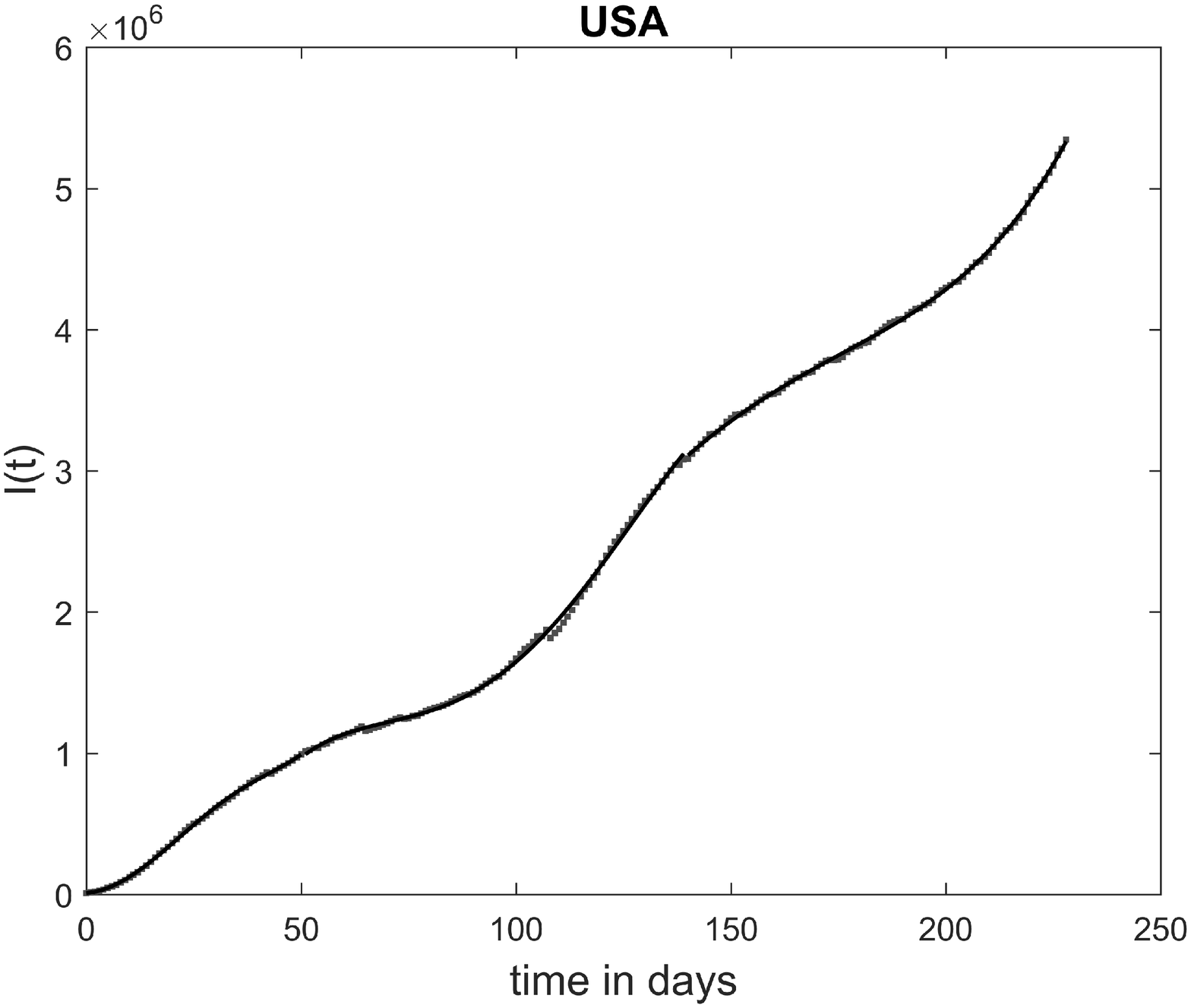}&
		\includegraphics[scale=0.25]{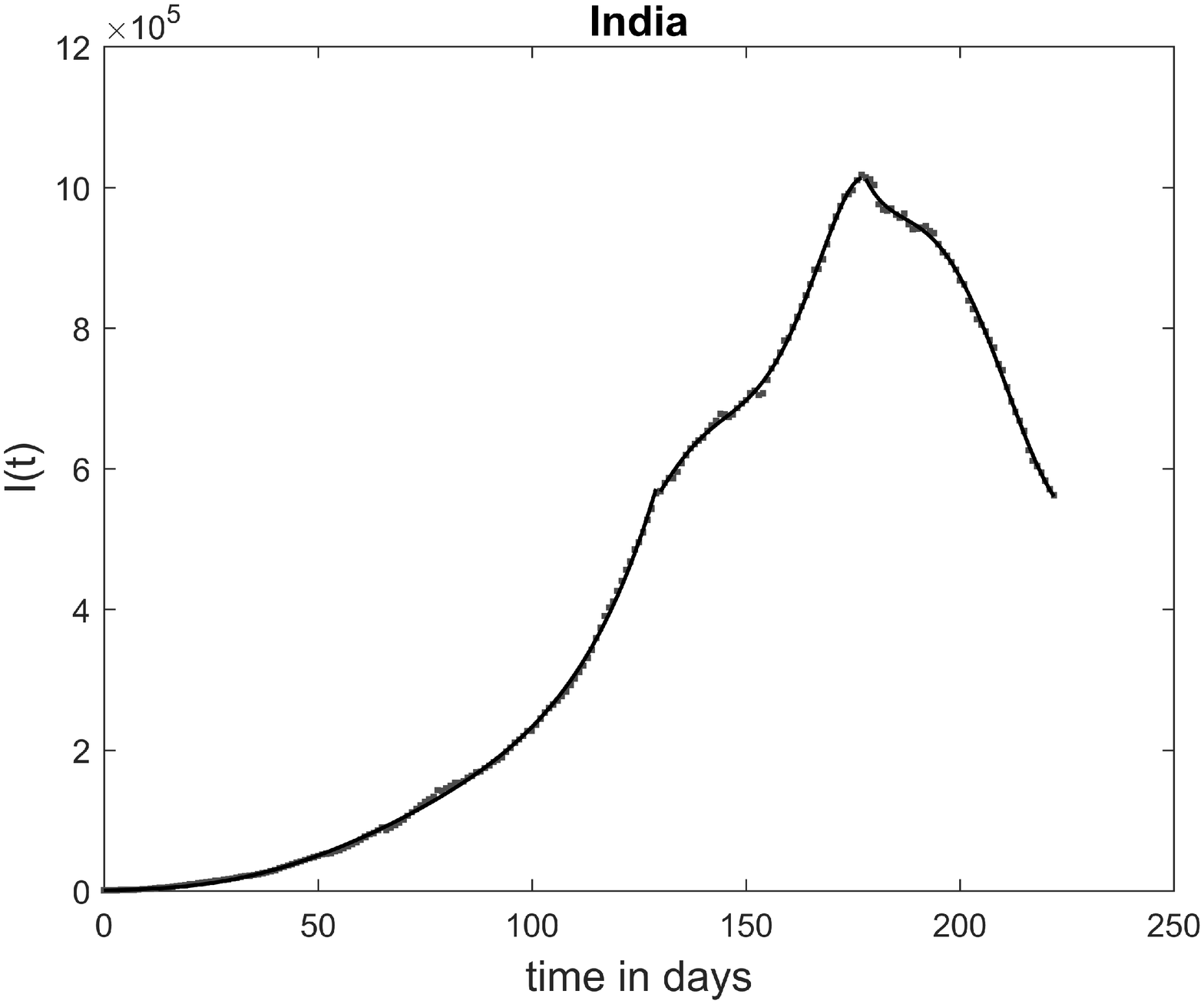}\\
		\includegraphics[scale=0.25]{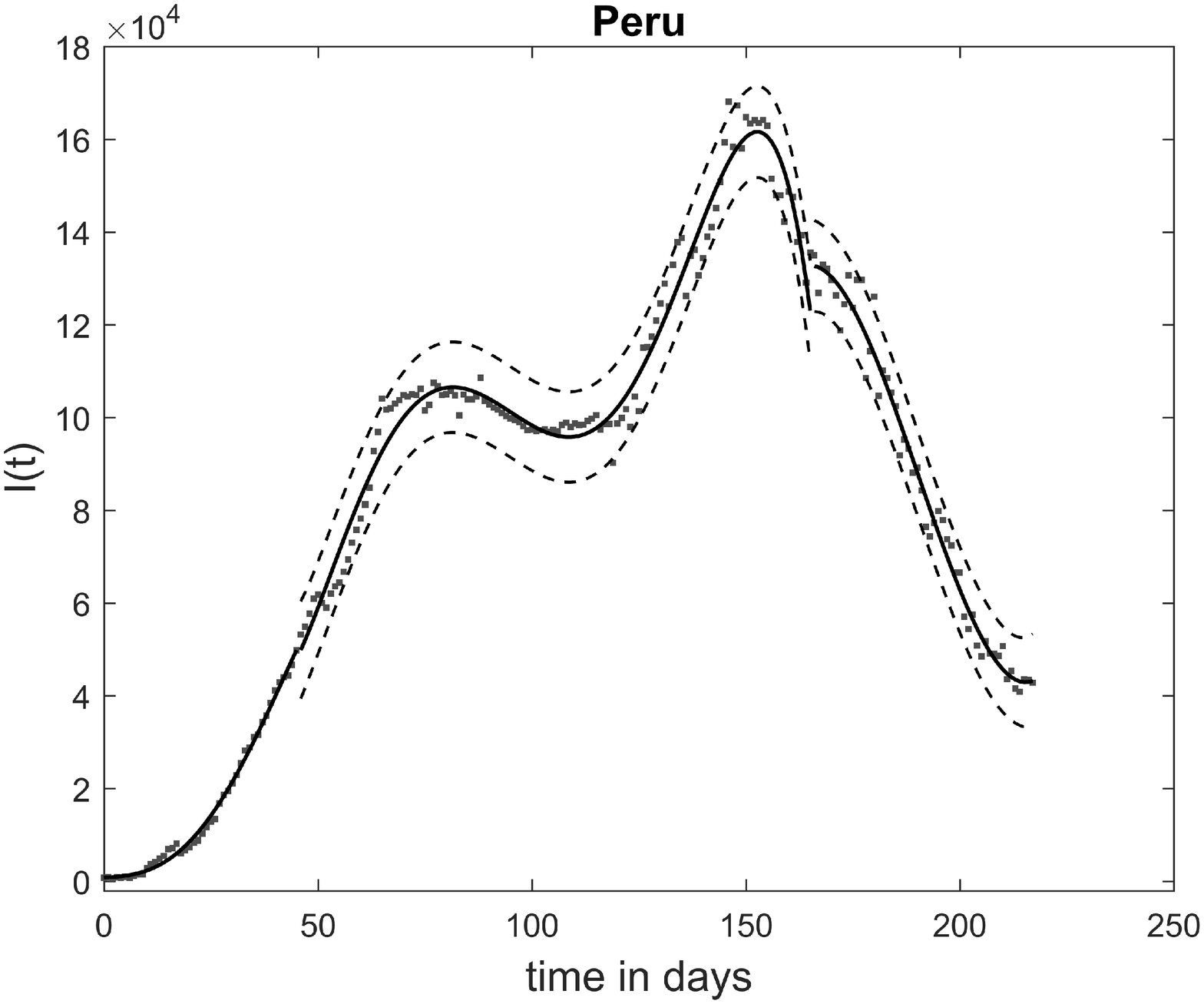}&
		\includegraphics[scale=0.25]{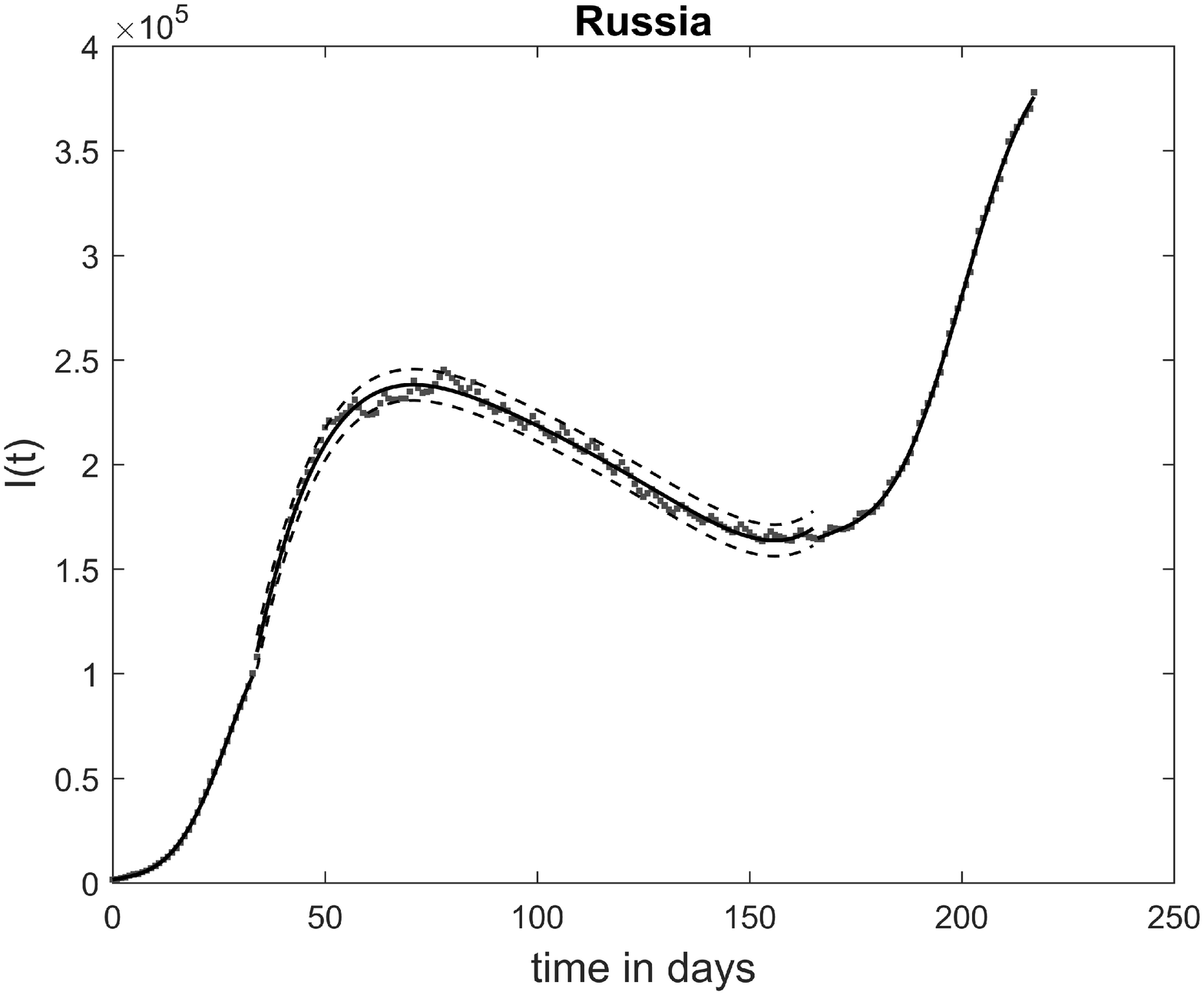}\\
		\includegraphics[scale=0.25]{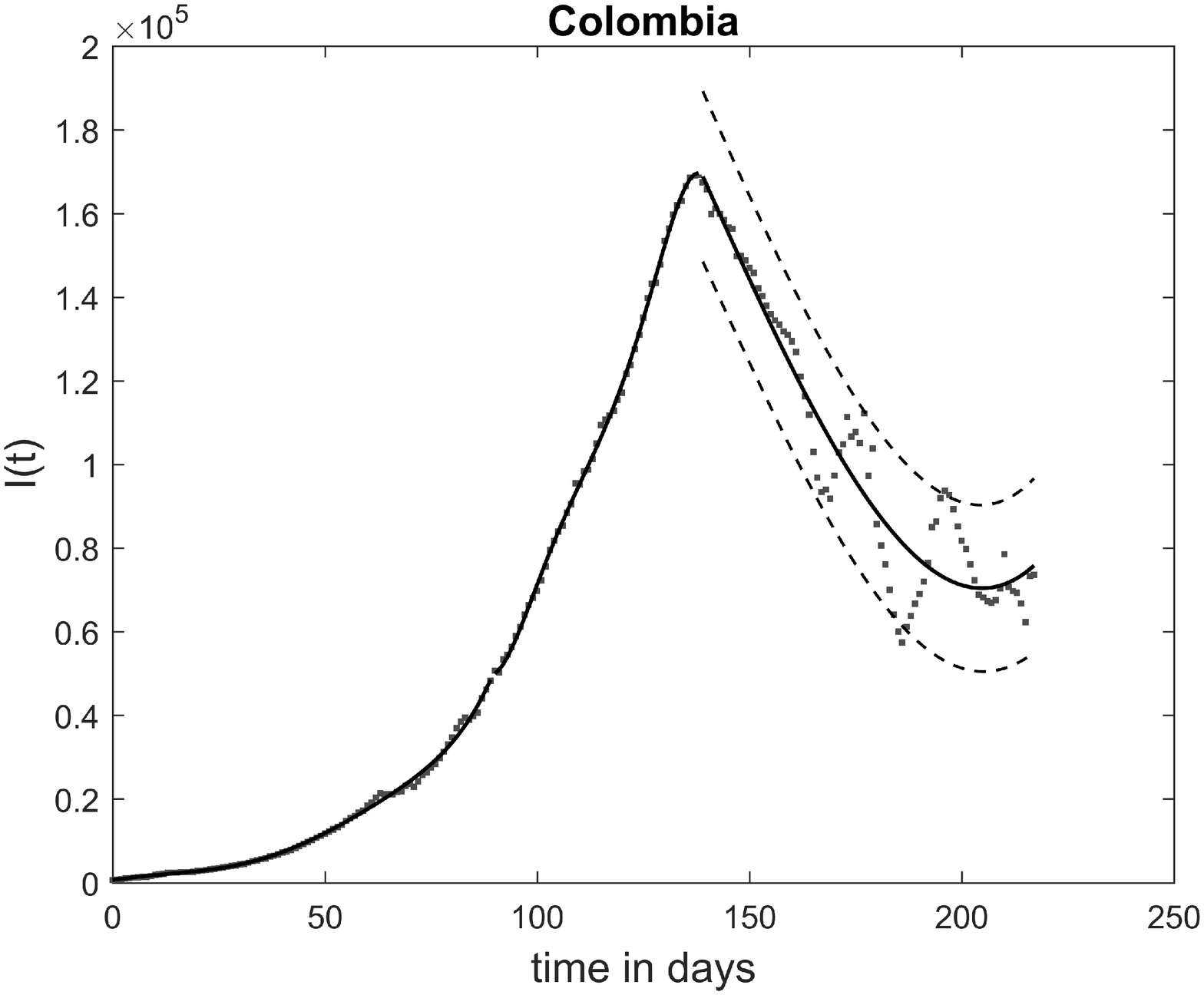}&
		\includegraphics[scale=0.25]{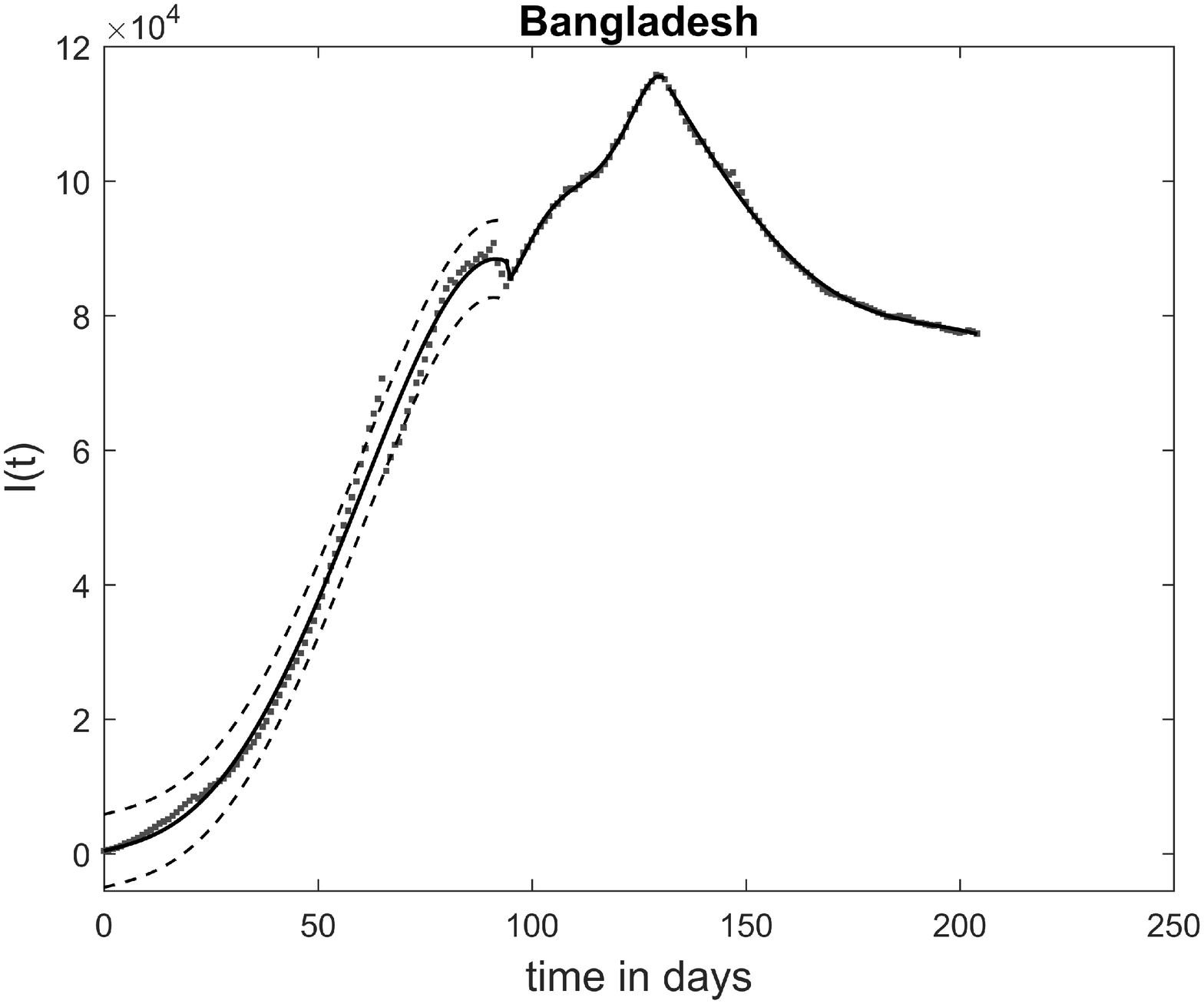}\\
		\includegraphics[scale=0.25]{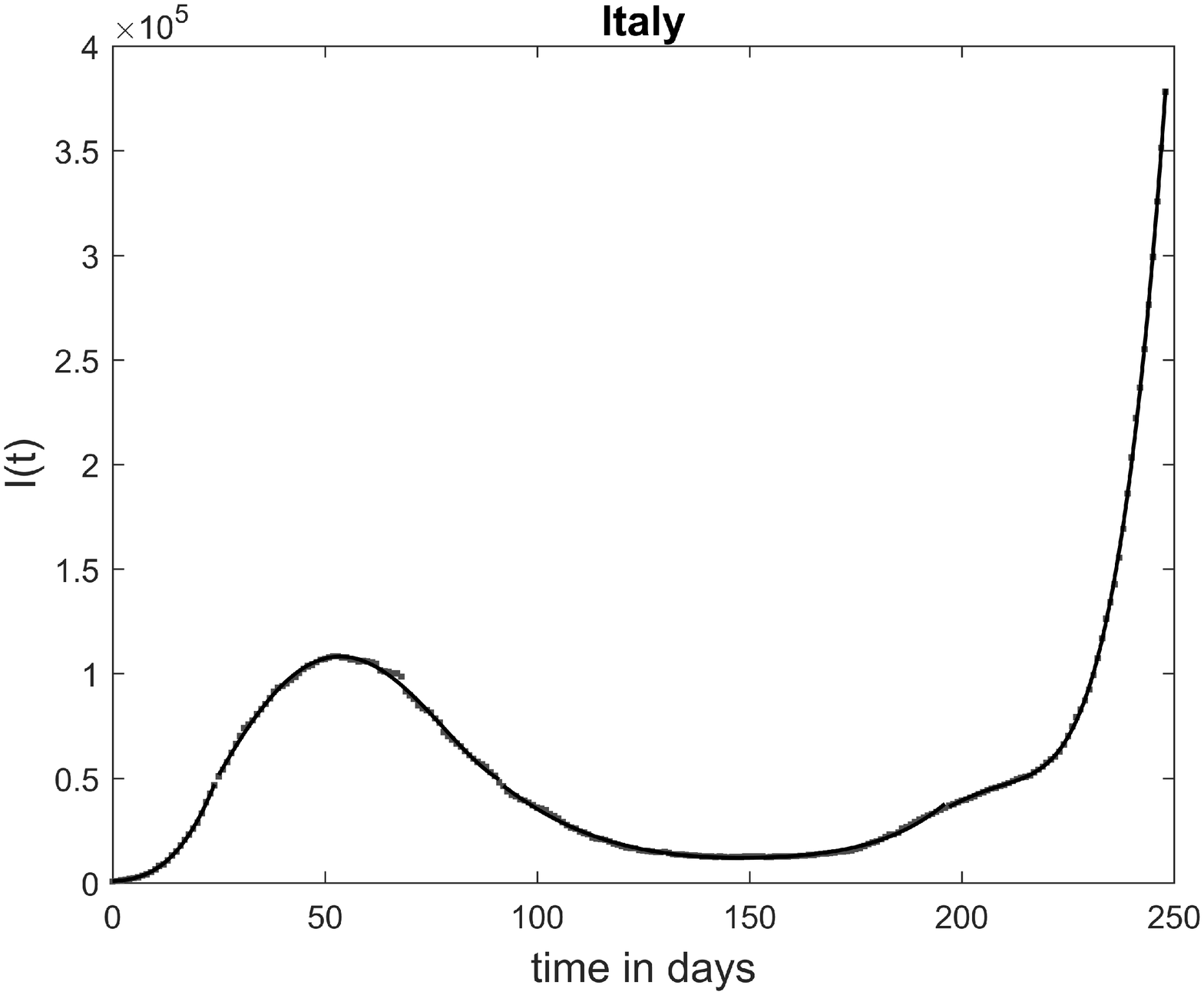}&
		\includegraphics[scale=0.25]{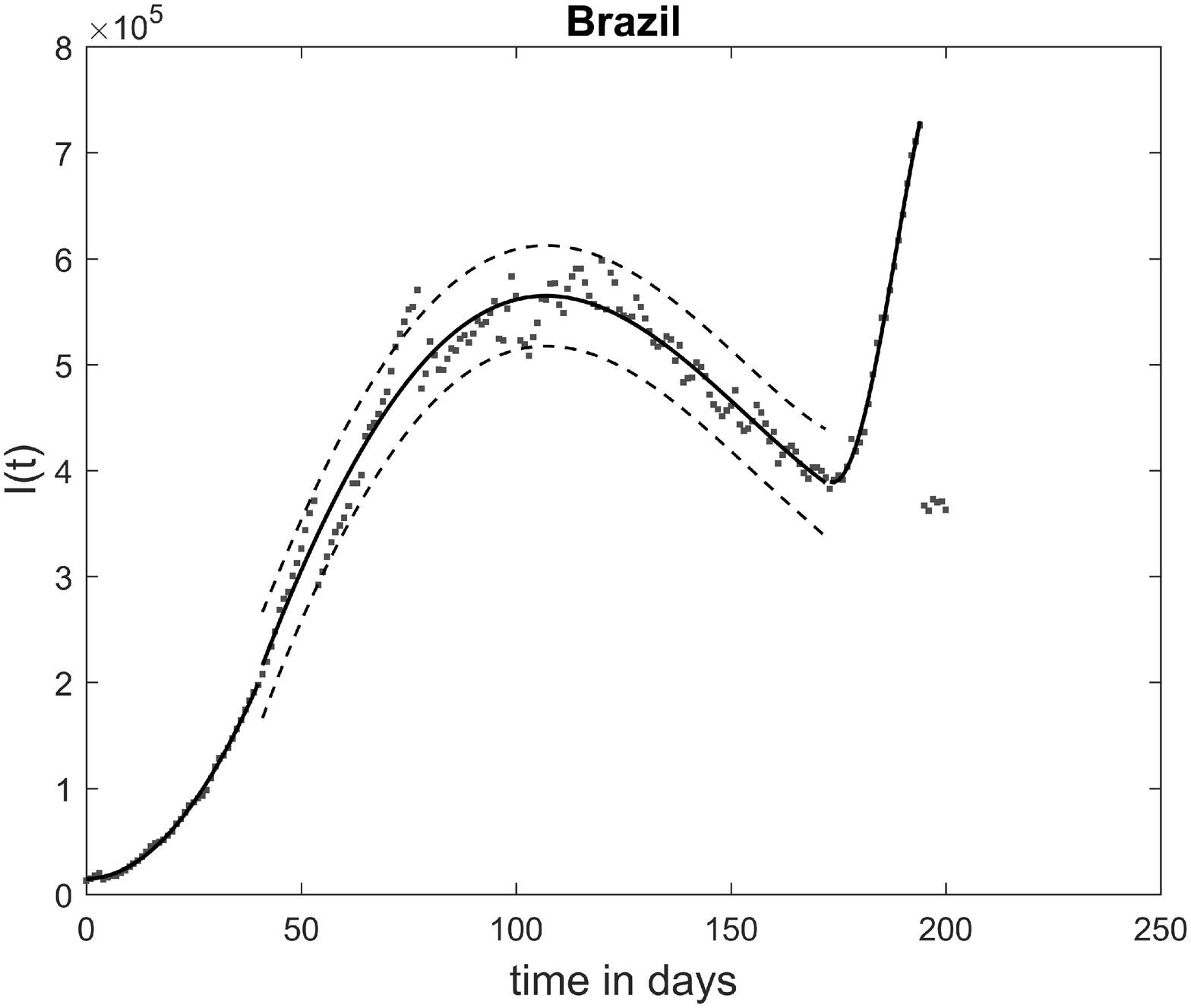}
	\end{tabular}
	\centering
	\caption{{\small Piece-wise continuous function fitting to the data of active cases
		($I$) for different countries.}\label{fig8}}
\end{figure}

\begin{figure}[H]
	\begin{tabular}{cc}
		\includegraphics[scale=0.25]{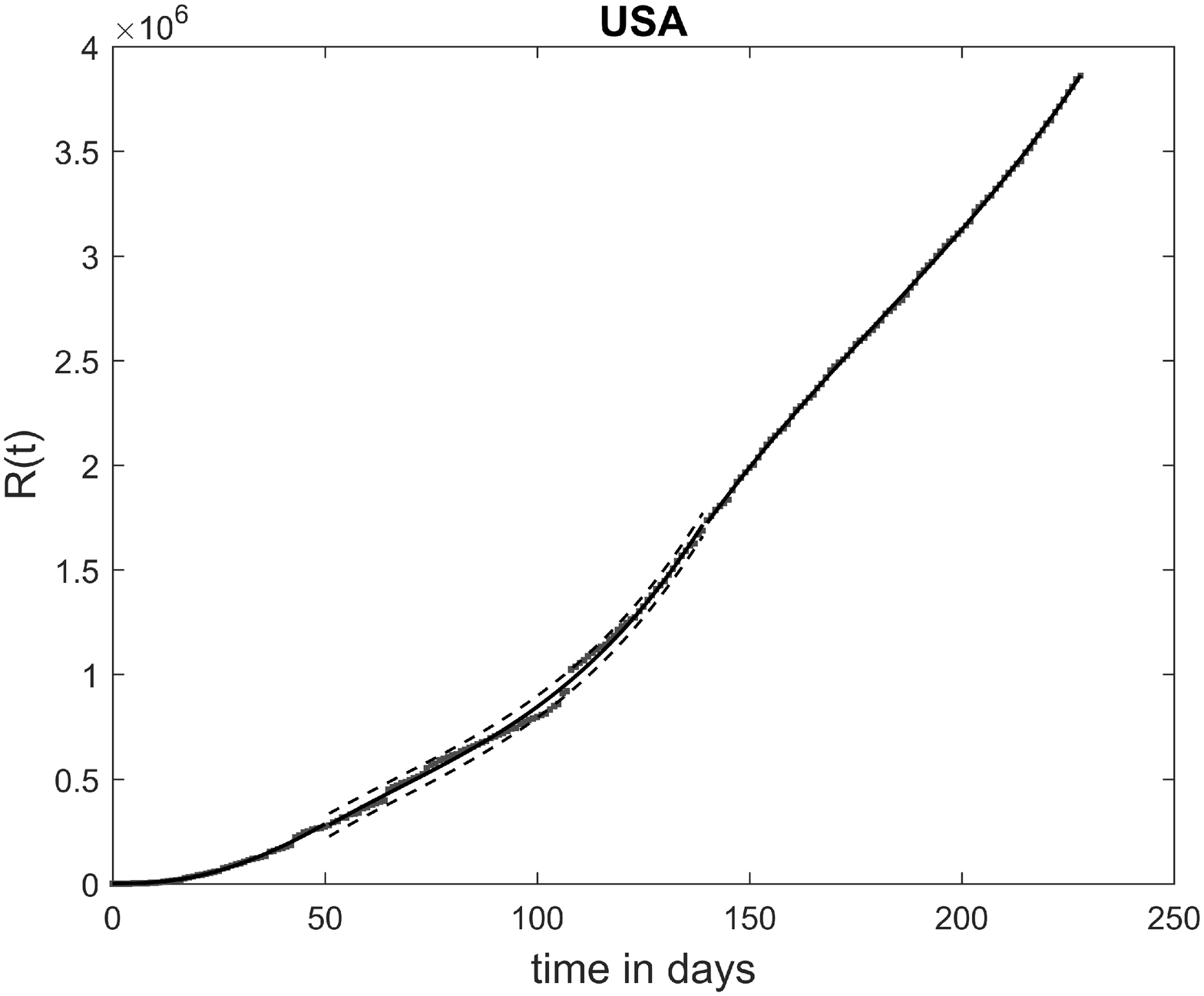}&
		\includegraphics[scale=0.25]{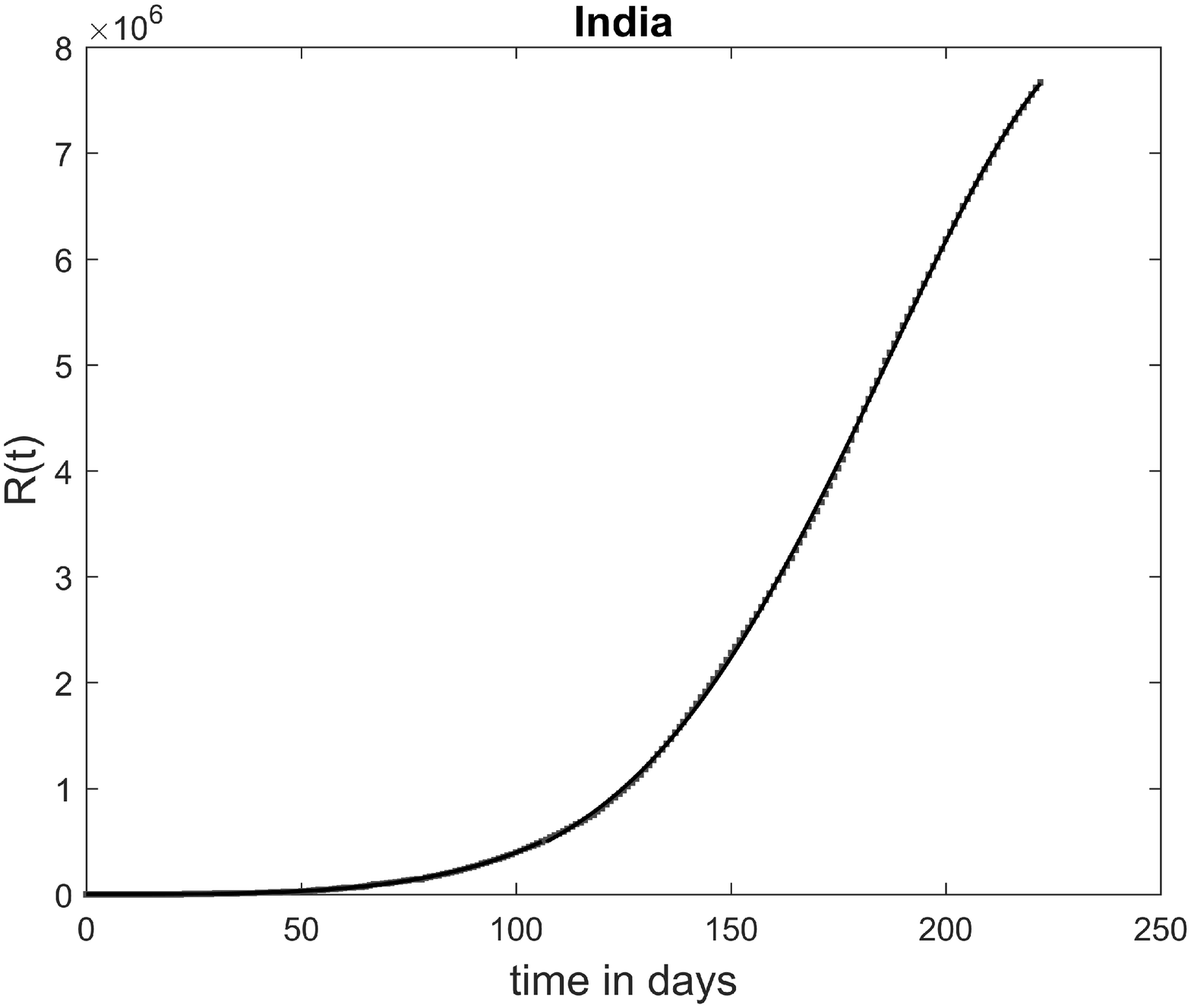}\\
		\includegraphics[scale=0.25]{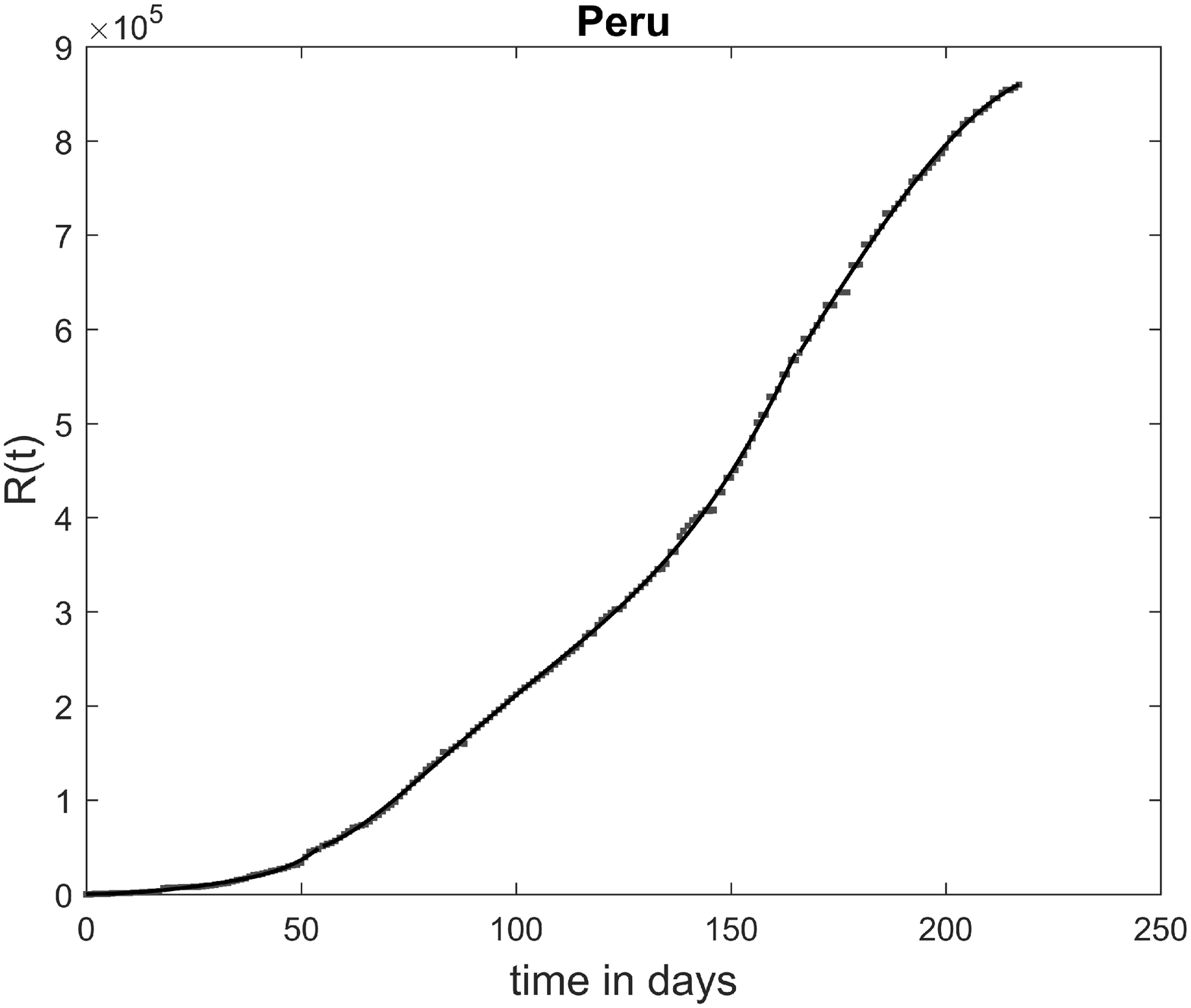}&
		\includegraphics[scale=0.25]{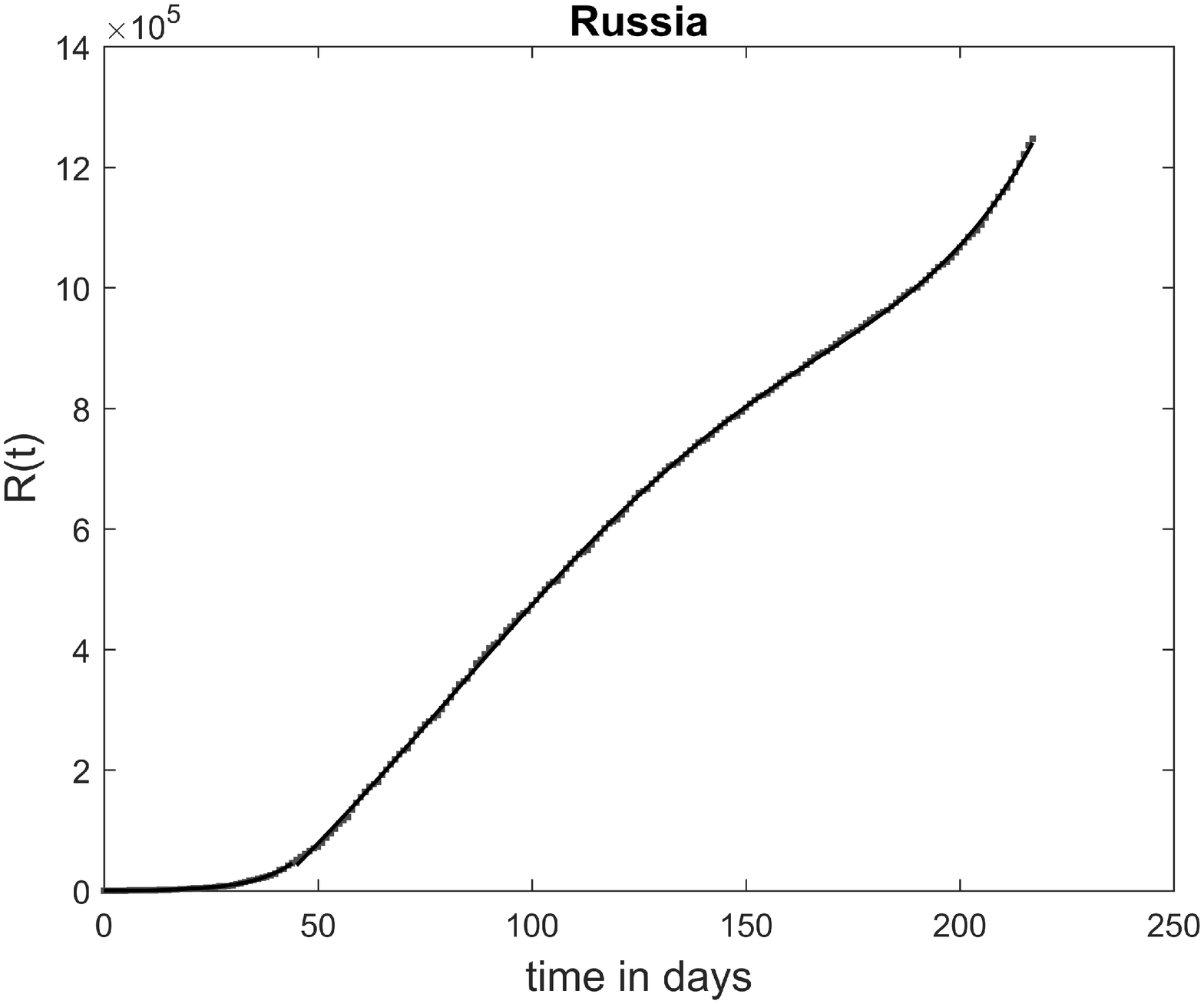}\\
		\includegraphics[scale=0.25]{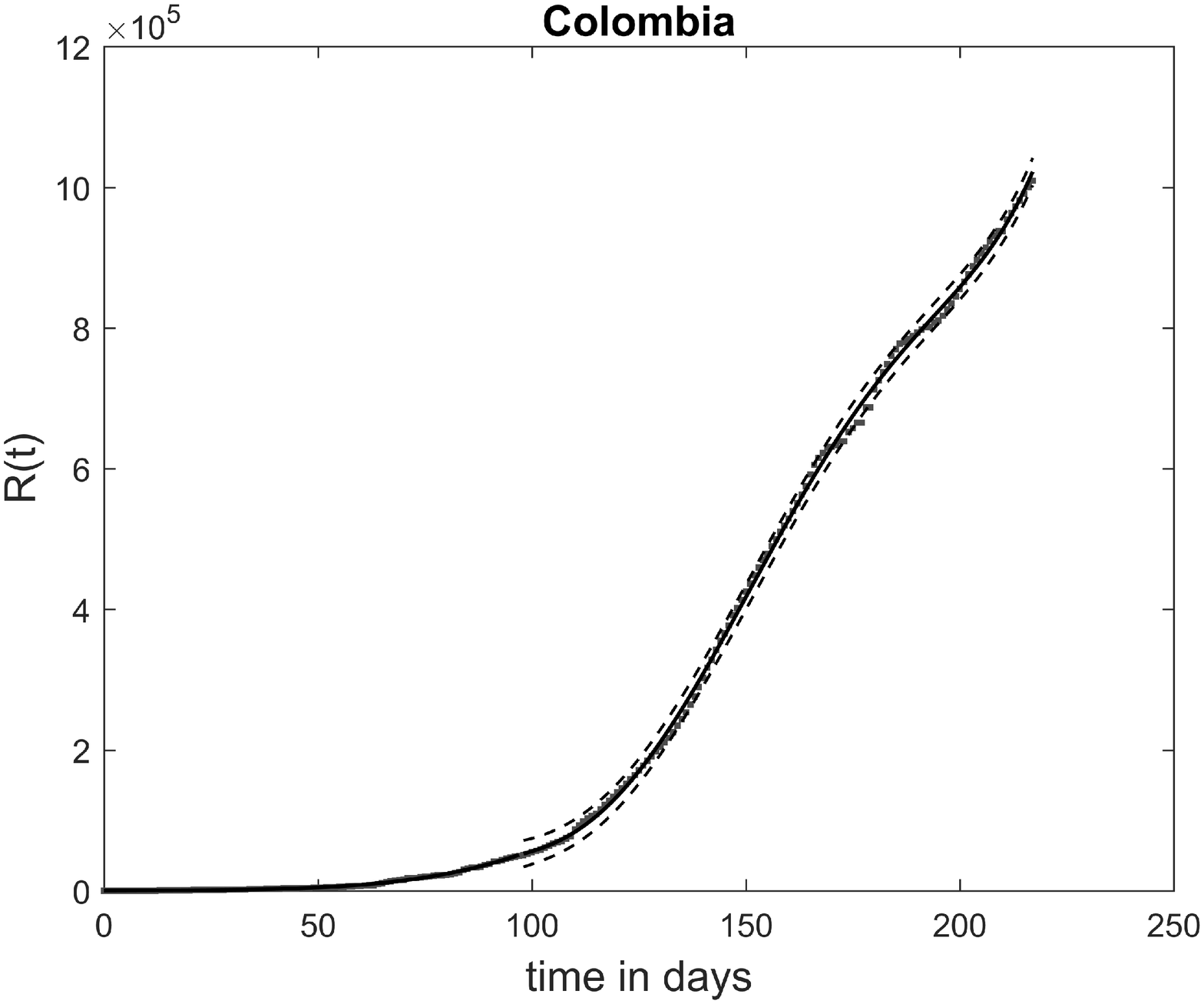}&
		\includegraphics[scale=0.25]{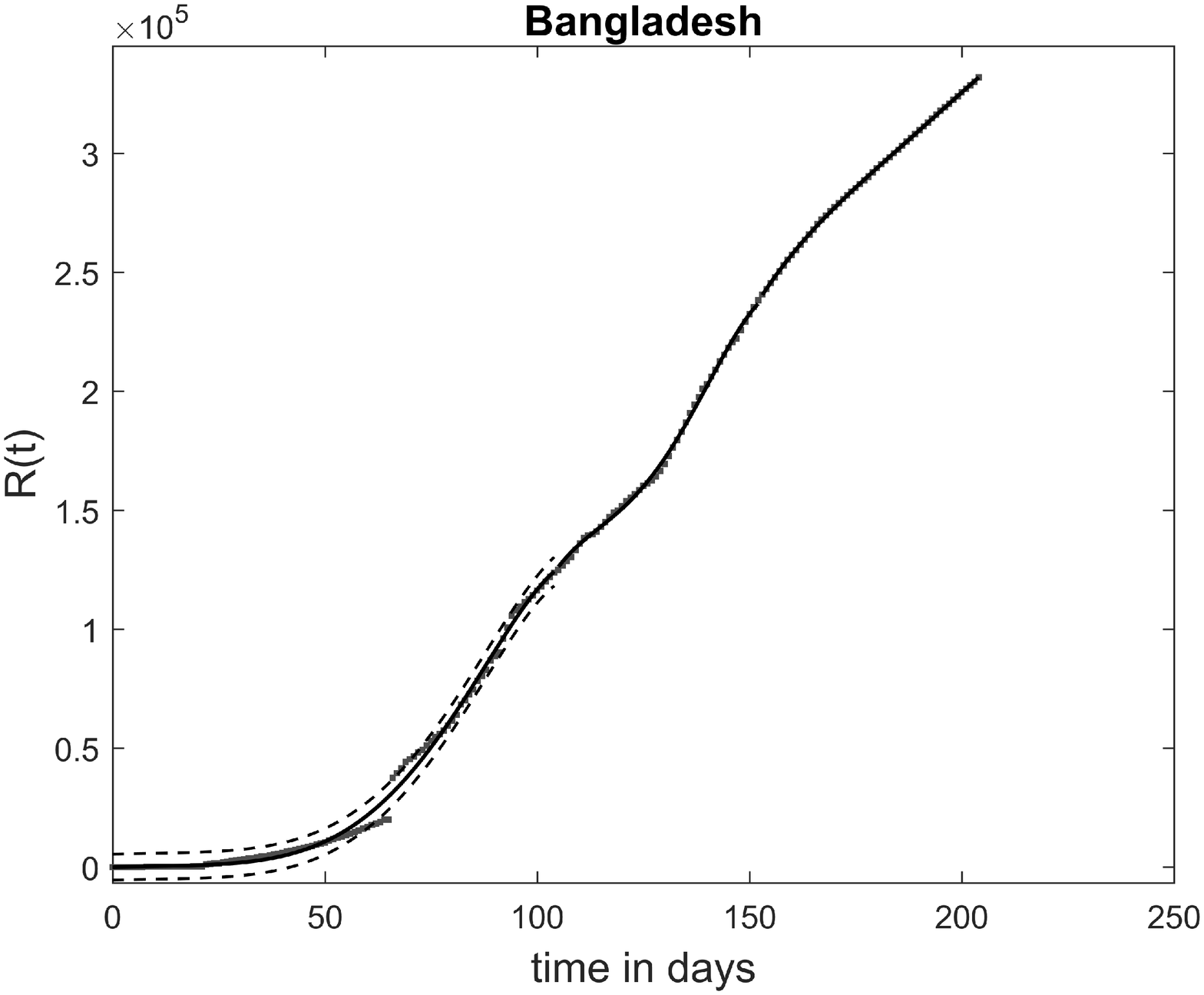}\\
		\includegraphics[scale=0.25]{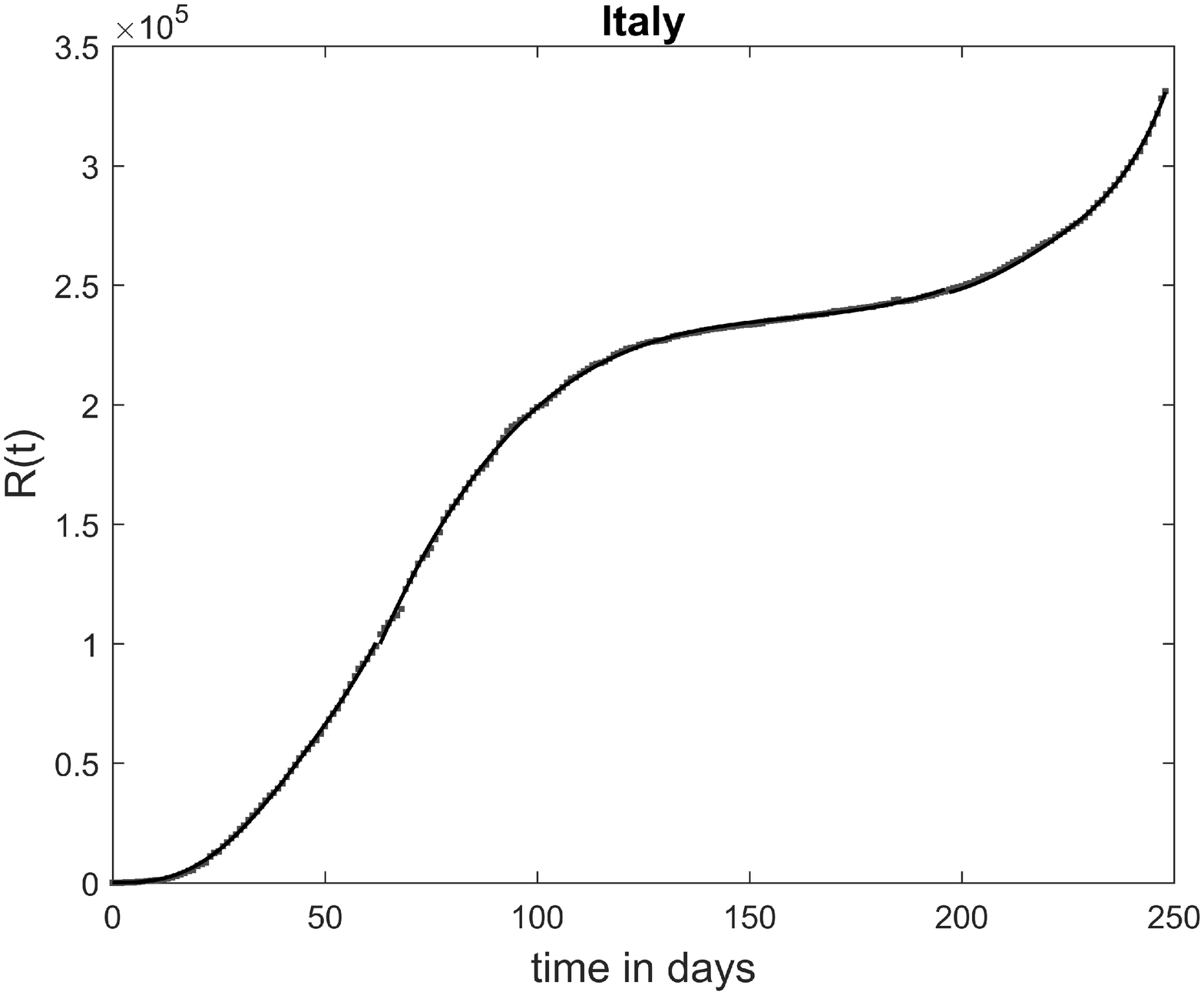}&
		\includegraphics[scale=0.25]{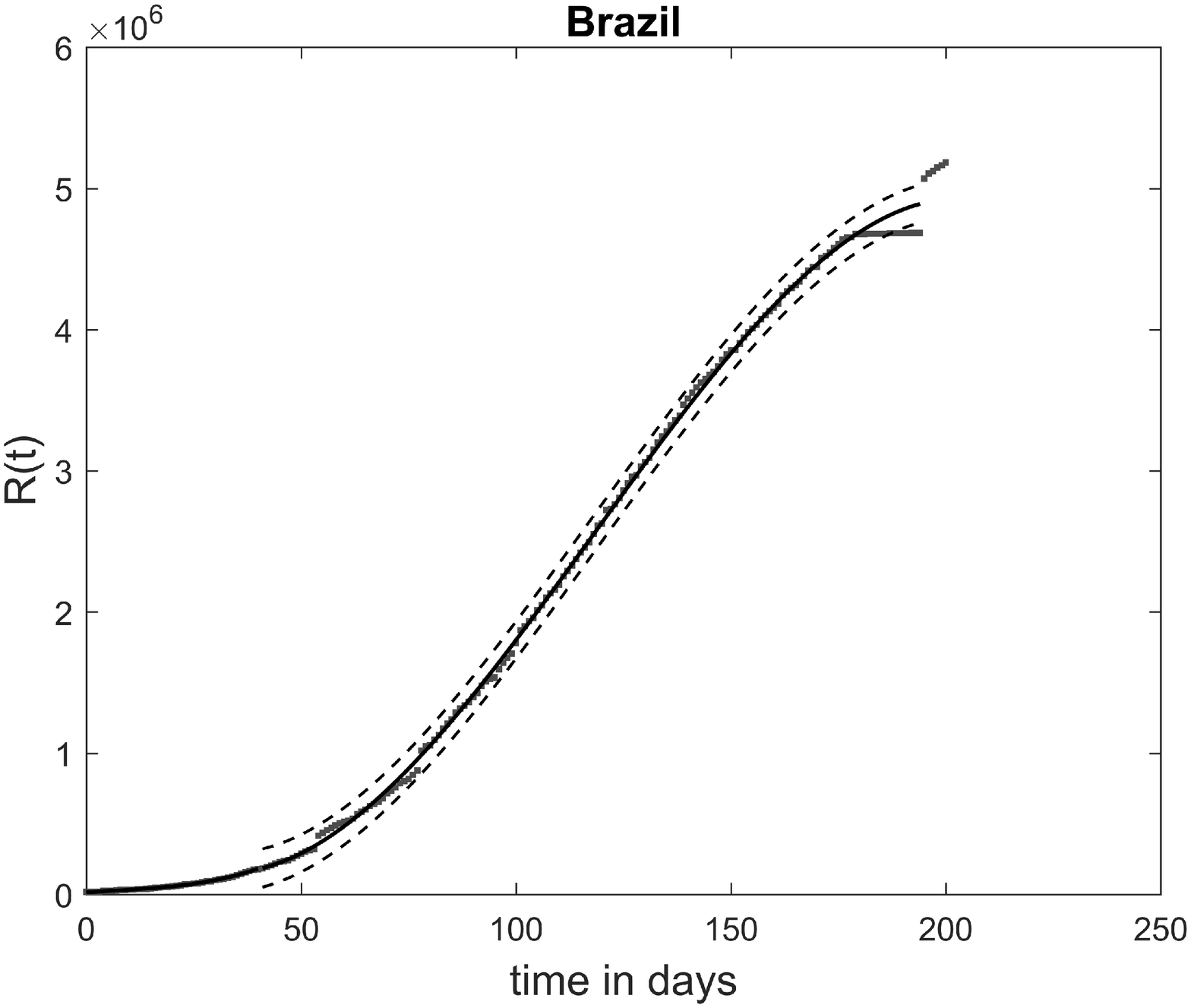}
	\end{tabular}
	\centering
	\caption{{\small Piece-wise continuous function fitting to the data of removed cases ($R$) for different countries.} \label{fig9}}
\end{figure}

\begin{figure}[H]
	\begin{tabular}{cc}
		\includegraphics[scale=0.25]{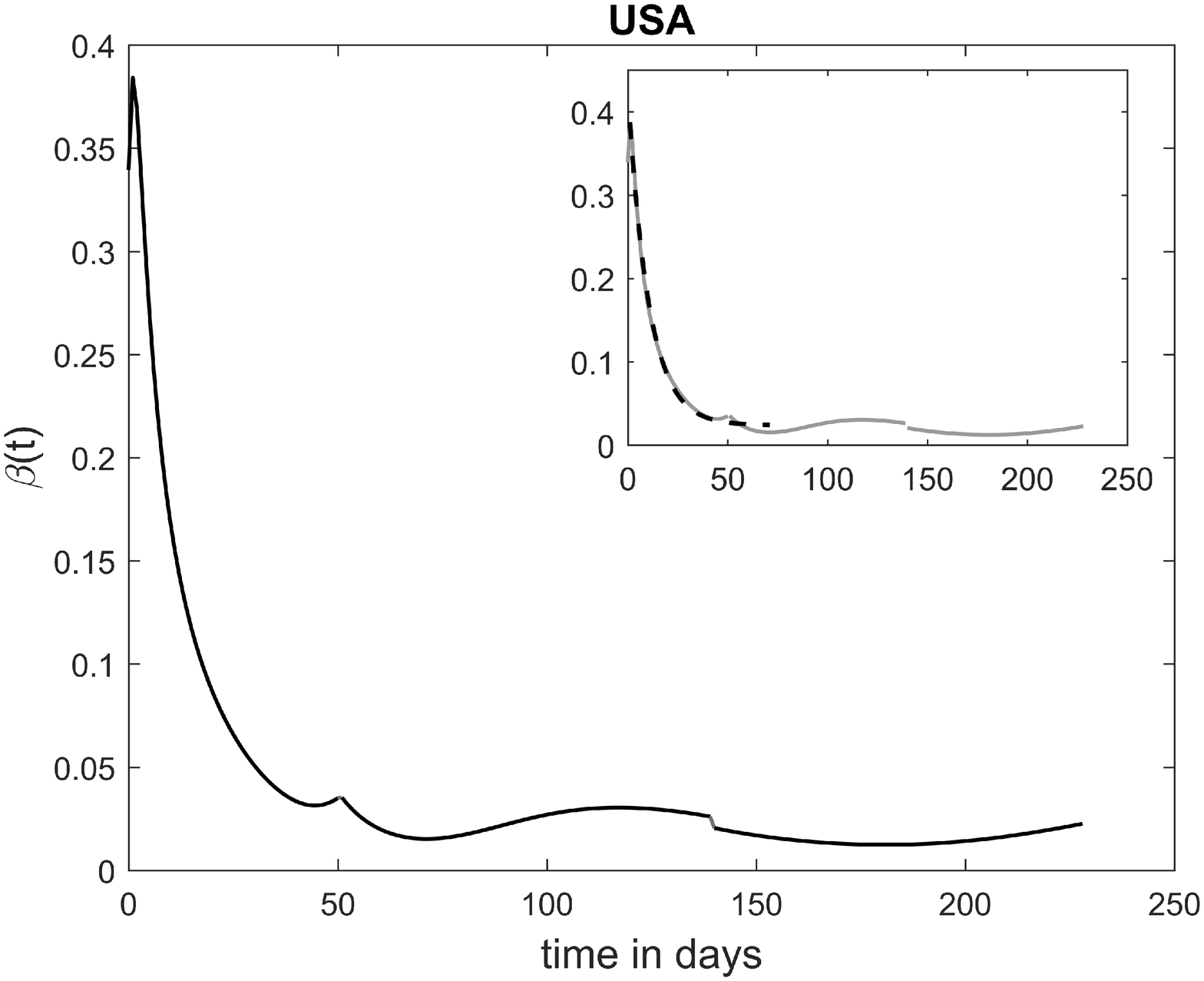}&
		\includegraphics[scale=0.25]{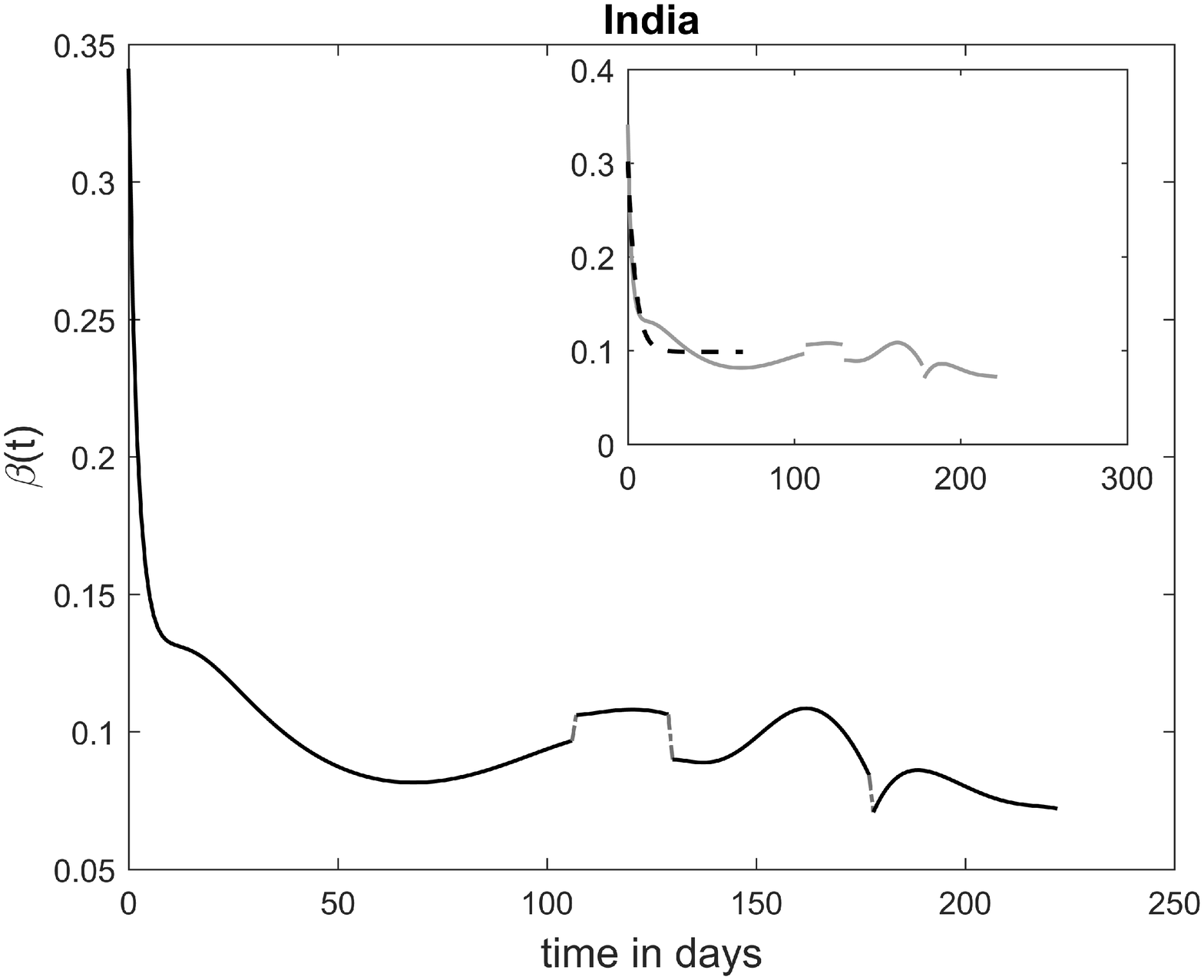}\\
		\includegraphics[scale=0.25]{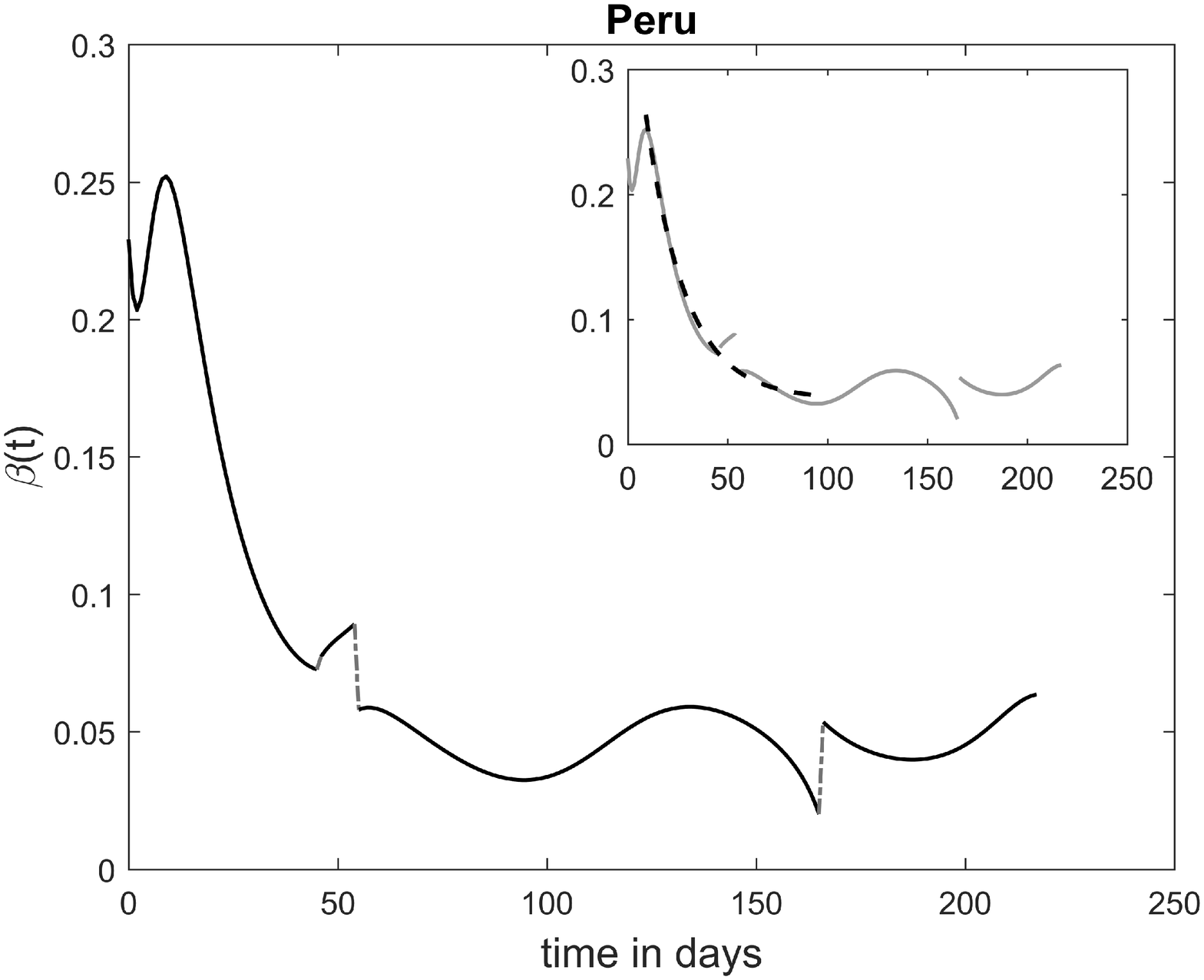}&
		\includegraphics[scale=0.25]{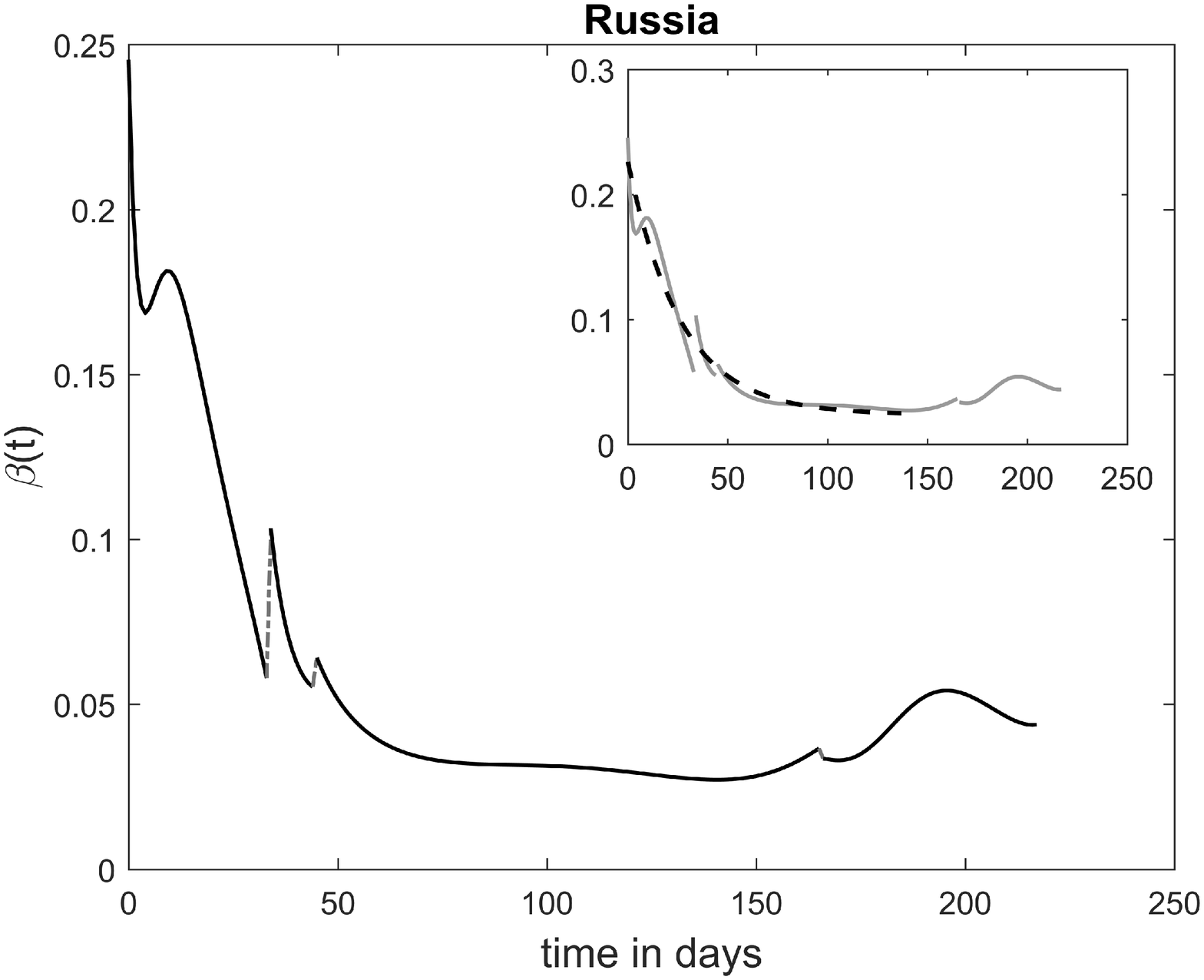}\\
		\includegraphics[scale=0.25]{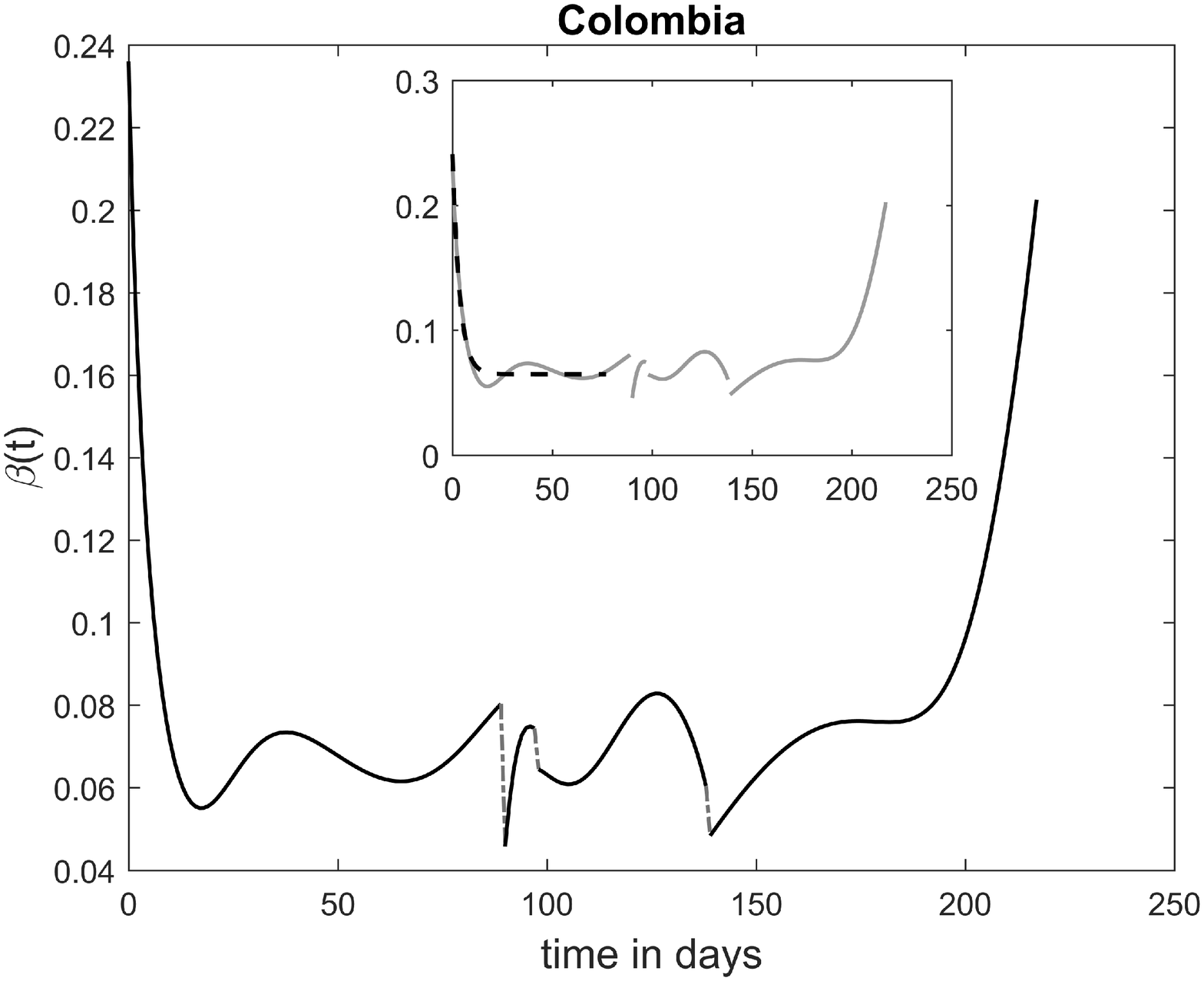}&
		\includegraphics[scale=0.25]{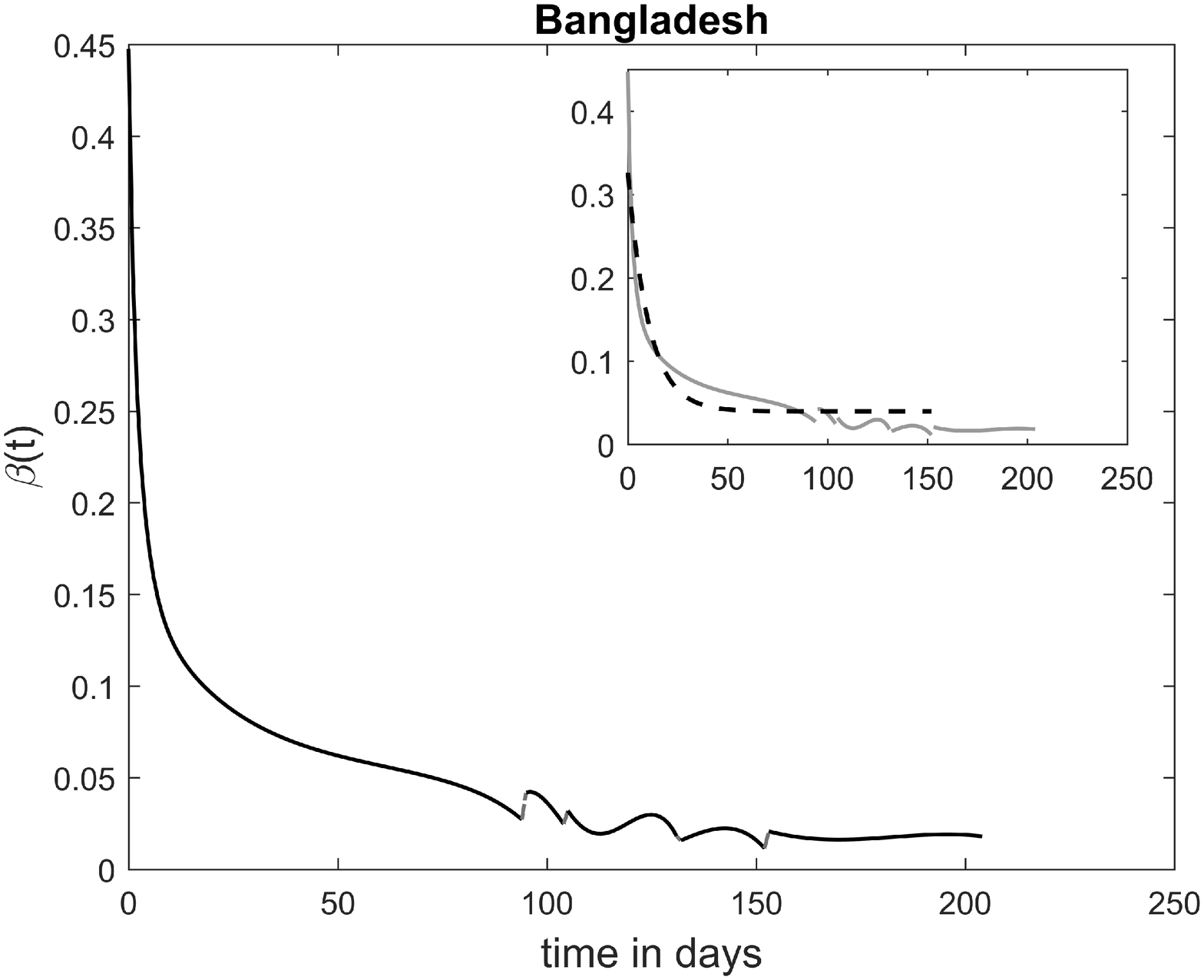}\\
		\includegraphics[scale=0.25]{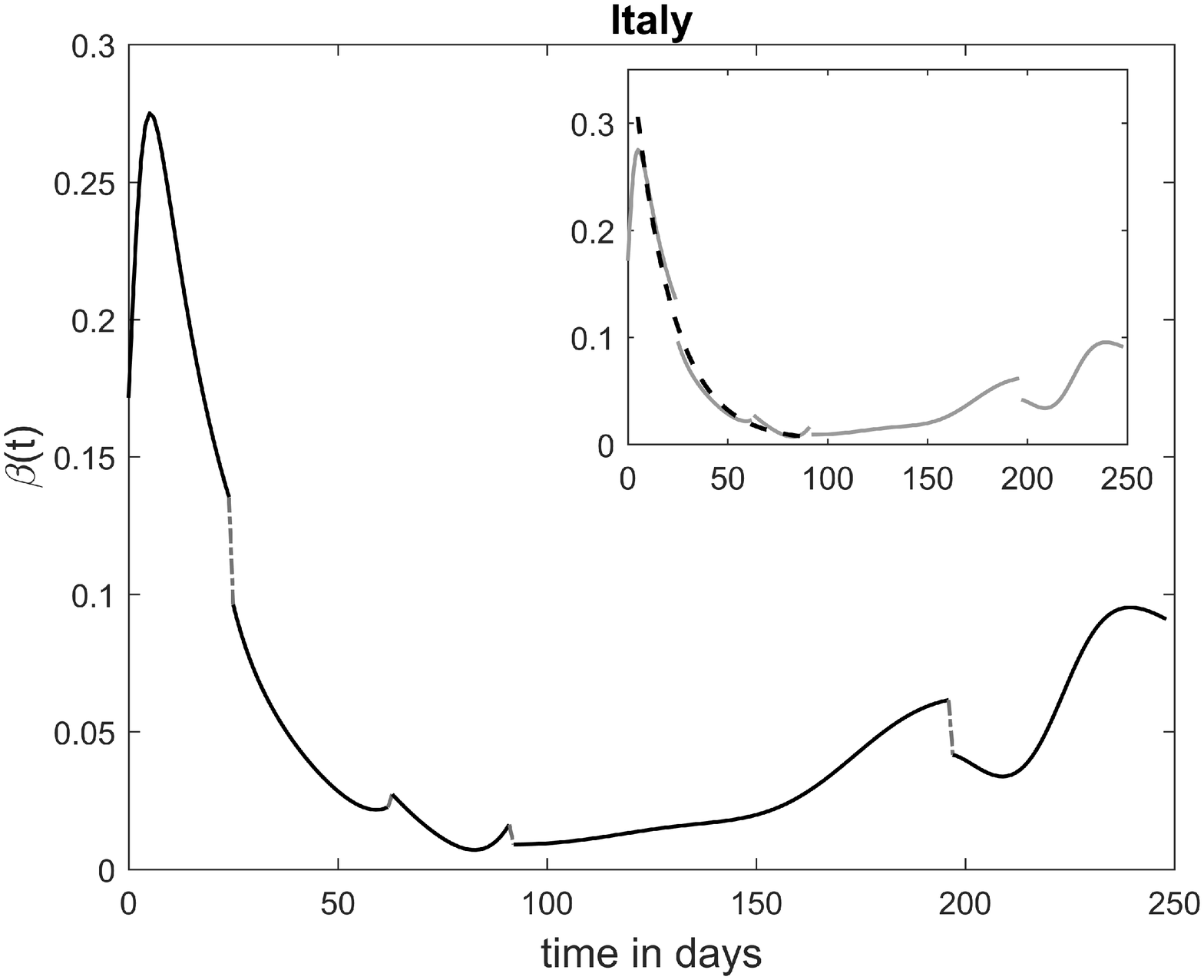}&
		\includegraphics[scale=0.25]{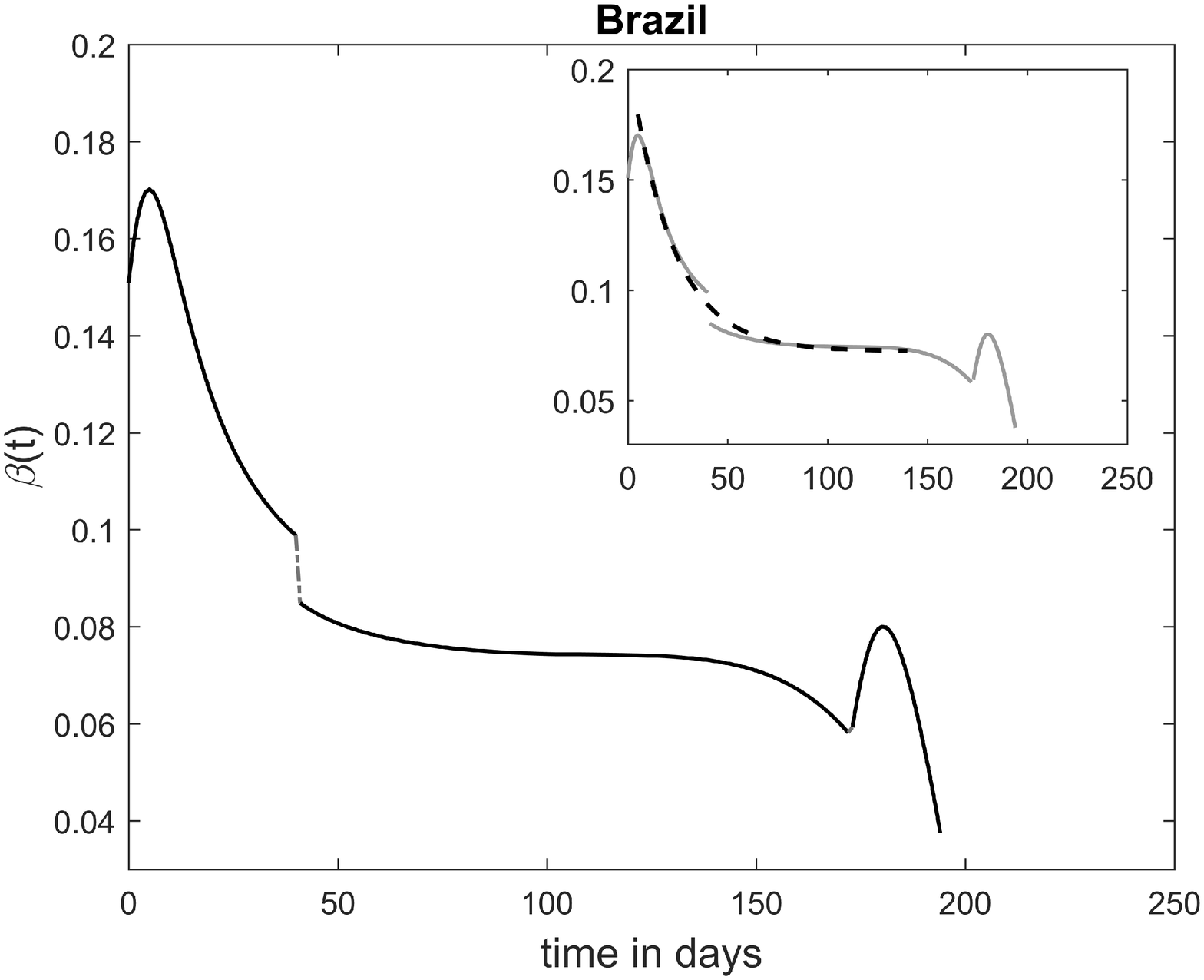}
	\end{tabular}
	\centering
	\caption{{\small Time variation of infection rate ($\beta$) 
		of various countries. Insets of the infection rate plots show the
		part in which we fit it exponentially.} \label{fig10}}
\end{figure}

\begin{figure}[H]
	\begin{tabular}{cc}
		\includegraphics[scale=0.25]{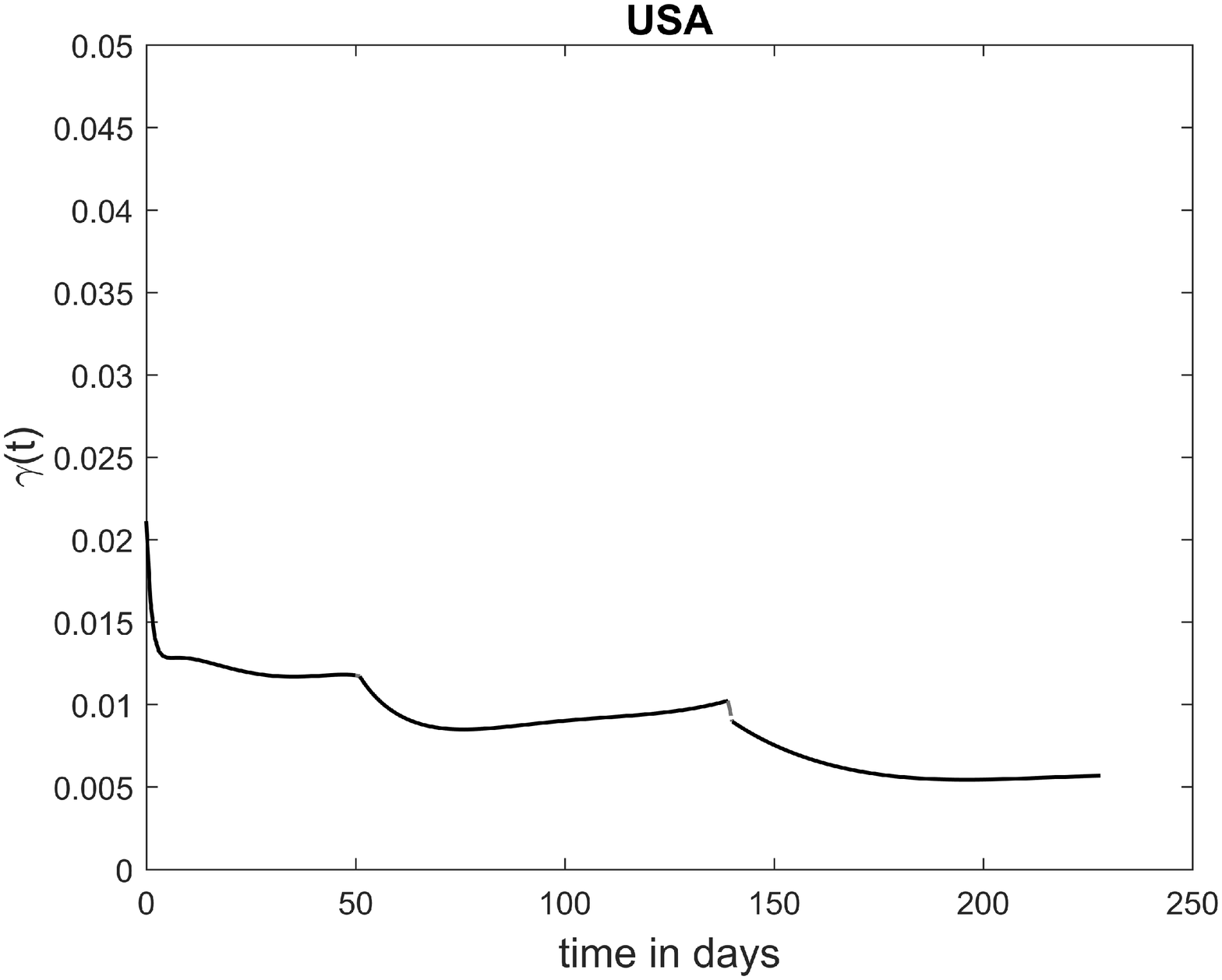}&
		\includegraphics[scale=0.25]{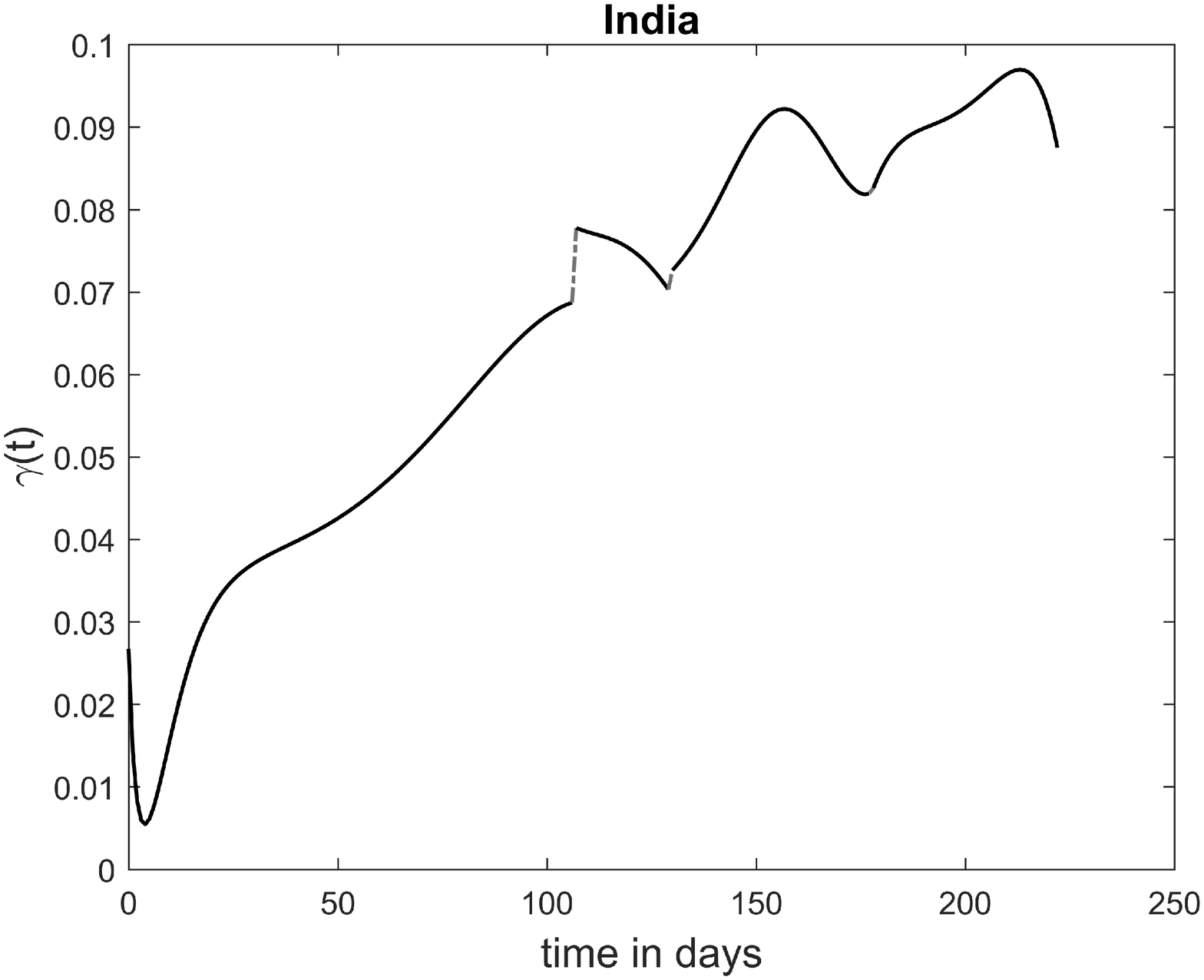}\\
		\includegraphics[scale=0.25]{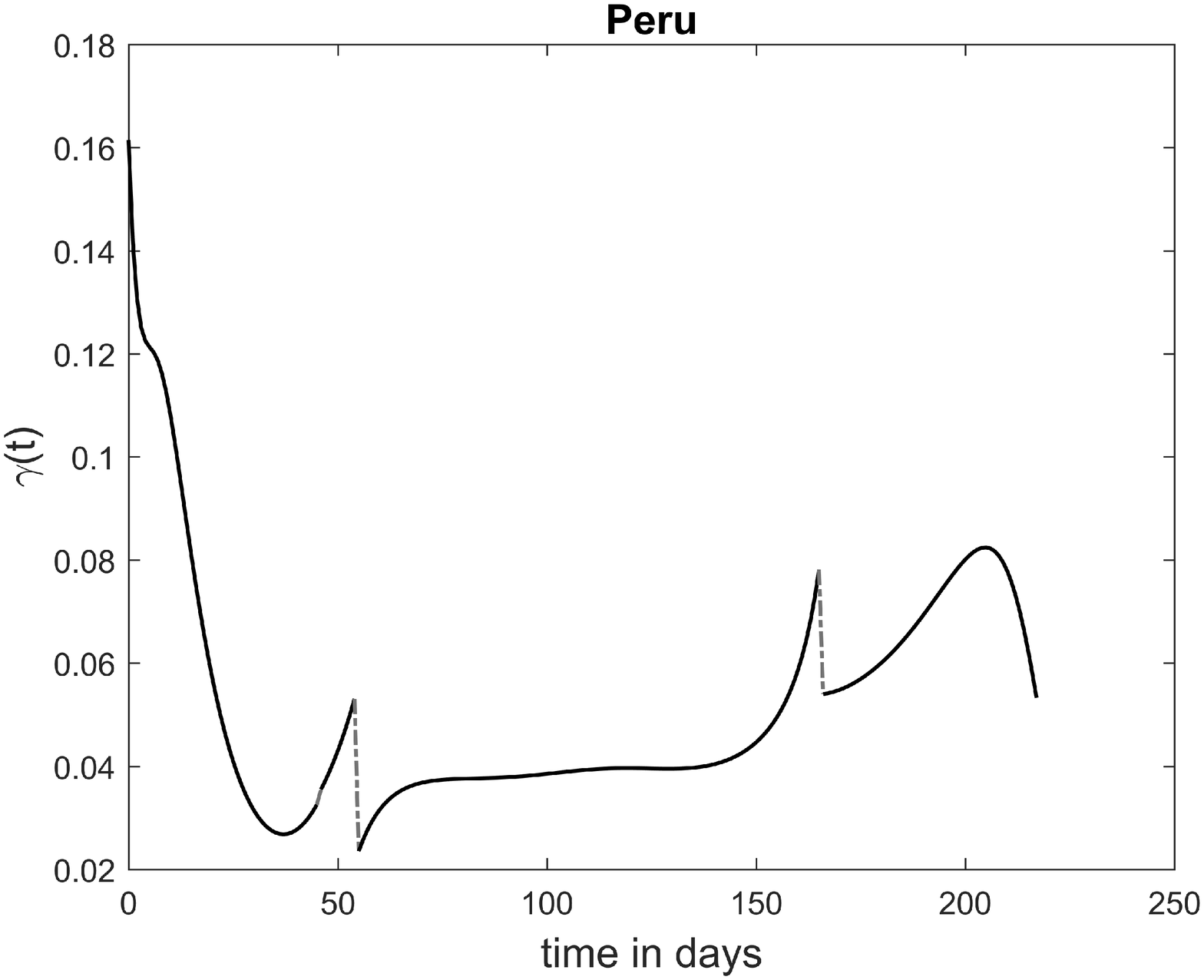}&
		\includegraphics[scale=0.25]{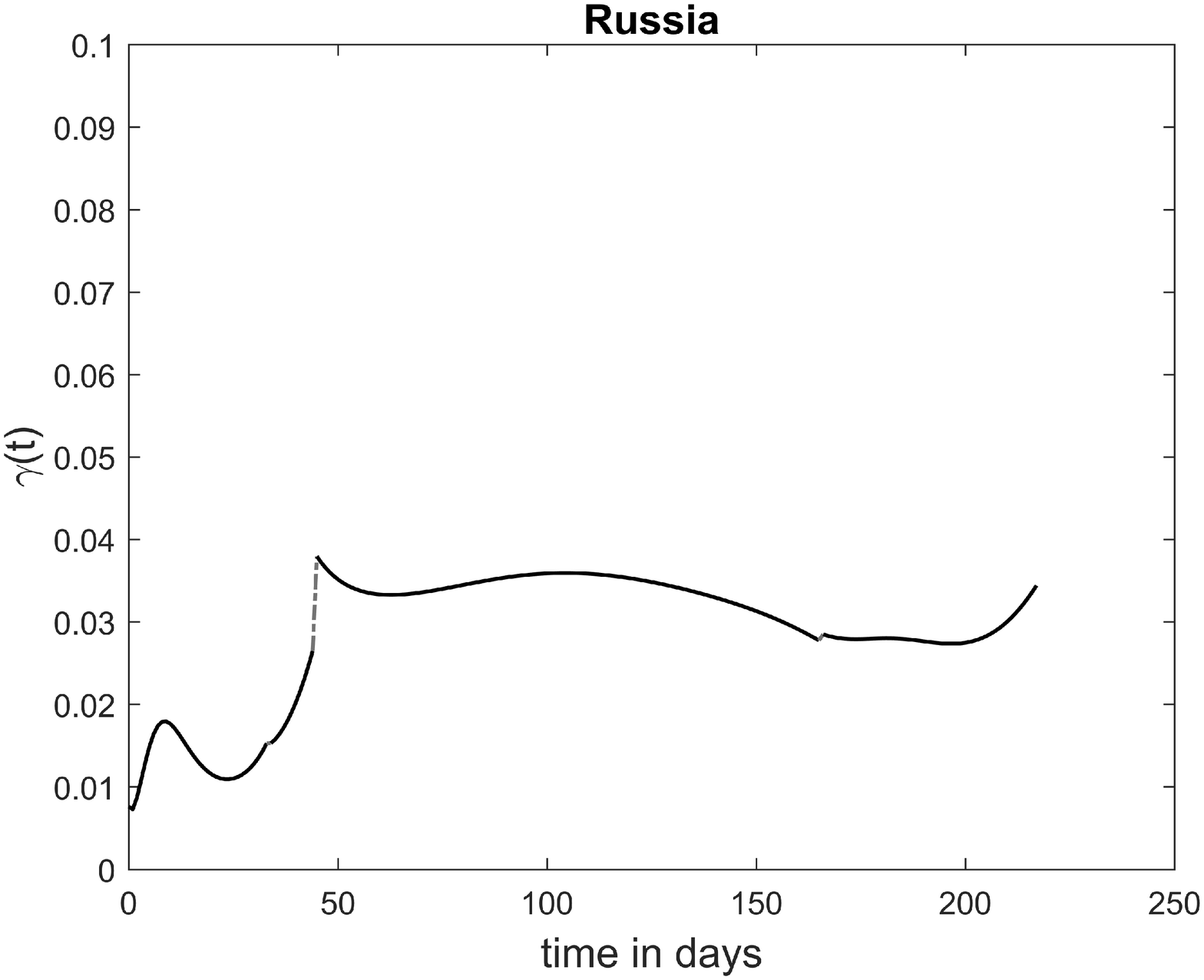}\\
		\includegraphics[scale=0.25]{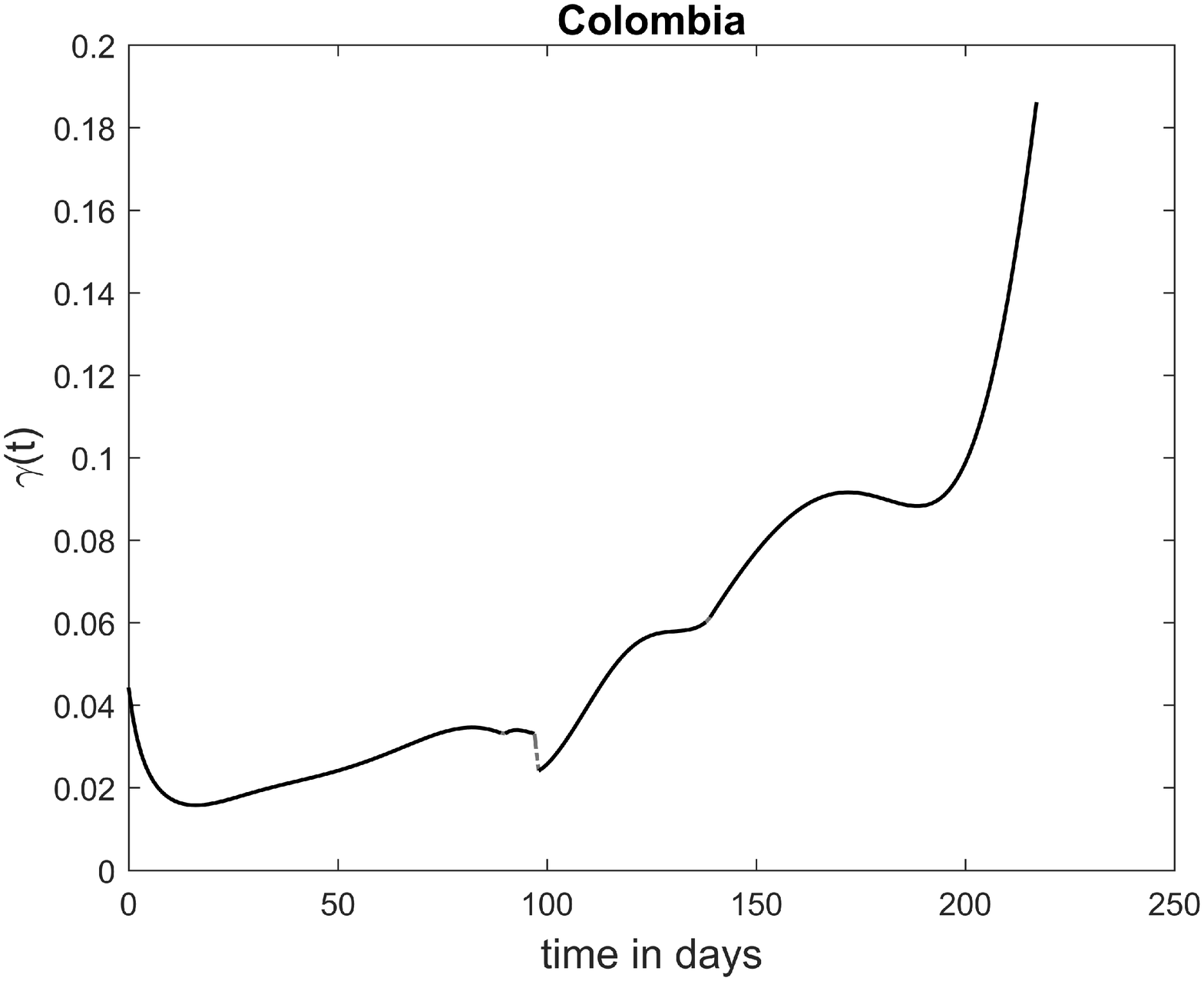}&
		\includegraphics[scale=0.25]{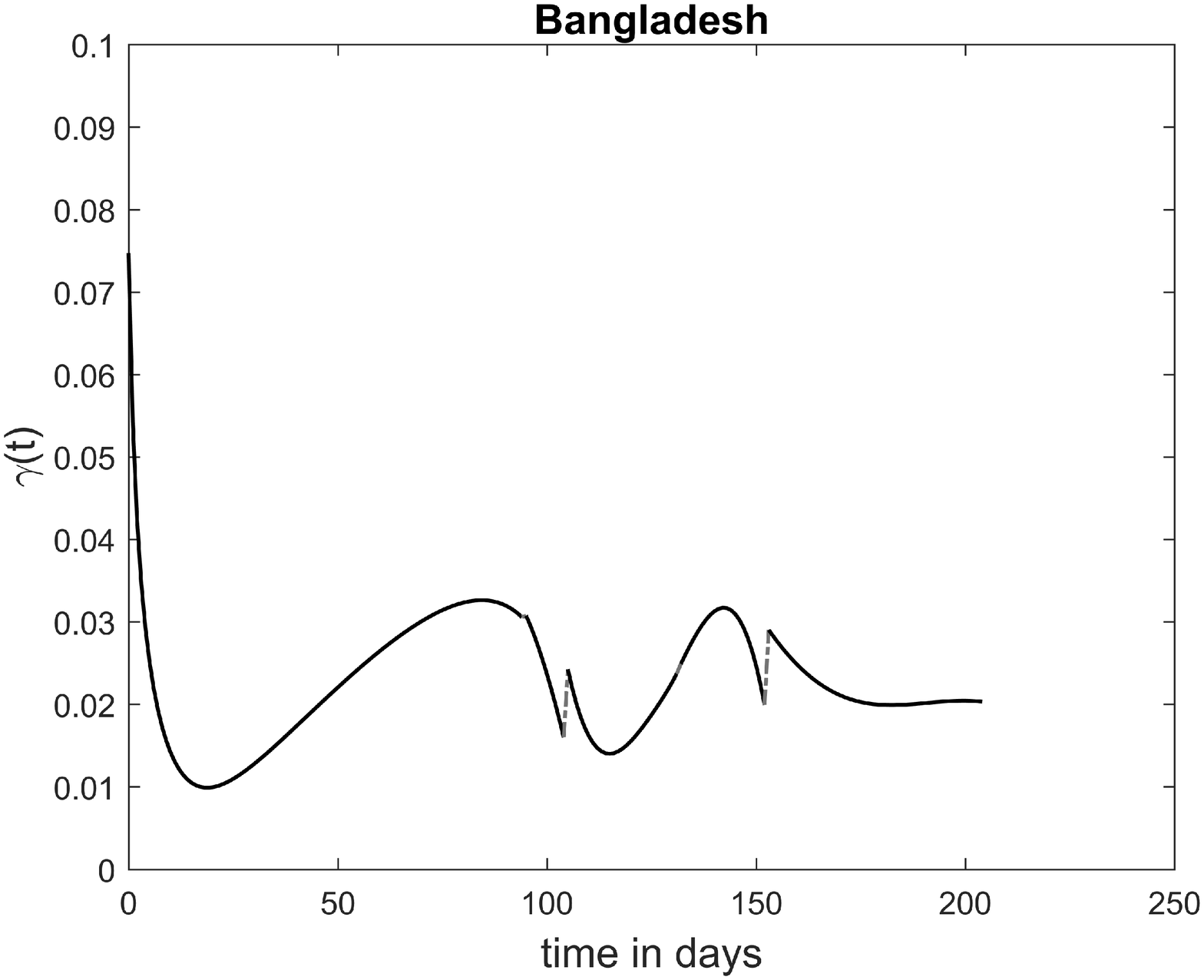}\\
		\includegraphics[scale=0.25]{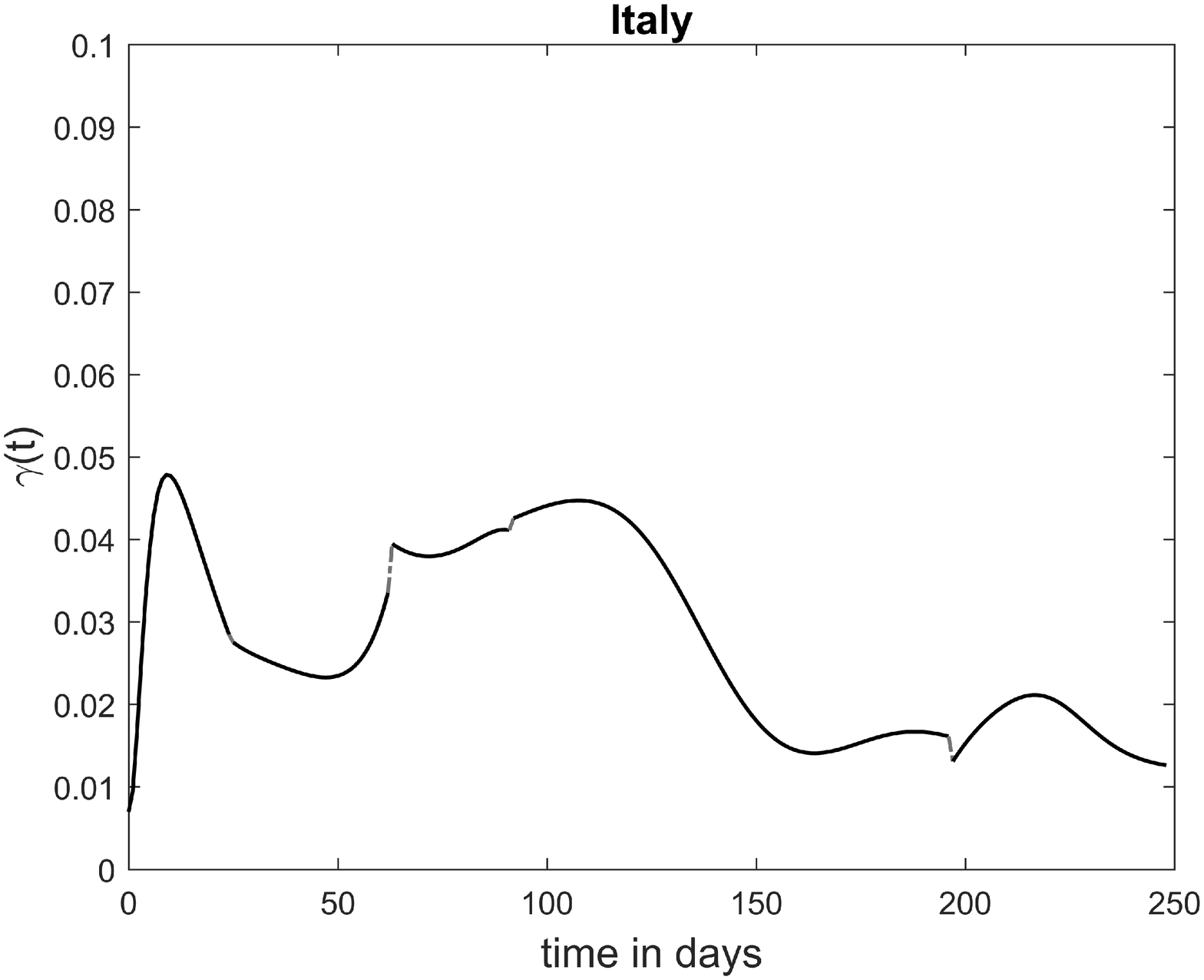}&
		\includegraphics[scale=0.25]{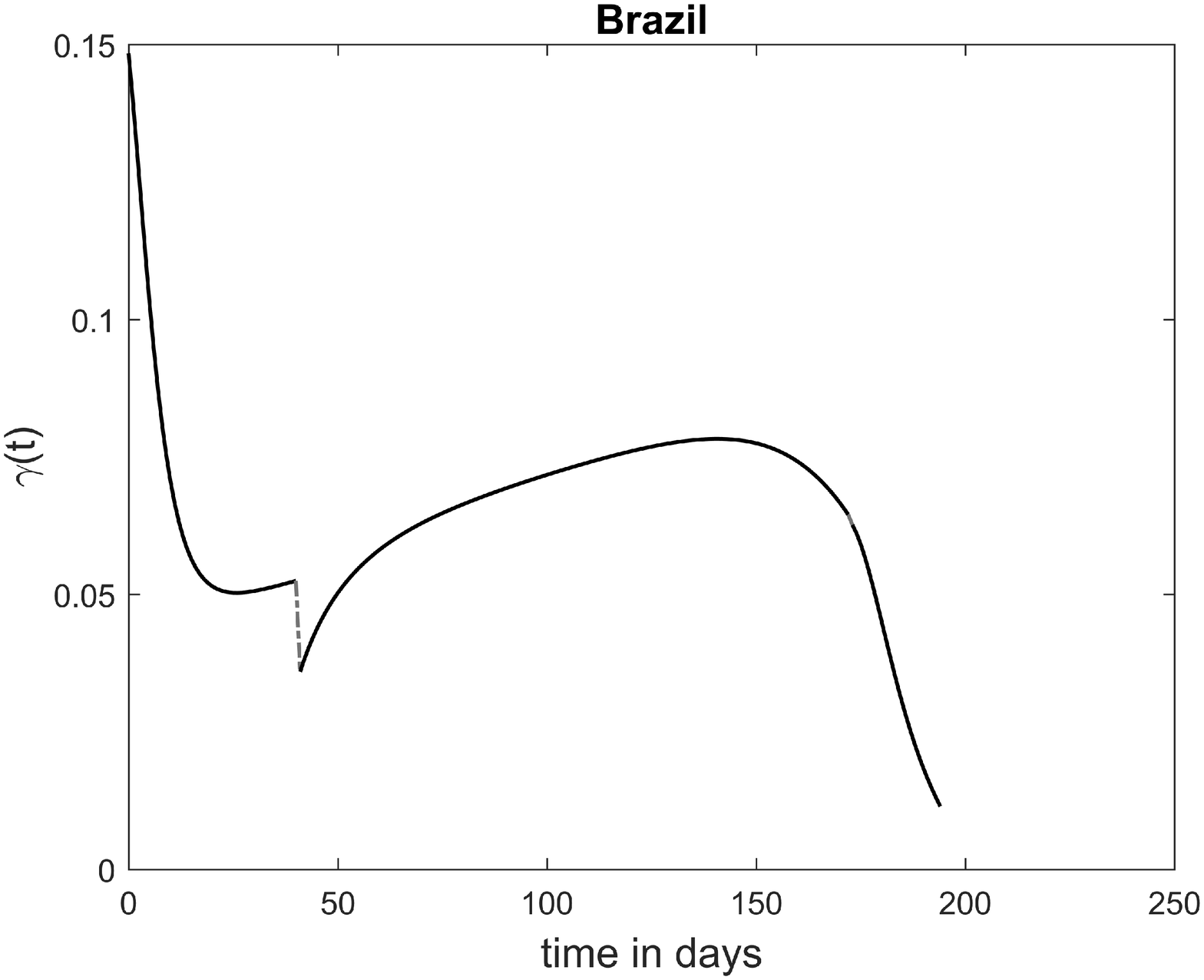}
	\end{tabular}
	\centering
	\caption{{\small Time variation of removed rate ($\gamma$) 
		of various countries.} \label{fig11}}
\end{figure}

Figures \ref{fig8} and \ref{fig9} show how the active cases and removed cases of various countries are varying with time. In figures \ref{fig8} and \ref{fig9}, the black dashed lines indicates 
the 95\% prediction interval of the
fitted data. We don't show this interval in some plots or some portions of the plot because the intervals are very
close to the best fit line.

Figure \ref{fig8} shows that the active case curves have different behaviors for different countries. Active
	cases curves are increasing for the countries like USA and
	are decreasing in countries like India, Bangladesh, Colombia. Also,
we find that a re-outbreak of the disease has started in Italy, Russia, Brazil
and has ended in Peru. Figures \ref{fig10} and \ref{fig11} show how the infection rate ($\beta$) and removed
rate ($\gamma$) vary with time in different countries. In these figures, the gray dashed lines represents
the discontinuities that come because of the piece-wise continuous
fitting of $I$ and $R$ data and not from the model. So, the discontinuities do not have any physical significance. 
The variation of removed rate is different for different countries.
But removed rates are varying in the order of $\sim10^{-2}$. Besides,
we can roughly say that the removed rates are linearly varying with
time. Countries like Russia, Bangladesh, Italy, and Brazil have removed rate
which is approximately constants with time. Also for India and Colombia,
we can approximate that the removed rate is increasing linearly with
time. In the USA removed rate is approximately decreasing linearly
with time. In the case of Peru, we will also get a linear region in
a particular period of time.

The variation of $\beta$ is also different for different countries. However, the initial variation of $\beta$ for
different countries are approximately the same. We can see that initially,
all the $\beta$s have an approximately exponential decrease. Here
we take an exponentially decreasing function $f(t)=ae^{-b(t-t_{0})}+c$.
We fit this function to that portion of the $\beta$ in which it is approximately decreasing exponentially. 

\begin{table}[H]
	\tbl{This table shows the parameter values of the fitted function in different
		countries. This also shows the time range in which $\beta$ is decreasing
		exponentially.}
	{\begin{tabular}{@{}cccccc@{}} \toprule
			Country\hphantom{000} & a & b & c & $t_{0}$ in day & time  range in days \\ \colrule
			USA\hphantom{000000} & 0.3639 & 0.0958 & 0.02372 & 1 & \hphantom{0}1-71 \\
			India\hphantom{00000} & 0.2032 & 0.2231 & 0.09874 & 0 & \hphantom{0}0-69 \\
			Peru\hphantom{000000} & 0.2278 & 0.05027 & 0.03617 & 9 & \hphantom{0}9-95 \\
			Russia\hphantom{0000} & 0.2024 & 0.03751 & 0.02375 & 0 & 0-137 \\
			Colombia\hphantom{00} & 0.1762 & 0.2803 & 0.06477 & 0 & \hphantom{0}0-77 \\
			Bangladesh & 0.2865 & 0.09588 & 0.03955 & 0 & 0-152 \\
			Italy\hphantom{00000} & 0.3026 & 0.05244 & 0.003245 & 5 & \hphantom{0}5-94 \\ 
			Brazil\hphantom{00000} &0.1074&0.04652&0.07227&10&5-140\\ \botrule
		\end{tabular} \label{ta2}}
\end{table}

Table \ref{ta2} shows the parameter values of the fitted function
in different countries. We can see that initially, the fitted function
has values of order $\sim10^{-1}$ for all of the countries. Then
the values of this function quickly decreased to an order of $\sim10^{-2}$
 (except Italy). So, we can see that there
lies a universal feature among the infection rates ($\beta$) for
a limited period of time.

\begin{figure}[H]
	\begin{tabular}{cc}
		\includegraphics[scale=0.25]{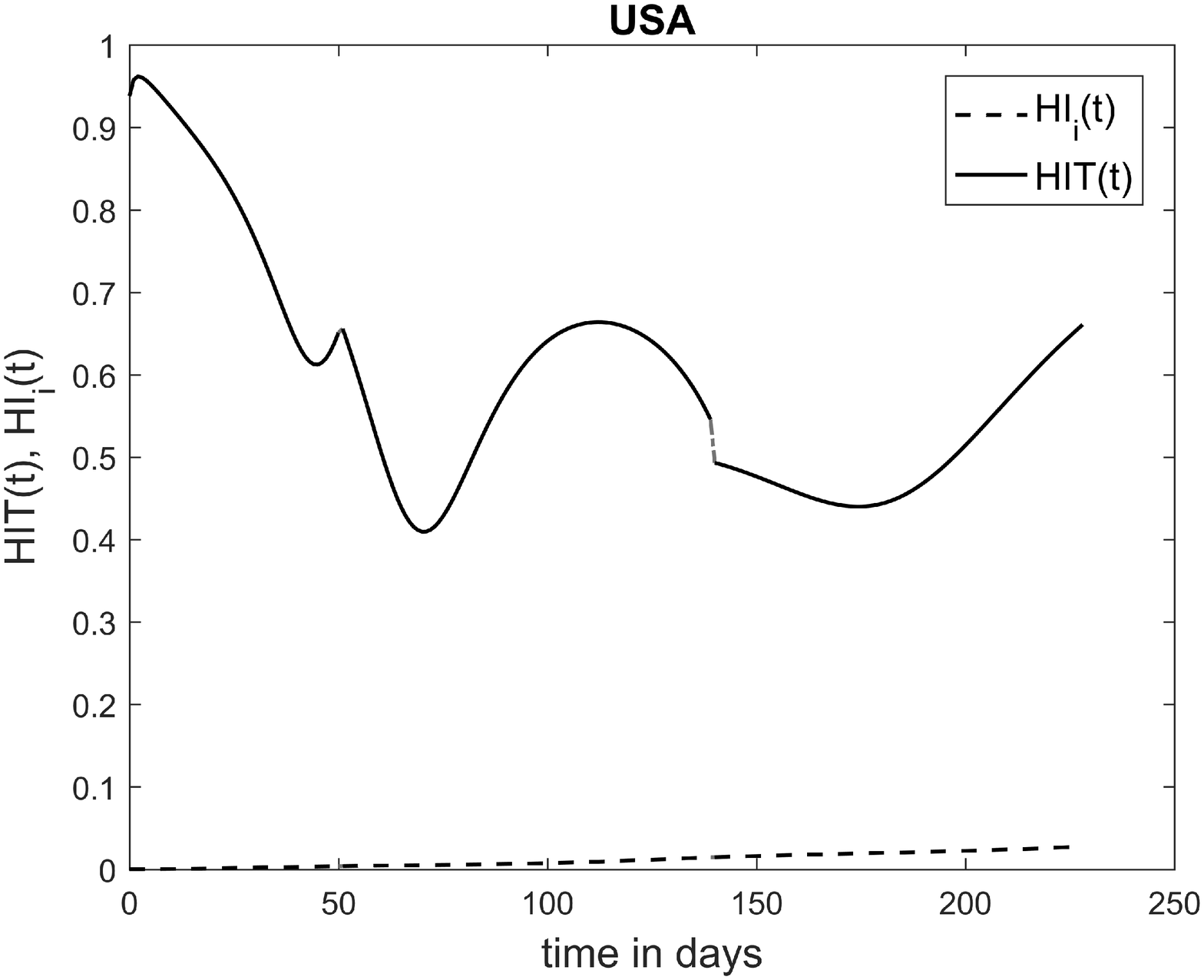}&
		\includegraphics[scale=0.25]{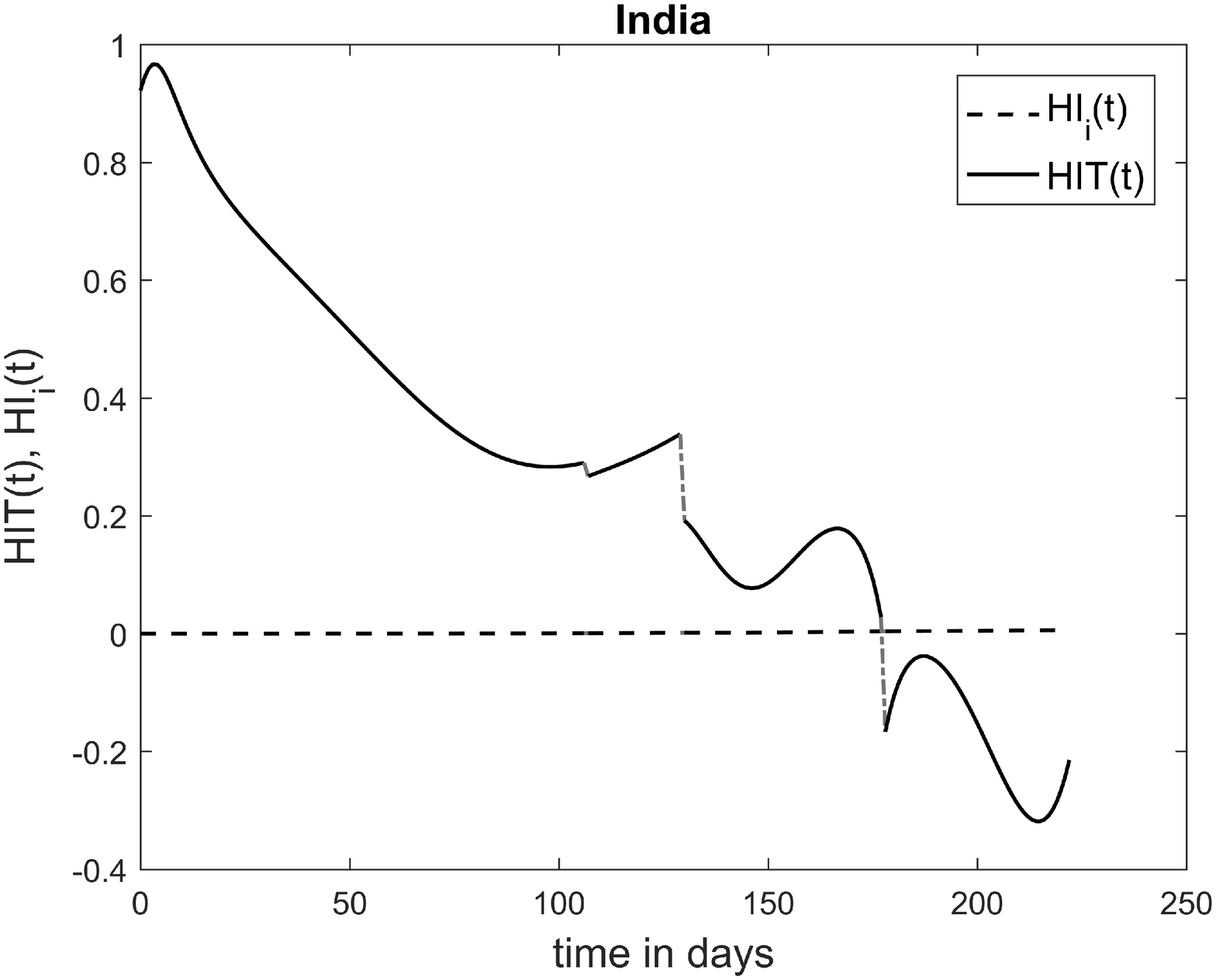}\\
		\includegraphics[scale=0.25]{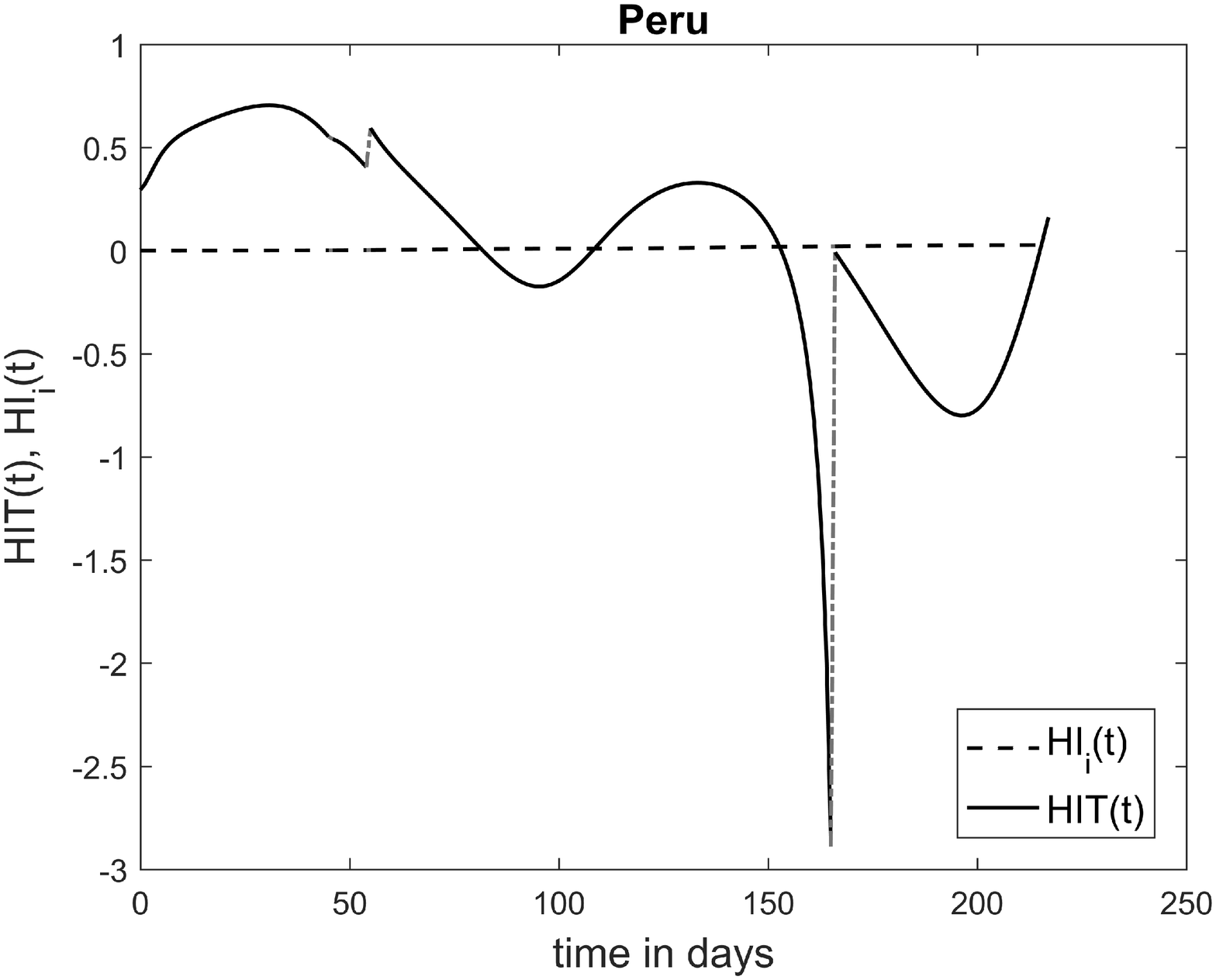}&
		\includegraphics[scale=0.25]{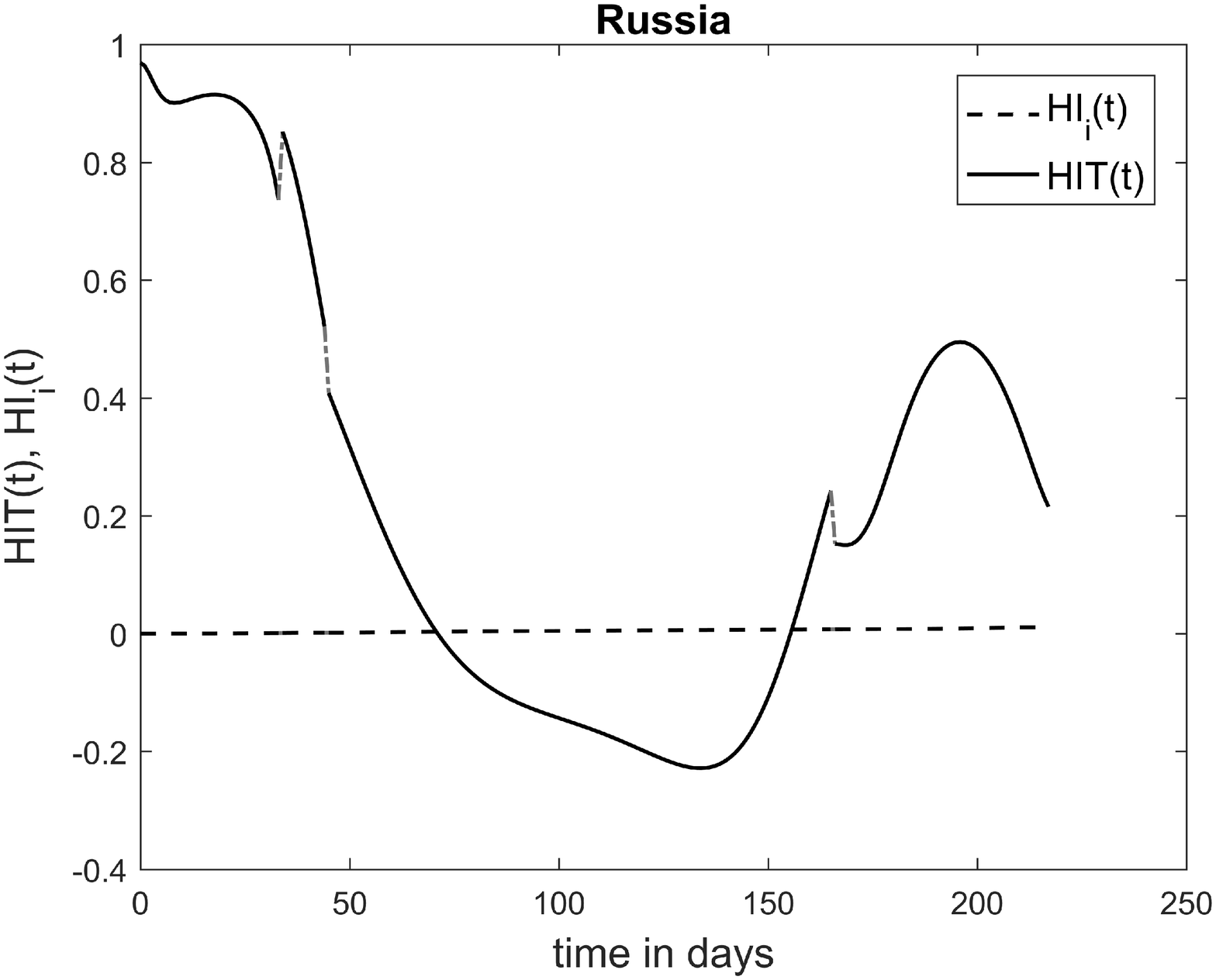}\\
		\includegraphics[scale=0.25]{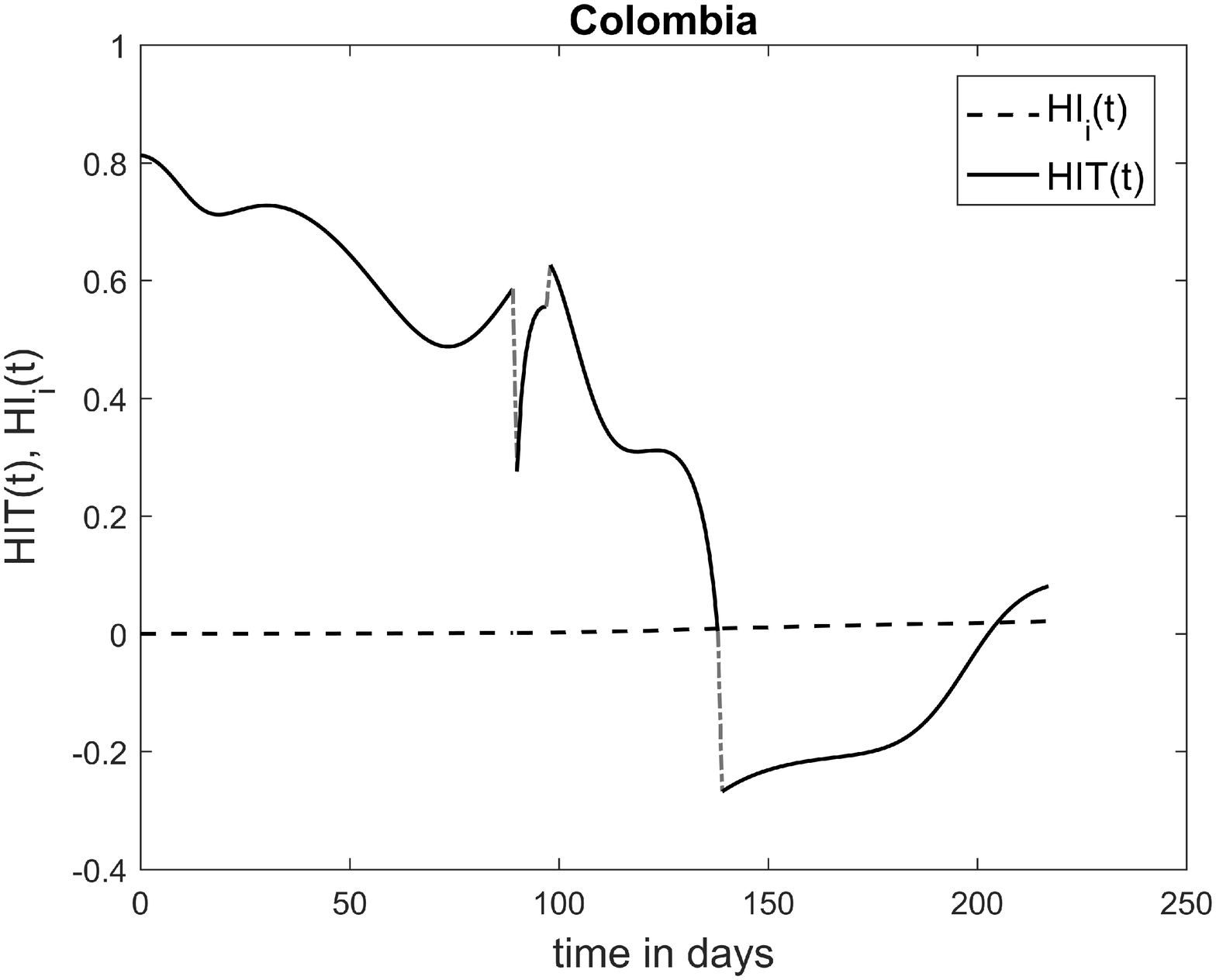}&
		\includegraphics[scale=0.25]{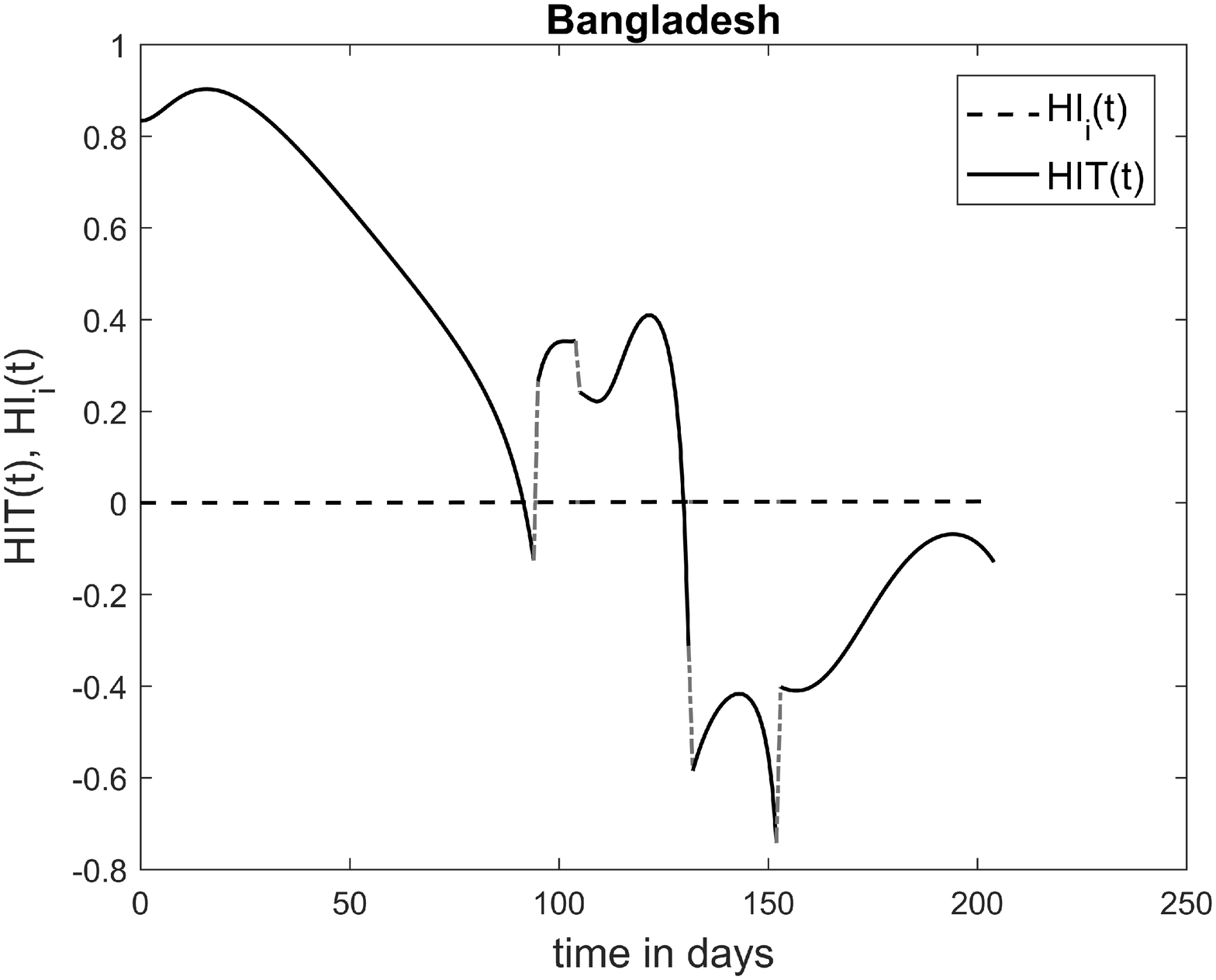}\\
		\includegraphics[scale=0.25]{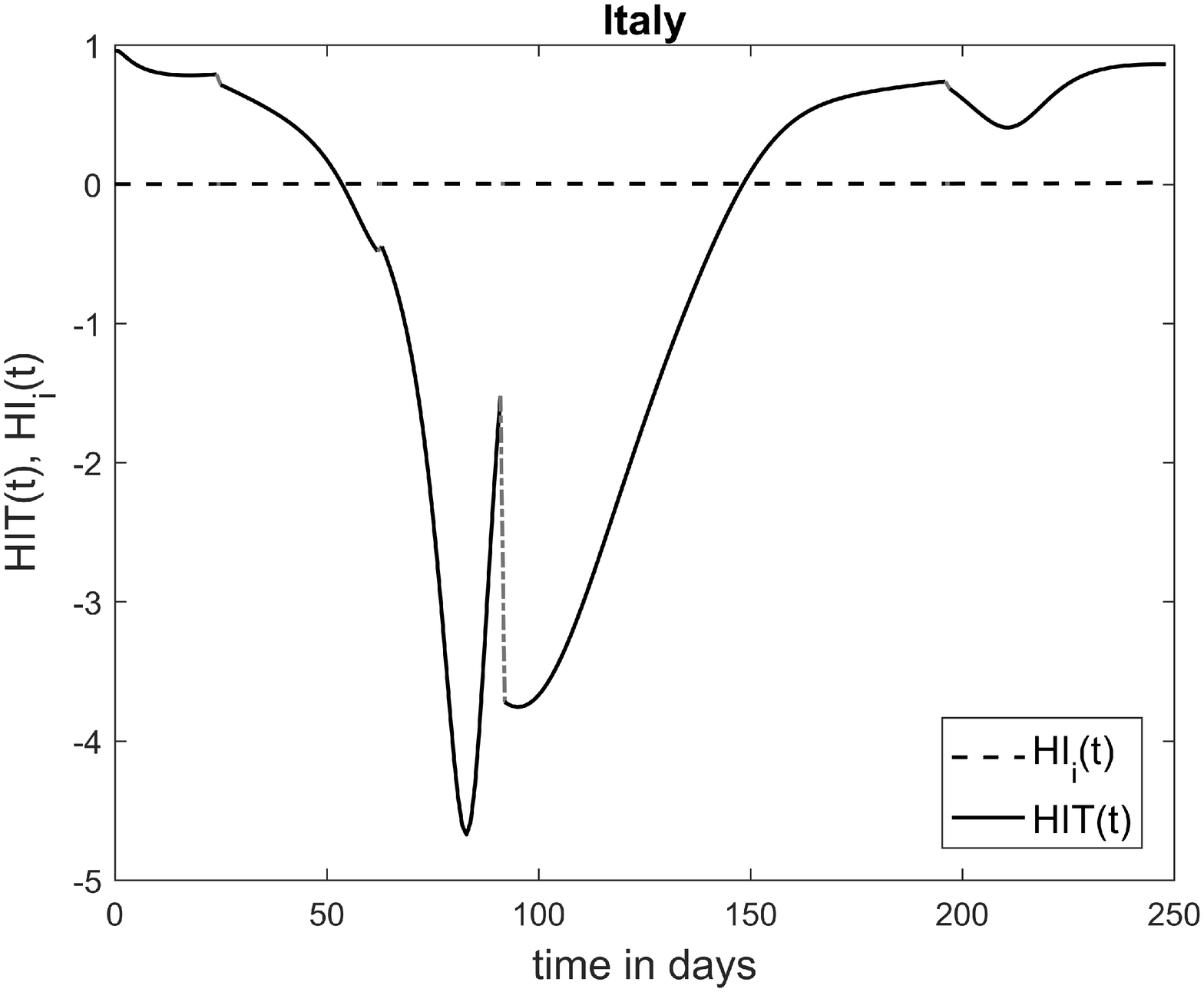}&
		\includegraphics[scale=0.25]{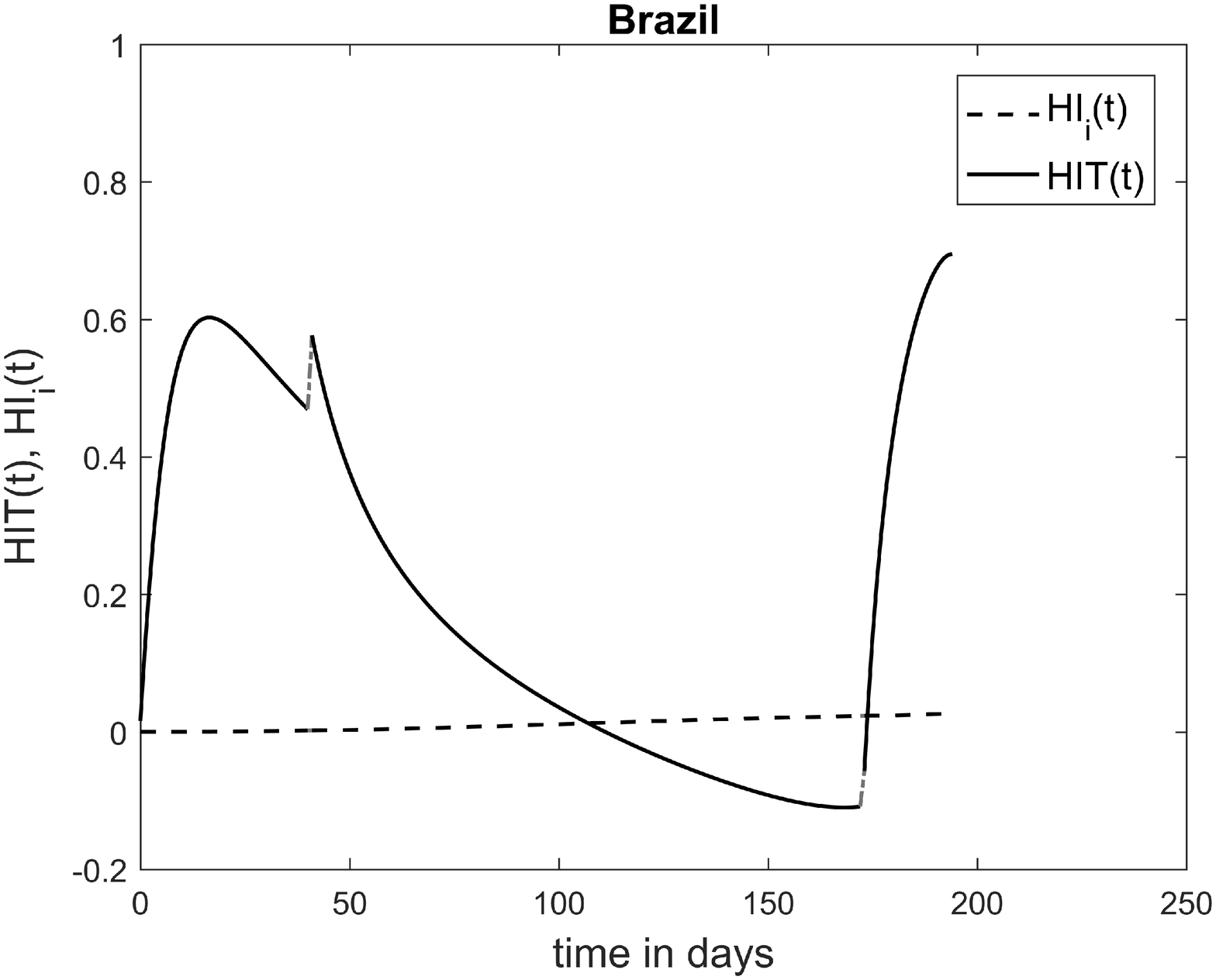}
	\end{tabular}
	\centering
	\caption{{\small Time variation of instantaneous herd immunity ($HI_{i}$), herd immunity
		threshold ($HIT$) of
		various countries.} \label{fig12}}
\end{figure}

\begin{figure}[H]
	\begin{tabular}{cc}
		\includegraphics[scale=0.25]{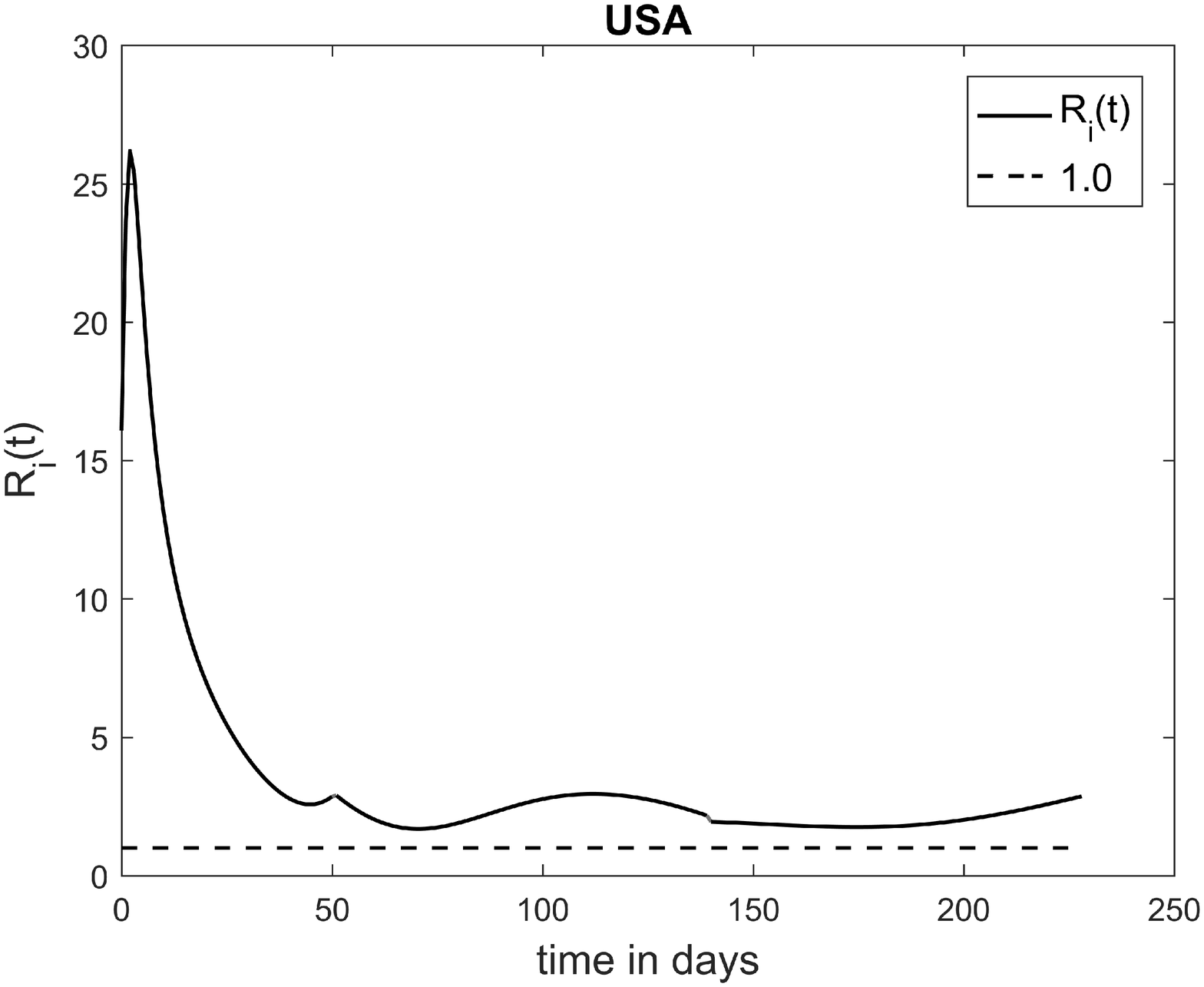}&
		\includegraphics[scale=0.25]{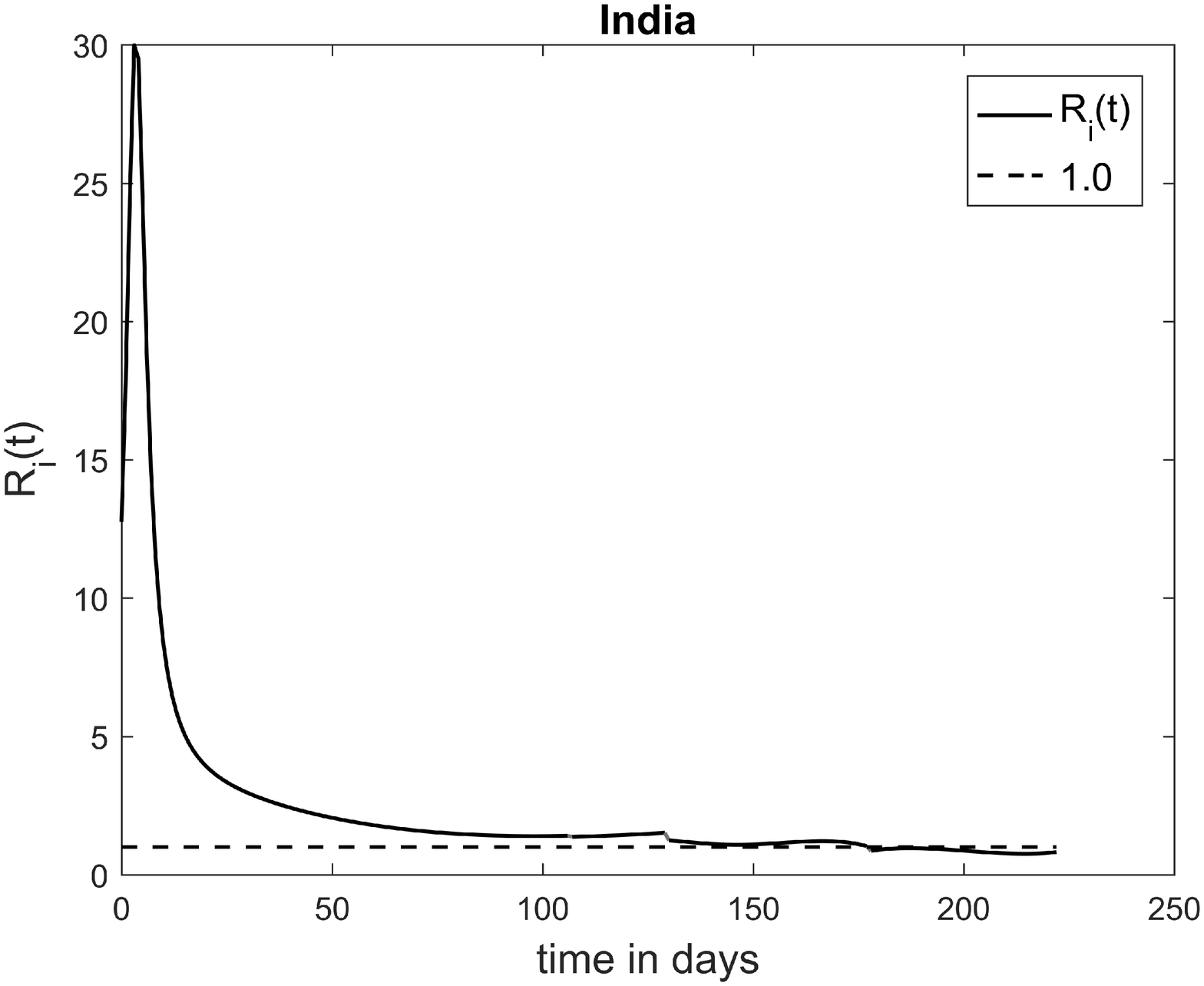}\\
		\includegraphics[scale=0.25]{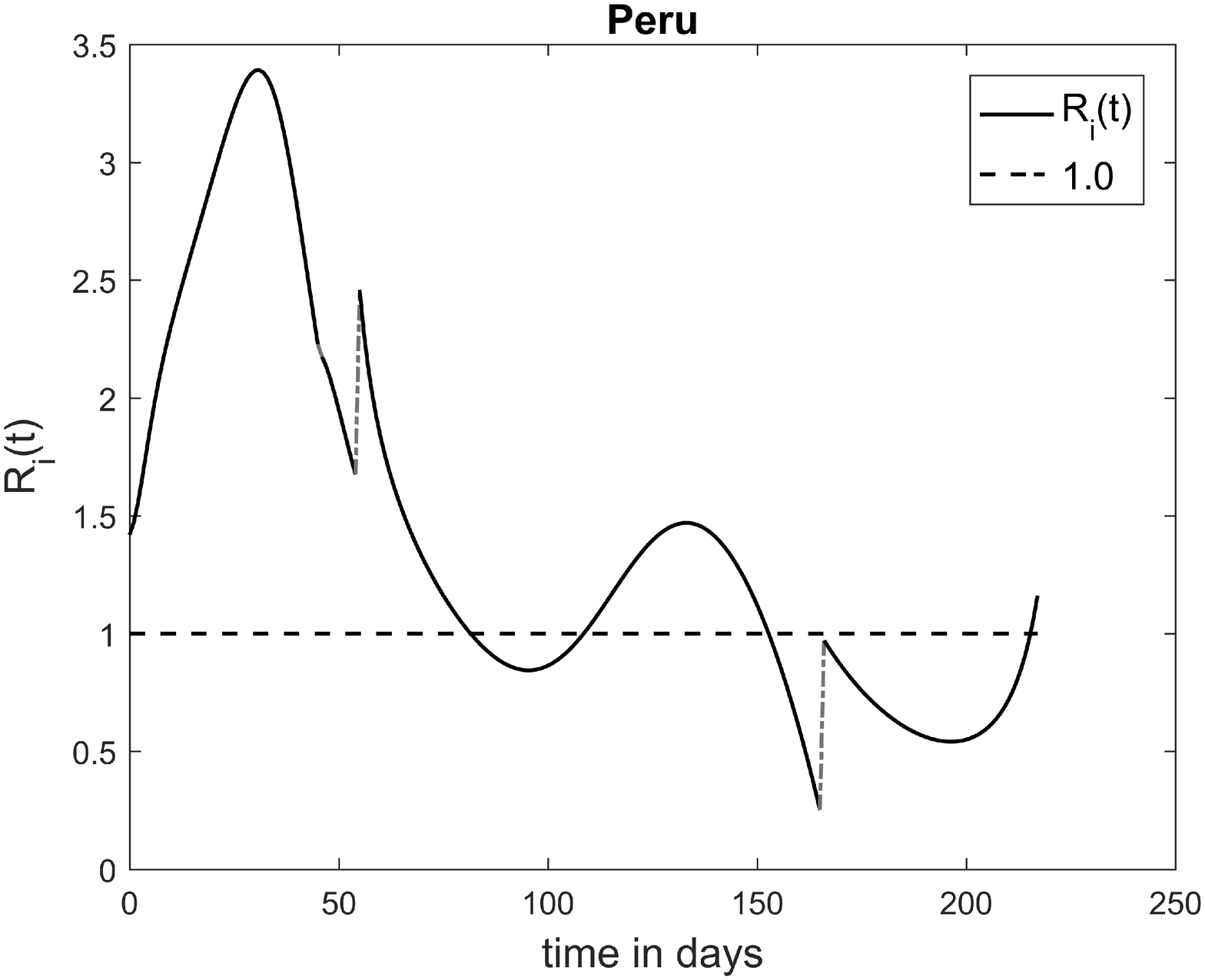}&
		\includegraphics[scale=0.25]{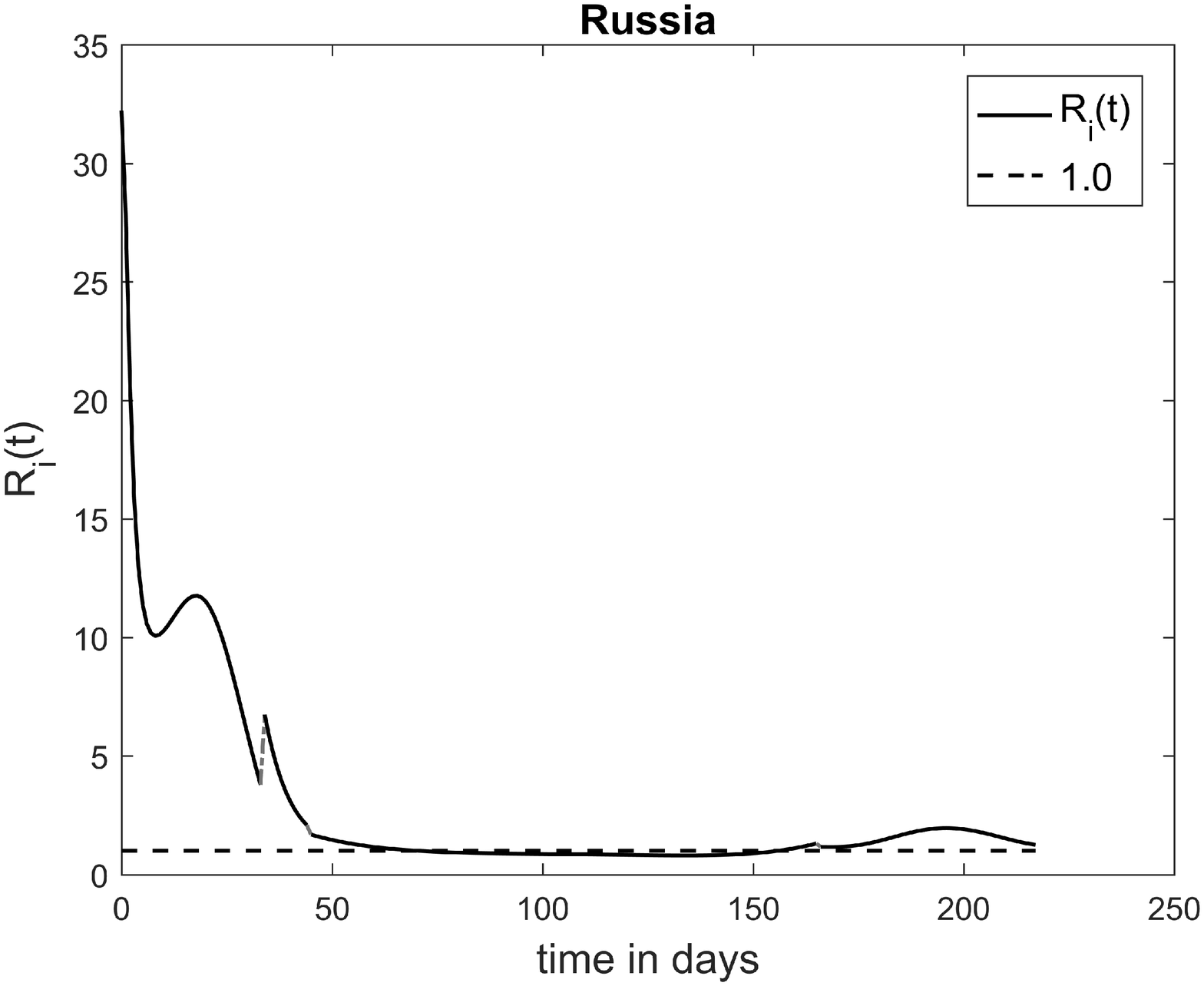}\\
		\includegraphics[scale=0.25]{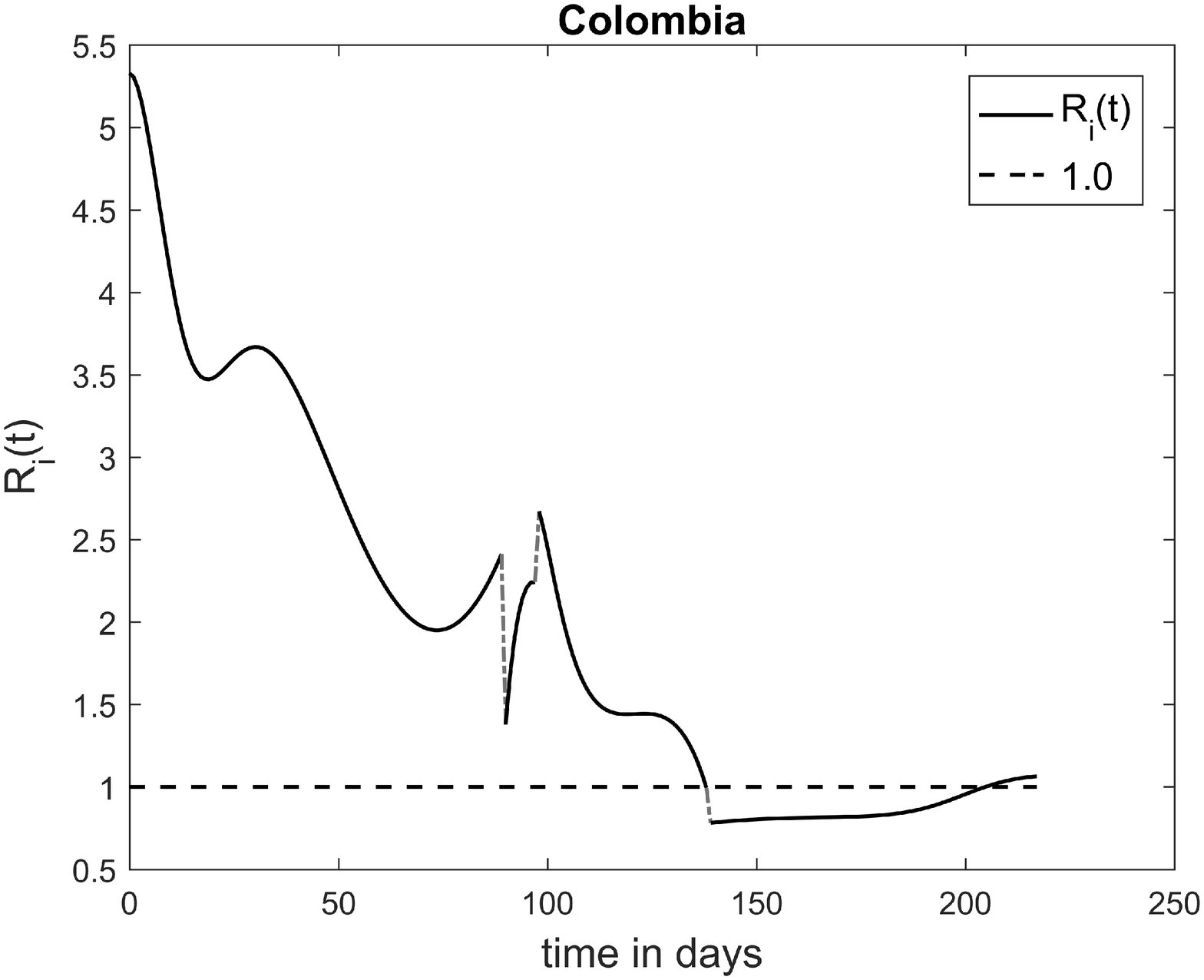}&
		\includegraphics[scale=0.25]{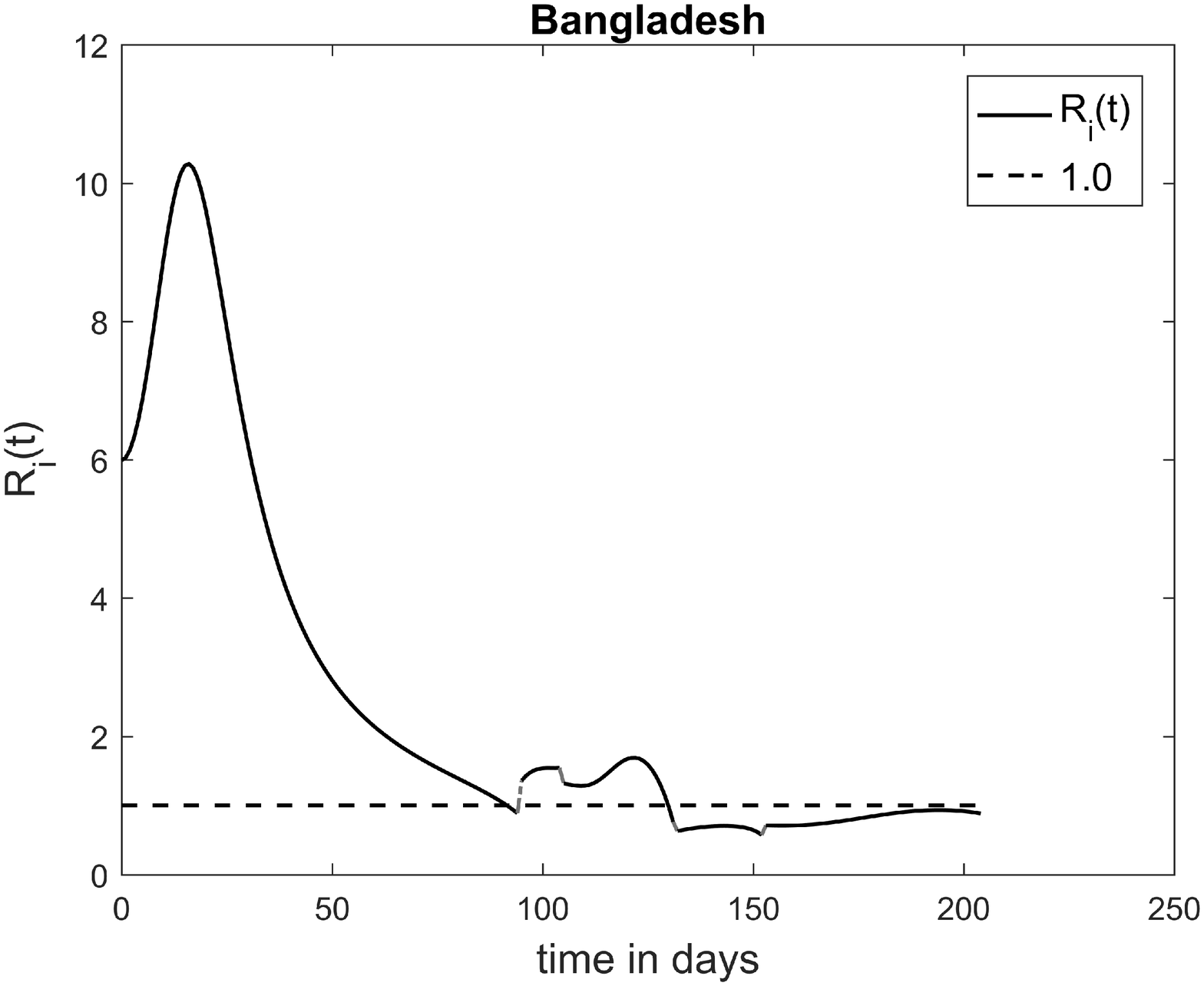}\\
		\includegraphics[scale=0.25]{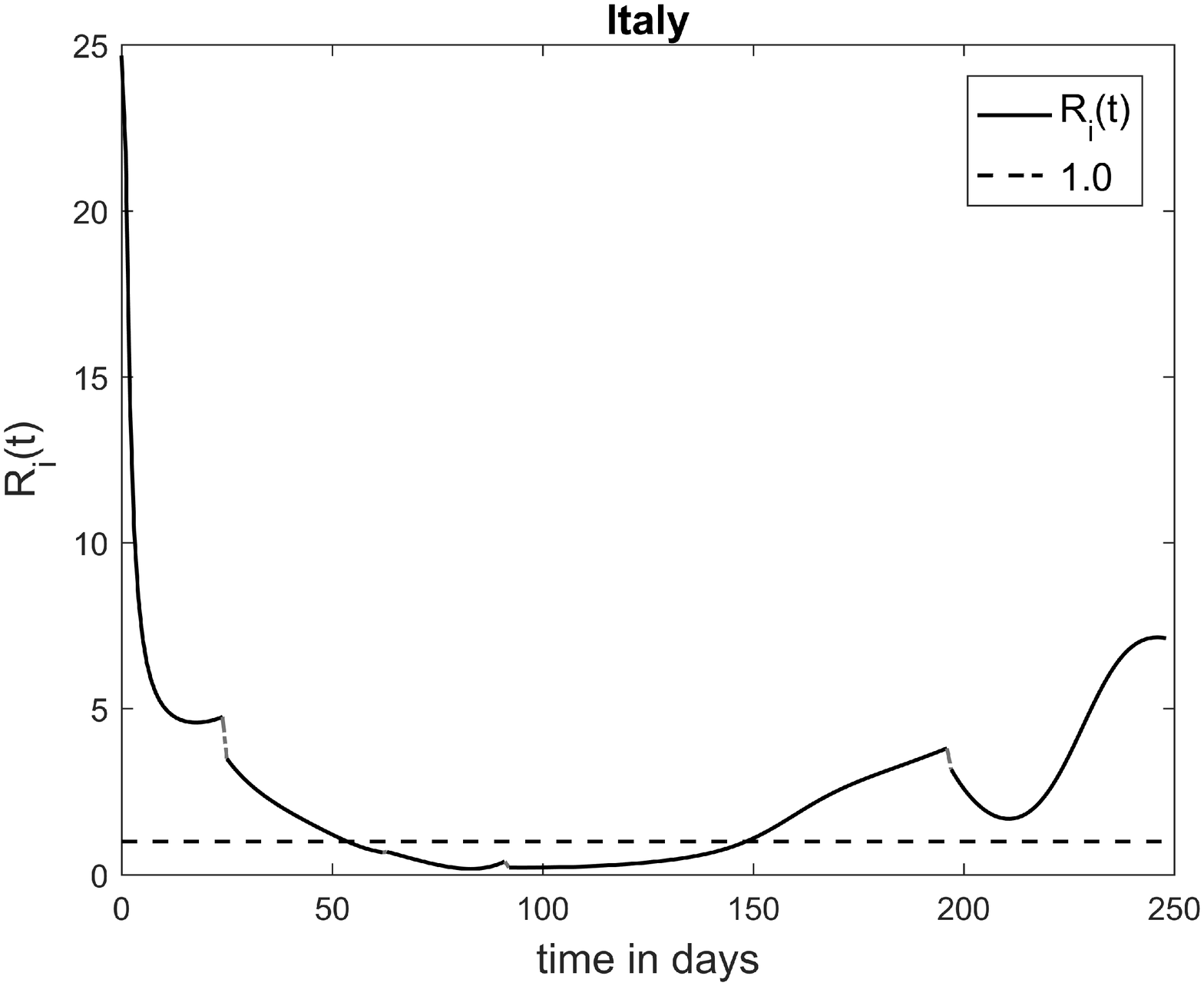}&
		\includegraphics[scale=0.25]{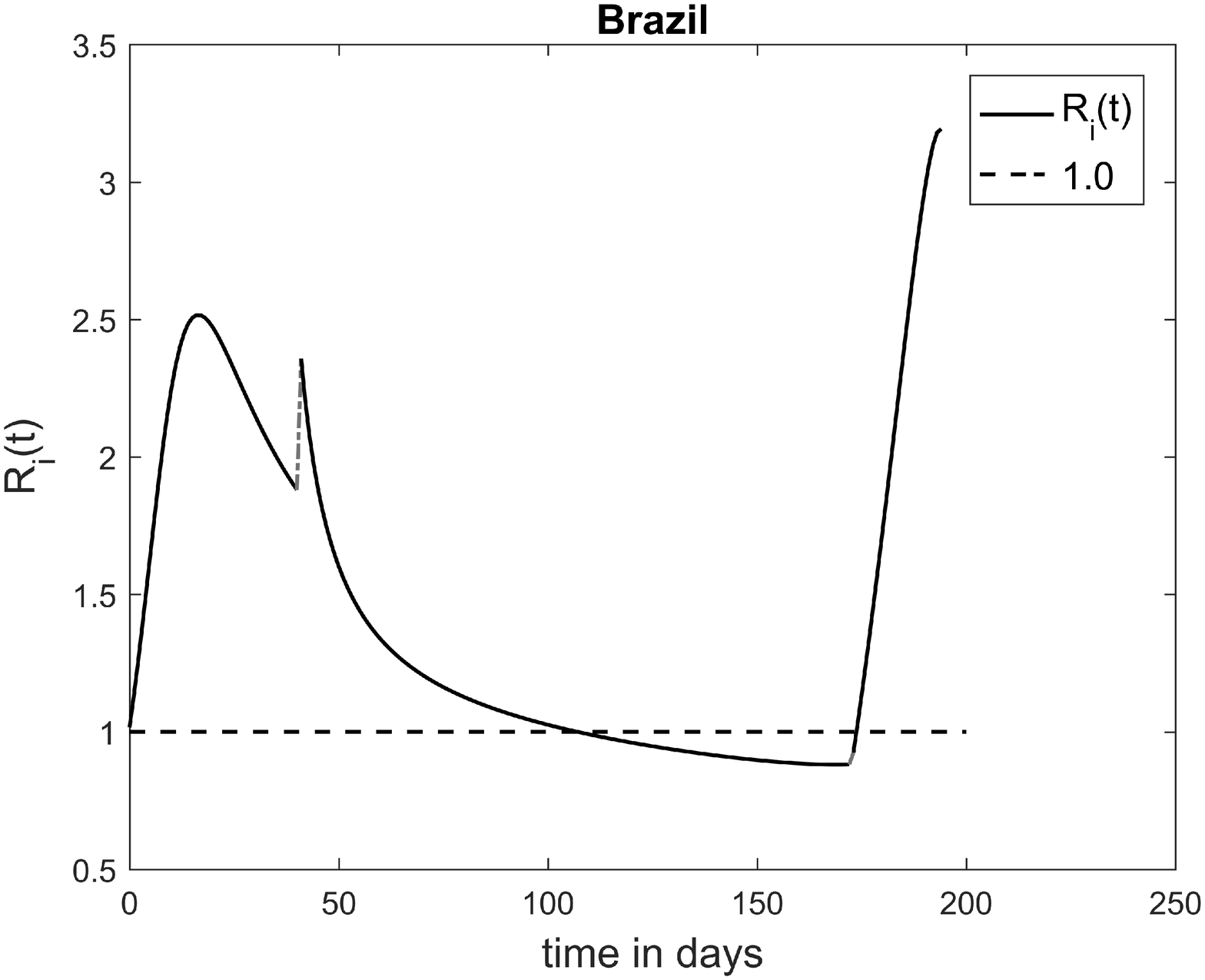}
	\end{tabular}
	\centering
	\caption{{\small Time variation of effective reproduction number ($R_{i}$) of
		various countries.} \label{fig13}}
\end{figure}

Figures \ref{fig12} and \ref{fig13} show how herd immunity ($HI_{i}(t)$), herd
immunity threshold ($HIT(t)$), and effective reproduction number
($R_{i}(t)$) vary with time in different countries. In figures \ref{fig12} and \ref{fig13}, the gray dashed lines again represent
the discontinuities that appear because of the same reason as discussed earlier. We can see
that herd immunity of the population is very close to zero and this
is true for all of the countries. This happens
because the total number of infected cases in a country is very small
compared to the country's total population. Now, there are many reasons behind the small number of infected people
compared to the country's population. First of all, as we mentioned earlier, countries take various actions and restrictions
to reduce the spread of the disease. Besides that, we only work with the reported cases. However, there can be many unreported
cases due to the lack of testing or awareness, which are not taken care of in our model.\cite{underreporting_M, underreporting, Asymp_M, Asymp_2 } Under-repoting
and asymptomatic cases can affect our work and results. This is one of the major drawbacks of our model.

The variation of the herd immunity threshold is different between the countries. This happens because of the different variations of both $\beta$ and $\gamma$ between the countries. Table \ref{ta3} shows
the maximum values of herd immunity threshold ($HIT_{max}$) in different
countries. We can see that the values of $HIT_{max}$ lie between 0.9 to 1.0, except for Peru, Colombia and Brazil. 

\begin{table}[H]
	\tbl{Maximum values of herd immunity threshold ($HIT_{max}$) in different
		countries.}
	{\begin{tabular}{@{}cc@{}} \toprule
			Country\hphantom{000} & $HIT_{max}$ \\ \colrule
			USA\hphantom{000000} & 0.9618  \\
			India\hphantom{00000} & 0.9688  \\
			Peru\hphantom{000000} & 0.7055  \\
			Russia\hphantom{0000} & 0.9690  \\
			Colombia\hphantom{00} & 0.8123  \\
			Bangladesh  & 0.9027 \\
			Italy\hphantom{00000} & 0.9595  \\
			Brazil\hphantom{00000}&0.6027-0.6951\\ \botrule
		\end{tabular} \label{ta3}}
	
\end{table}

Figure \ref{fig13} shows that, initial value of  $R_{i}(t)$ varies so much. $R_{i}(t)$ is not directly calculated
from the data. It is estimated from the model. Now, in our model, initial value of  $R_{i}(t)$ is approximately
related to the $HIT_{max}$. Also, $R_{i}(t)$ is sensitively depends on the $HIT_{max}$. So, when $HIT_{max}$ varies 
from $\sim0.7$ to $0.97$ then $R_{i}(t)$ varies from $\sim3.3$ to $33.3$.

From figures \ref{fig12} and \ref{fig13} we can infer:
\begin{itemize}
\item In USA, $HI_{i}(t)<HIT(t)$ and $R_{i}(t)>1$.
So, infection is still increasing for these countries.

\item Whereas in countries like Colombia, Bangladesh, and India $HI_{i}(t)>HIT(t)$
and $R_{i}(t)<1$. So, the infection is in a downward trend.

\item In Italy, Russia, and Brazil, $HIT(t)$ and $R_{i}(t)$
goes below $HI_{i}(t)$ and 1.0 respectively. So, the infection has started
to decrease. But after some time $HIT(t)$ and $R_{i}(t)$ goes above
$HI_{i}(t)$ and 1.0. Hence the infection starts to rise again and the re-outbreak occurs. A similar thing happens for Peru.
\end{itemize}

In some plots of herd immunity threshold, $HIT(t)$ goes into the negative region. 
This happens because at that time removed rate ($\gamma$) becomes larger than
the infection rate ($\beta$). We can also say that a negative herd
immunity threshold of a population means there is no need for any
immune people to prevent the disease. The disease will automatically
decrease with time.

So from these discussions, we can see that the herd immunity threshold
($HIT(t)$) is a major player than the herd immunity $HI_{i}(t)$
to decrease the COVID-19 infections. As, governments of the countries, intervene in the situation and imposed restrictions on the population,
so, infection rates are reduced. Thus, herd immunity threshold ($HIT(t)$)
is changed and becomes the main quantity which describes the decrease
of infections. Later, if this restrictions are eased
or lifted then a re-outbreak of the disease can happen (like Italy,
Russia, and Brazil).

\section{COVID-19 pandemic in India}

In this section, we will try to analyze the current COVID-19 pandemic situation in 
India. Also we will try to find the possibility to predict the future behavior of the COVID-19 pandemic by effective reproduction number ($R_{i}(t)$)
and herd immunity threshold ($HIT(t)$), as has been studied by several authors\cite{Forecst_covid_india,stat_forecast_India,Logistic_SIR_covid}.

 Figure \ref{fig13} shows that, in India, the effective reproduction number initially
decreases quickly from a large value. But then it starts to decrease
very slowly and becomes less than unity. Also, figure \ref{fig12}
shows that, in India, the herd immunity ($HI_{i}(t)$) of the population
is very low. But herd immunity threshold ($HIT(t)$) goes below
the herd immunity ($HI_{i}(t)$) of the population and then into the
negative region. Thus the infection is decreasing.

If, in future the herd immunity threshold ($HIT(t)$) and the effective
reproduction number ($R_{i}(t)$) remains below than the herd immunity
($HI_{i}(t)$) and 1.0 respectively, then the infection will decrease
continuously. However, the herd immunity of the population is very
low thus a re-outbreak can occur at any time in countries like Russia, Brazil, and Italy.
We can't speculate the time of the re-outbreak because we don't know
how the infection rate ($\beta$) and removed rate ($\gamma$) will
vary with time. So, only in future we will able to get the actual variation
of the COVID-19 pandemic in India and the rest of the world.

\section{Conclusion}

In this paper, we have modified the basic SIR model by considering
the infection rate ($\beta$) and the removed rate ($\gamma$) both
as time-dependent parameters. We have redefined the four important variables,
 basic reproduction number ($R_{0}$), effective reproduction number
($R_{i}(t)$), herd immunity ($HI_{i}(t)$) of the population, and
herd immunity threshold ($HIT$), according to the modifications that
are suggested in the SIR model. Since infection rate and removed rate
are both time-dependent parameters, the herd immunity threshold ($HIT$)
is also time-dependent. 

The salient points of this study are presented below :
\begin{itemize}
	\item The herd immunity threshold ($HIT$) is related to $\beta$ and $\gamma$
	as, $HIT(t)=1-\frac{\gamma(t)}{\beta(t)}$.
	Hence, $HIT(t)$ can be varied with the variation of $\beta$
	and $\gamma$. For example, herd immunity threshold ($HIT$) can be
	reduced by decreasing infection rate, $\beta(t)$ or by
	increasing removed rate, $\gamma(t)$.
	\item An epidemic will start decreasing when $HI_{i}(t)>HIT(t)$,
	which means herd immunity has greater value than the herd immunity
	threshold. Thus we propose two different ways of reducing the spread
	of a disease : (i) by increasing herd immunity of the population over
	the herd immunity threshold, and (ii) by decreasing the herd immunity
	threshold below the herd immunity of the population.
	\item We fit the COVID-19 data of various countries with piece-wise continuous
	functions. From these fitted functions, we have estimated the time
	variation of the infection rate ($\beta$) and the removed rate ($\gamma$)
	of the SIR model. Using these rate parameters ( $\beta$ and $\gamma$),
	we have estimated the variations of the effective reproduction number($R_{i}(t)$),
	the herd immunity of the population ($HI_{i}(t)$), and the herd immunity
	threshold ($HIT(t)$). 
	\item For this COVID -19 pandemic, we see that the herd immunity of the
	various countries', is close to zero. Also, we have found that the
	initial value of the herd immunity threshold is $\sim$0.9. However
	there are many countries, where COVID -19 pandemic is decreasing. This means, the herd immunity threshold went below the
	herd immunity at a certain time.
	\item We observe that the infection is currently decreasing in the countries
	like Colombia, Bangladesh, India and had decreased sufficiently like
	Italy, Russia. and Brazil. In these countries, the herd immunity threshold
	went below the herd immunity of the population at a certain time.
	We have also found that the removed rate $\gamma(t)$ does
	not change much in these countries. However, infection rate $\beta(t)$
	is greatly decreased from the initial value and thus herd immunity
	threshold ($HIT(t)$) is also reduced. We also observe
	that the infection rate $\beta(t)$ can be decreased by
	imposing restrictions, like lockdown, or increasing awareness about
	the disease.
	\item Hence, the countries are trying to prevent COVID -19 pandemic by the
	second method that is taking measures so that $HIT(t)$ goes below
	$HI_{i}(t)$. However, for the same reason the herd immunity of the
	population also remains low. Thus, a country might face a re-outbreak
	of the disease very easily. 
	\item In countries like Italy, Russia and Brazil, we see that there a re-outbreak
	of the COVID-19 disease has occurred now. Here, the herd immunity
	threshold has increased again due to the increment of the infection
	rate ($\beta$) or decrement of the removed rate ($\gamma$). We understand that infection rate ($\beta$) will increase if people start to neglect
	restrictions or if the government decides to lift lockdown. Thus imposing restrictions and lockdown is a good way to
	prevent the spread of the disease. However, there always remains a
	chance of re-outbreak of the disease and this is the major drawback
	of this method.
	\item In COVID-19 pandemic we see that the epidemic is prevented mainly
	by the second method, where herd immunity threshold ($HIT$) goes
	below the herd immunity ($HI_{i}(t)$) of the population. This makes
	herd immunity of the population ($HI_{i}(t)$) insignificant. 
\end{itemize}
In addition to these, we have also tried to find some universal features
between the variations of these components in different countries. 
\begin{itemize}
	\item We have found that initially there is an exponential drop in the infection
	rates of various countries. Infection rates decrease from a value
	of the order of $\sim10^{-1}$ to a value of order $\sim10^{-2}$
	exponentially in a very short period of time. 
	\item Also, we have found that the maximum value of herd immunity threshold
	in different countries has a value between 0.9 to 1.0.
	\item In the end, we have tried to discuss the upcoming COVID-19 behavior
	in India from the point of view of effective reproduction number and
	herd immunity threshold. We see that the COVID-19 pandemic is currently
	decreasing in India. This is because the herd immunity threshold ($HIT$)
	has gone below the herd immunity ($HI_{i}(t)$) of the population.
	Since herd immunity ($HI_{i}(t)$) of the population is close to zero,
	we predict that with the smallest of fluctuations, there is a possibility
	of re-outbreak in India.
\end{itemize}
We would like to conclude that COVID-19 pandemic might be a curse
on humanity, but we can defeat it together. We will hope that our
future will be COVID-19 free very soon.

\section*{Acknowledgment}

We would like to thank Dr. Indrani Bose, Dr. Tapati Dutta, and Dr.
Sujata Tarafdar for their useful comments and suggestions. We also
like to thank the Department of Physics, St. Xavier's College, Kolkata
for providing support during this work. Last but not the least, the authors would like to acknowledge the anonymous
referee for his/her valuable comments and suggestions.

\appendix
\section{Fitted functions}

The fitted functions of the active cases and the removed cases for
the various countries are given below. 

\textbf{\underline{USA}}

For USA the fit function for active cases is,

\begin{eqnarray}
	\begin{array}{lcl}
		a+b\times t+c\times t^{2}+e\times t^{4}+f\times t^{5} & 0\leq t\leq50.\\
		a1+b1\times (t-50)+c1\times (t-50)^{2}+d1\times (t-50)^{3} & 50<t\leq139\\
		\hspace{2cm}+e1\times (t-50)^{4}\\
		a2+b2\times (t-139)+c2\times (t-139)^{2}+d2\times (t-139)^{3}  & 139<t\leq228 \\
		\hspace{2cm}+e2\times (t-139)^{4}	
	\end{array}
	\label{app1}
\end{eqnarray}

The value of the parameters are,

$a=8873,b=2823,c=907.4,e=-0.5671,f=0.006771$.

$a1=9.688\times 10^{5},b1=2.373\times 10^{4},c1=-899.5,d1=18.74,e1=-0.09631$.

$a2=3.083\times 10^{6},b2=2.676\times 10^{4},c2-218.1,d2=0.4034,e2=0.0209$.

The fit function for removed cases is,

\begin{eqnarray}
	\begin{array}{lcl}
		a11+b11\times t+d11\times t^{3}+e11\times t^{4}+f11\times t^{5} & 0\leq t\leq50\\
		a21+b21\times (t-50)+c21\times (t-50)^{2}+d21\times (t-50)^{3} & 50<t\leq139\\
		a31+b31\times (t-139)+c31\times (t-139)^{2}+d31\times (t-139)^{3} & 139<t\leq228
	\end{array}
	\label{app2}
\end{eqnarray}

The value of the parameters are,

$a11=296,b11=187.3,d11=5.973,e11=-0.1104,f11=0.0006998$.

$a21=2.685\times 10^{5},b21=1.175\times 10^{4},c21=-74.96,d21=1.412$.

$a31=1.695\times 10^{6},b31=2.818\times 10^{4},c31=-152.3,d31=1.229$.
\vspace{8pt}

\textbf{\underline{India}}

For India the fit function for active cases is,

\begin{eqnarray}
	\begin{array}{lcl}
		a+b\times t+d\times t^{3}+e\times t^{4}+f\times t^{5}+g\times t^{6} & 0\leq t\leq129\\
		a1+b1\times (t-129)+d1\times (t-129)^{3}+e1\times (t-129)^{4} & 129<t\leq177\\
		\hspace{2cm}+f1\times (t-129)^{5}\\
		a2+b2\times (t-177)+c2\times (t-177)^{2}+d2\times (t-177)^{3} & 177<t\leq223\\
		\hspace{2cm}+e2\times (t-177)^{4}+f2\times (t-177)^{5}
	\end{array}
	\label{app3}
\end{eqnarray}

The value of the parameters are,

$a=486,b=152.8,d=0.4716,e=-0.001239,f=-4.67\times 10^{-5},g=3.364\times 10^{-7}$.

$a1=5.575\times 10^{5},b1=9798,d1=-24.74,e1=1.089,f1=-0.012$.

$a2=1.025\times 10^{6},b2=-1.459\times 10^{4},c2=1265,d2=-58.71,e2=0.9071,f2=-0.004006$.

The fit function for removed cases is,

\begin{eqnarray}
	\begin{array}{lcl}
		a11+b11\times t+c11\times t^{2}+d11\times t^{3}+f11\times t^{5} & 0\leq t\leq106\\
		a21+b21\times (t-106)+c21\times (t-106)^{2} & 106<t\leq223\\
		\hspace{2cm}+d21\times (t-106)^{3}+e21\times (t-106)^{4}
	\end{array}
	\label{app4}
\end{eqnarray}

The value of the parameters are,

$a11=50,b11=13,c11=-2.287,d11=0.2378,f11=1.789\times 10^{-5}$.

$a21=4.724\times 10^{5},b21=2.139\times 10^{4},c21=262.5,d21=5.476,e21=-0.04073$.

\vspace{8pt}

\textbf{\underline{Peru}}

For Peru the fit function for active cases is,

\begin{eqnarray}
	\begin{array}{lcl}
		a+b\times t+d\times t^{3}+e\times t^{4} & 0\leq t\leq45\\
		a1+b1\times (t-45)+c1\times (t-45)^{2}+d1\times (t-45)^{3} & 45<t\leq165\\
		\hspace{2cm}+e1\times (t-45)^{4}+f1\times (t-45)^{5}\\
		a2+b2\times (t-165)+c2\times (t-165)^{2}+d2\times (t-165)^{3}; & 165<t\leq217
	\end{array}
	\label{app5}
\end{eqnarray}

The value of the parameters are,

$a=818,b=55.29,d=1.089,e=-0.01283$.

$a1=4.787\times 10^{4},b1=1948,c1=71.62,d1=-3.747,e1=0.04906,f1=-0.0001964$.

$a2=1.328\times 10^{5},b2=-0.1234,c2=-106.4,d2=1.41$.

The fit function for removed cases is,

\begin{eqnarray}
	\begin{array}{lcl}
		a11+b11\times t+c11\times t^{2}+d11\times t^{3}+e11\times t^{4}+f11\times t^{5} & 0\leq t\leq54\\
		a21+b21\times (t-121)+c21\times (t-121)^{2}+d21\times (t-121)^{3} & 54<t\leq165\\
		\hspace{2cm}+e21\times (t-121)^{4}+f21\times (t-121)^{5}\\
		a31+b31\times (t-165)+c31\times (t-165)^{2}+d31\times (t-165)^{3} & 165<t\leq217
	\end{array}
	\label{app6}
\end{eqnarray}

The value of the parameters are,

$a11=34,b11=132.1,c11=-5.678,d11=1.2,e11=-0.03648,f11=0.0003907$.

$a21=4.938\times 10^{4},b21=1445,c21=117.9,d21=-2.429,e21=0.02103,f21=-5.689\times 10^{-5}$.

$a31=5.687\times 10^{5},b31=7160,c31=3.441,d31=-0.6423$.

\vspace{8pt}

\textbf{\underline{Russia}}

For Russia the fit function for active cases is,

\begin{eqnarray}
	\begin{array}{lcl}
		a+b\times t+d\times t^{3}+e\times t^{4}+f\times t^{5} & 0\leq t\leq33\\
		a1+b1\times (t-33)+c1\times (t-33)^{2}+d1\times (t-33)^{3} & 33<t\leq165\\
		\hspace{2cm}+e1\times (t-33)^{4}+f1\times (t-33)^{5}\\
		a2+b2\times (t-165)+d2\times (t-165)^{3}+e2\times (t-165)^{4} & 165<t\leq217\\
		\hspace{2cm}+f2\times (t-165)^{5}+g2\times (t-165)^{6}
	\end{array}
	\label{app7}
\end{eqnarray}

The value of the parameters are,

$a=1462,b=347.6,d=1.824,e=0.1441,f=-0.003847$.

$a1=1.003\times 10^{5},b1=1.025\times 10^{4},c1=-279.3,d1=3.551,e1=-0.02304,f1=6.014\times 10^{-5}$.

$a2=1.638\times 10^{5},b2=809.1,d2=-1.995,e2=0.3368,f2=-0.008406,g2=5.989\times 10^{-5}$.

The fit function for removed cases is,

\begin{eqnarray}
	\begin{array}{lcl}
		a11+b11\times t+d11\times t^{3}+e11\times t^{4}+f11\times t^{5} & 0\leq t\leq44\\
		a21+b21\times (t-44)+c21\times (t-44)^{2}+e21\times (t-44)^{4} & 44<t\leq217\\
		\hspace{2cm}+f21\times (t-44)^{5}
	\end{array}
	\label{app8}
\end{eqnarray}

The value of the parameters are,

$a11=72,b11=11.13,d11=0.7236,e11=-0.0298,f11=0.0005763$.

$a21=3.494\times 10^{4},b21=7121,c21=20.55,e21=-0.003328,f21=1.51\times 10^{-5}$.

\vspace{8pt}

\textbf{\underline{Colombia}}

For Colombia the fit function for active cases is,

\begin{eqnarray}
	\begin{array}{lcl}
		a+b\times t+d\times t^{3}+e\times t^{4}+f\times t^{5}+g\times t^{6} & 0\leq t\leq89\\
		a1+c1\times (t-89)^{2}+d1\times (t-89)^{3}+e1\times (t-89)^{4} & 89<t\leq138\\
		\hspace{2cm}+f1\times (t-89)^{5}\\
		a2+b2\times (t-138)+d2\times (t-138)^{3} & 138<t\leq217
	\end{array}
	\label{app9}
\end{eqnarray}

The value of the parameters are,

$a=664.5,b=127.4,d=-0.2445,e=0.01288,f=-0.000192,g=9.507\times 10^{-7}$.

$a1=4.967\times 10^{4},c1=341,d1=-19.47,e1=0.4733,f1=-0.004023$.

$a2=1.711\times 10^{5},b2=-2264,d2=0.1695$.

The fit function for removed cases is,

\begin{eqnarray}
	\begin{array}{lcl}
		a11+b11\times t+e11\times t^{4}+f11\times t^{5} & 0\leq t\leq97\\
		a21+b21\times (t-97)+c21\times (t-97)^{2}+d21\times (t-97)^{3} & 97<t\leq217\\
		\hspace{2cm}+f21\times (t-97)^{4}+g21\times (t-97)^{5}
	\end{array}
	\label{app10}
\end{eqnarray}

The value of the parameters are,

$a11=20,b11=29.43,e11=0.0004539,f11=9.232\times 10^{-7}$.

$a21=5.106\times 10^{4},b21=1494,c21=42.39,d21=3.275,f21=-0.05163,g21=0.0002101$.

\vspace{8pt}

\textbf{\underline{Bangladesh}}

For Bangladesh the fit function for active cases is,

\begin{eqnarray}
	\begin{array}{lcl}
		a+b\times t+d\times t^{3}+e\times t^{4} & 0\leq t\leq94\\
		a1+b1\times (t-94)+c1\times (t-94)^{2}+d1\times (t-94)^{3} & 94<t\leq131\\
		\hspace{2cm}+e1\times (t-94)^{4}+f1\times (t-94)^{5}\\
		a2+b2\times (t-131)+d2\times (t-131)^{3}+e2\times (t-131)^{4} & 131<t\leq204
	\end{array}
	\label{app11}
\end{eqnarray}

The value of the parameters are,

$a=416,b=155.1,d=0.4037,e=-0.003359$.

$a1=8.476\times 10^{4},b1=721.6,c1=132.6,d1=-14.52,e1=0.532,f1=-0.006333$.

$a2=1.149\times 10^{5},b2=-1050,d2=0.2373,e2=-0.001881$.

The fit function for removed cases is,

\begin{eqnarray}
	\begin{array}{lcl}
		a11+b11\times t+f11\times t^{5}+g11\times t^{6} & 0\leq t\leq104\\
		a21+b21\times (t-104)+c21\times (t-104)^{2} & 104<t\leq152\\
		\hspace{2cm}+d21\times (t-104)^{3}+e21\times (t-104)^{4}\\
		a31+b31\times (t-152)+c31\times (t-152)^{2} & 152<t\leq204\\
		\hspace{2cm}+d31\times (t-152)^{3}+e31\times (t-152)^{4}
	\end{array}
	\label{app12}
\end{eqnarray}

The value of the parameters are,

$a11=66,b11=31.08,f11=4.687\times 10^{-5},g11=-3.551\times 10^{-7}$.

$a21=1.238\times 10^{5},b21=2559,c21=-121,d21=4.979,e21=-0.05306$.

$a21=2.376\times 10^{5},b21=2819,c21=-48.14,d21=0.8254,e21=-0.005229$.

\vspace{8pt}

\textbf{\underline{Italy}}

For Italy the fit function for active cases is,

\begin{eqnarray}
	\begin{array}{lcl}
		a+b\times t+c\times t^{2}+d\times t^{3} & 0\leq t\leq24\\
		a1+b1\times (t-24)+c1\times (t-24)^{2}+d1\times (t-24)^{3} & 24<t\leq91\\
		\hspace{2cm}+e1\times (t-24)^{4}\\
		a2+b2\times (t-91)+c2\times (t-91)^{2}+d2\times (t-91)^{3} & 91<t\leq196\\
		\hspace{2cm}+e2\times (t-91)^{4}+f2\times (t-91)^{5}\\
		a3+b3\times (t-196)+d3\times (t-196)^{3}+e3\times (t-196)^{4} & 196<t\leq248
	\end{array}
	\label{app13}
\end{eqnarray}

The value of the parameters are,

$a=862.2,b=169.3,c=28.79,d=2.225$.

$a1=4.795\times 10^{4},b1=3586,c1=-23.34,d1=-1.408,e1=0.01436$.

$a2=4.792\times 10^{4},b2=-1598,c2=21.16,d2=0.003922,e2=-0.002203,f2=1.469e-05$.

$a3=3.527\times 10^{4},b3=1036,d3=-2.201,e3=0.08203$.

The fit function for removed cases is,

\begin{eqnarray}
	\begin{array}{lcl}
		a11+b11\times t+d11\times t^{3}+f11\times t^{5}+g11\times t^{6} & 0\leq t\leq62\\
		a21+b21\times (t-62)+c21\times (t-62)^{2} & 62<t\leq196\\
		\hspace{2cm}+d21\times (t-62)^{3}+e21\times (t-62)^{4}\\
		a31+b31\times (t-196)+c31\times (t-196)^{2} & 196<t\leq248\\
		\hspace{2cm}+e31\times (t-196)^{4}+f31\times (t-196)^{5}
	\end{array}
	\label{app14}
\end{eqnarray}

The value of the parameters are,

$a11=62,b11=5,d11=1.051,f11=-0.0003997,g11=3.796\times 10^{-6}$.

$a21=9.586\times 10^{4},b21=4112,c21=-43.59,d21=0.1829,e21=-0.0001741$.

$a31=2.465\times 10^{5},b31=432.2,c31=21.27,e31=-0.01229,f31=0.0002477$.

\vspace{8pt}

\textbf{\underline{Brazil}}

For Brazil the fit function for active cases is, 

\begin{eqnarray}
	\begin{array}{lcl}
		a+b\times t+c\times t^{2} & 0\leq t\leq40\\
		a1+b1\times (t-40)+c1\times (t-40)^{2}+d1\times (t-40)^{3} & 40<t\leq172\\
		\hspace{2cm}+e1\times (t-40)^{4}\\
		a2+b2\times (t-172)+c2\times (t-172)^{2}+e2\times (t-172)^{4} & 172<t\leq194\\
	\end{array}
	\label{app15}
\end{eqnarray}

The value of the parameters are,

$a=1.452\times 10^{4},b=34.86,c=114.2$.

$a1=2.056\times 10^{5},b1=1.069\times 10^{4},c1=-69.16,d1=-0.3049,e1=0.002232$.

$a2=3.925\times 10^{5},b2=-4398,c2=1277,e2=-0.7863$.

The fit function for removed cases is,

\begin{eqnarray}
	\begin{array}{lcl}
	a11+b11\times t+c11\times t^{2}+d11\times t^{3} & 0\leq t\leq40\\
	a21+b21\times (t-40)+c21\times (t-40)^{2}+d21\times (t-40)^{3} & 40<t\leq194\\
	\end{array}
	\label{app16}
\end{eqnarray}

The value of the parameters are,

$a11=1.4\times 10^{4},b11=2155,c11=-54.36,d11=2.628$.

$a21=1.785\times 10^{5},b21=6869,c21=452.7,d21=-1.939$.

\bibliographystyle{ws-ijmpc}
\nocite{*}
\bibliography{References}
\end{document}